\documentclass[a4paper,fleqn,usenatbib]{mnras}

\usepackage{newtxtext,newtxmath}
 
\usepackage[T1]{fontenc}
\usepackage{ae,aecompl}

\usepackage{graphicx}	
\usepackage{amsmath}	
\usepackage{amssymb}	
\usepackage{multirow}

\usepackage[utf8]{inputenc}
\usepackage[export]{adjustbox}
\usepackage{wrapfig}
\usepackage{dutchcal}
\usepackage{braket}
\usepackage{grffile} 
\usepackage{xspace} 

\usepackage{pdflscape}
\usepackage{booktabs}

\usepackage{ulem}

\graphicspath{ {images/} }

\newcommand{\lya}{\mbox{$\rmn{Ly}\alpha$}}
\newcommand{\Lya}{Ly$\alpha$\xspace}

\newcommand{\LLya}{$L_{\rm Ly\alpha}$\xspace}

\newcommand{\flareon}{\texttt{FLaREON}}
\newcommand{\lyart}{\texttt{LyaRT}}
\newcommand{\zelda}{\texttt{zELDA}}
\newcommand{\lasd}{\texttt{LASD}}

\newcommand{\ThinShell}{Thin Shell}

\newcommand{\kms}{\,\ifmmode{\mathrm{km}\,\mathrm{s}^{-1}}\else km\,s${}^{-1}$\fi\xspace}
\newcommand{\vexp}{$V_{\rm exp}$}
\newcommand{\nh}{$N_{\rm H}$}
\newcommand{\ta}{$\tau_a$}
\newcommand{\ew}{$EW_{\rm in}$}
\newcommand{\w}{$W_{\rm in}$}

\newcommand{\dlt}{$\Delta \lambda _{\rm True}$}

\newcommand{\wg}{$W_{\rm g}$}
\newcommand{\dl}{$\Delta \lambda_{\rm Pix}$}
\newcommand{\sn}{$S/N_p$}

\newcommand{\llya}{$L_{\rm Ly\alpha}$} 
\newcommand{\ewl}{$EW_{\rm Ly\alpha}$}

\newcommand{\outr}{Outflow:R\xspace}
\newcommand{\outm}{Outflow:M\xspace}
\newcommand{\inm}{Inflow:M\xspace}

\title [Ly$\alpha$ line profile fitting]{zELDA: fitting Lyman-alpha line profiles using deep learning.   }

\author[S. Gurung-L\'opez. et al.]{
Siddhartha Gurung-L\'opez$^{1,2,3,4}$,\thanks{E-mail: sidgurung@cefca.es}
Max Gronke,$^{5,6}$
Shun Saito,$^{3,7}$
Silvia Bonoli,$^{8}$ 
\newauthor
and \'Alvaro A. Orsi.$^{9}$
\\
$^{1}$ Observatori Astron\`omic, Universitat de Val\`encia, C/ Catedr\'atico Jos\'e Beltran, 2, 46980 Paterna (Val\`encia), Spain\\
$^{2}$ Departament d'Astronomia i Astrof\'isica, Universitat de Val\`encia, 46100-Burjassot, Val\`encia, Spain\\
$^{3}$ Institute for Multi-messenger Astrophysics and Cosmology, Department of Physics\\
Missouri University of Science and Technology, 
1315 N. Pine St., Rolla MO 65409, USA\\
$^{4}$ Centro de Estudios de F\'isica del Cosmos de Arag\'on, Plaza San Juan 1, piso 2, Teruel, 44001, Spain. \\
${}^{5}$ Department of Physics \& Astronomy, Johns Hopkins University, 
    Bloomberg Center, 3400 N. Charles St., Baltimore, MD 21218, USA\\
${}^{6}$ Hubble fellow\\
$^{7}$Kavli Institute for the Physics and Mathematics of the Universe (WPI), Todai Institutes for Advanced Study,\\
the University of Tokyo, Kashiwanoha, Kashiwa, Chiba 277-8583, Japan\\
$^{8}$ DIPC, Manuel Lardizabal Ibilbidea, 4, 20018 San Sebastian, Spain. \\
$^{9}$ PlantTech Research Institute Limited. South British House, 4th Floor, 35 Grey Street, Tauranga 3110, New Zealand\\
}

\date{Accepted XXX. Received YYY; in original form ZZZ}

\pubyear{2020}

\begin{document}
\label{firstpage}
\pagerange{\pageref{firstpage}--\pageref{lastpage}}
\maketitle

\begin{abstract}

We present \texttt{zELDA}  (redshift Estimator for Line profiles of Distant Lyman-Alpha emitters), an open source code to fit Lyman-$\alpha$ (Ly$\alpha$) line profiles. The main motivation is to provide the community with an easy to use and fast tool to analyze Ly$\alpha$ line profiles uniformly to improve the understating of Ly$\alpha$ emitting galaxies. \texttt{zELDA} is based on line profiles of the commonly used `shell-model' pre-computed with the full Monte Carlo radiative transfer code \texttt{LyaRT}. Via interpolation between these spectra and the addition of noise, we assemble a suite of realistic Ly$\alpha$ spectra which we use to train a deep neural network.We show that the neural network can predict the model parameters to high accuracy (e.g., $\lesssim 0.34$ dex HI column density for $R\sim 12000$) and thus allows for a significant speedup over existing fitting methods.As a proof of concept, we demonstrate the potential of \texttt{zELDA} by fitting 97 observed Ly$\alpha$ line profiles from the \texttt{LASD} data base. Comparing the fitted value with the measured systemic redshift of these sources, we find that Ly$\alpha$ determines their rest frame Ly$\alpha$ wavelength with a remarkable good accuracy of $\sim 0.3$\AA{} ($\sim 75\; {\rm km/s}$).Comparing the predicted outflow properties and the observed Ly$\alpha$ luminosity and equivalent width, we find several possible trends. For example, we find an anticorrelation between the Ly$\alpha$ luminosity and the outflow neutral hydrogen column density, which might be explained by the radiative transfer process within galaxies. 
\end{abstract}




\begin{figure*} 
        \includegraphics[width=6.9in]{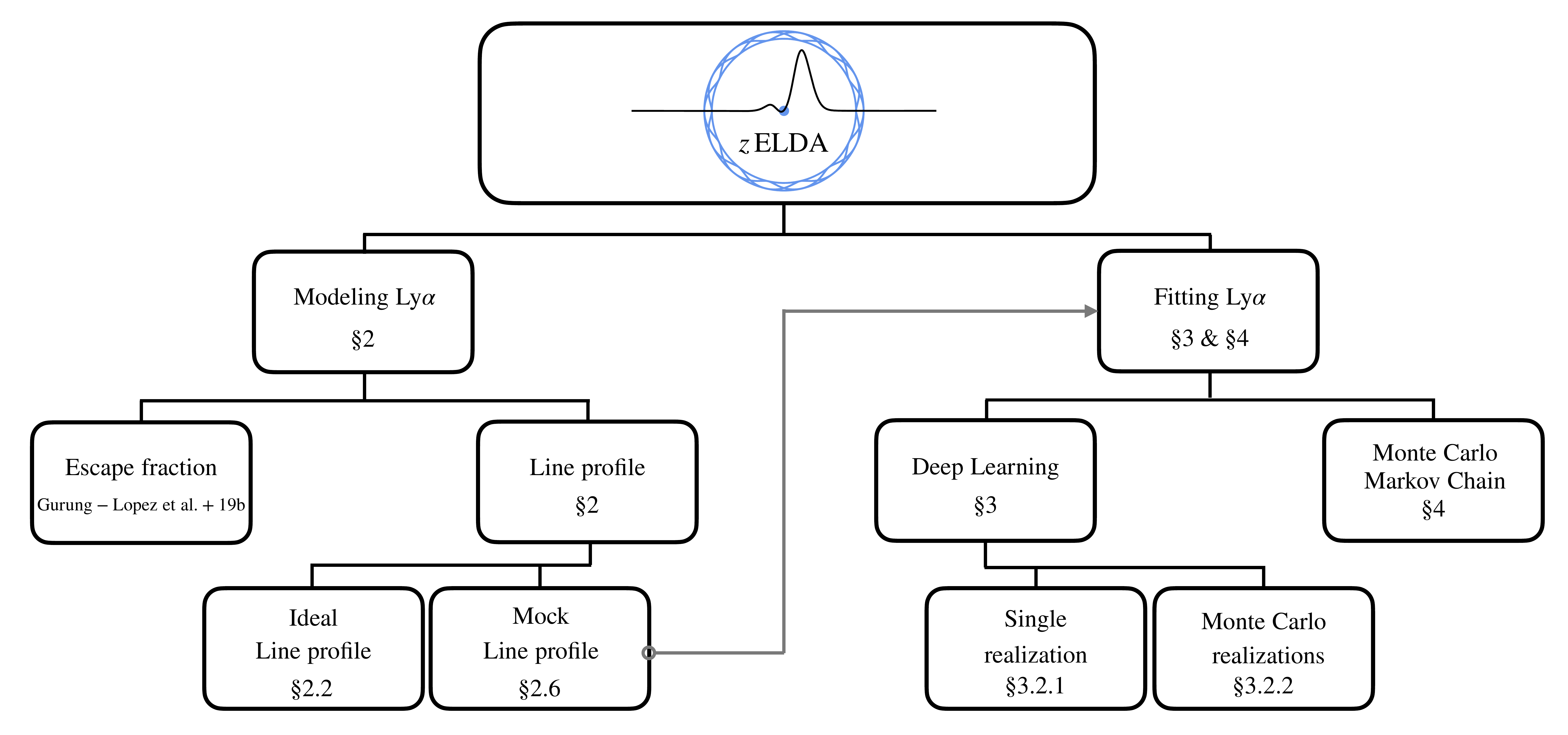}%
        \caption{ Sketch of \zelda's content. \zelda\ source code is publicly available at \url{https://github.com/sidgurun/Lya_zelda} while installation instructions and tutorials can be found at  \url{https://zelda.readthedocs.io/en/latest/}. }
        \label{fig:zelda_sketch}
        \end{figure*}

\section{Introduction}\label{sec:intro}
The Lyman-$\alpha$ (\Lya) emission line of neutral hydrogen plays a prominent role in astrophysics. 
Since it is the first transition of the most abundant element in the Universe, it is extremely bright and, thus, used in large galaxy surveys as well as to detect even the most distant galaxies \citep[for a recent review, see,][]{OuchiObservationsLymana2020}.
Specifically, surveys such as {\it Hobby-Eberly Telescope Dark Energy Experiment} (HETDEX, $\sim 0.8$ million \Lya emitting galaxies at $1.9<z<3.5$; \citealp{hill08,Farrow_2021,Weiss_2021}),  \textit{Systematic Identification of LAEs for Visible Exploration and Reionization Research Using Subaru HSC} (SILVERRUSH, $\sim 2,000$ at $6<z<7$; \citealp{Ouchi2018a,Kakuma_2019}), \textit{MUSE WIDE} ($\sim 500$ at $3\lesssim z\lesssim 6$; \citealp{Herenz2017,Urrutia2019A&A...624A.141U,Caruana_2020}) or the \textit{Javalambre Photometric Local Universe Survey} \citep[J-PLUS, $\sim 14,500$ at $2\lesssim z\lesssim 3.3$;][]{spinoso_2020}  have increased the pure number of detect \Lya emitting galaxies at every redshift by orders of magnitude.

Beyond a pure tool for detecting galaxies, the \Lya line is, however, also an invaluable tracer of cold gas composition and kinematics. 
This is because \Lya is a resonant line which means that \Lya photons get absorbed and re-emitted by neutral hydrogen atoms. 
In fact, the re-emission process occurs on such a short time scale ($\sim 10^{-8}\,$s), this process is usually referred to a \textit{scattering}. 
Since for typical \Lya emitting galaxies, the hydrogen column density is $N_{\rm HI}\sim 10^{17}-10^{20}\,\mathrm{cm}^{-2}$, and the scattering cross section at line center (for gas with $T\sim 10^4\,$K) is $\sigma\sim 6\times 10^{-14}\,\mathrm{cm}^{2}$, \Lya photons typically scatter thousands of times before they reach the observer \citep[for a review, see,][]{dijkstra17}.
Scatterings occur because of the density and kinematics of the neutral gas at that point, and each scattering alters the \Lya photon's frequency (mostly due to Doppler boosting).
This implies firstly that the redshift of the emergent \Lya line is not corresponding to the true systemic redshift of the source $z_{\rm sys}$. 
Thus, estimating $z_{\rm sys}$ using \Lya is more complex than with other nebular emission lines such as $\rm H\alpha$. 
The community has made a great effort to learn how to estimate the systemic redshift solely from \lya\ line profile \citep[e.g.][]{steidel10,Rudie_2012,Verhamme:2018aa,GurungLopez_2019b,Byrohl_2019,Runnholm_2020}. 
This is particularly important for measuring galaxy clustering, as the redshift is crucial to determine the 3D position of a source in the Universe. 
In \cite{gurung_2020b} we explore the usage of neural networks to extract the systemic redshift from \lya{} line profiles, obtaining accurate results on simulated data. 
This work is a continuation of our previous study. Here, we extend upon this and use deep learning to model the full line shape including the systemic redshift of the source.

The second implication of the complex resonant radiative transfer is that, information about the density and kinematic structure of the cold gas is embedded in the \Lya observables such as the \Lya spectra, surface brightness profiles and polarization.
This is particularly interesting because this cold gas plays a key role in a range of astrophysical processes -- but is often hard to probe otherwise.
For instance, in the circumgalactic medium, the cold gas is a reservoir of gas for future star formation, and can trace in- and outflows of galaxies \citep{Bresolin_2017,Tumlinson2017}. 
In this context, the detection of glowing \Lya halos surrounding star-forming galaxies has opened a new pathway to probe this cold gas directly \citep{steidel11,Wisotzki2016}. 
Furthermore, the study of the variation of \Lya spectra in space which provides insight into the connection between galaxies and their surrounding medium \citep{Rauch_2015,Leclercq_2017,Erb_2018} -- a direction of research which has been facilitated by integral field spectrographs such as \textit{MUSE}  \citep{bacon10} and \textit{KCWI} \citep{Martin_2010}.

Another important application of \Lya observables is as a proxy for ionizing photon escape. 
As both types of radiation are directly susceptible to intervening neutral hydrogen -- but the direct detection of Lyman-continuum (LyC) photons is hard (or at $z\gtrsim 4$ impossible due to the increasingly neutral IGM), \Lya plays a deciding role both for observational as well as theoretical studies focusing on ionizing escape mechanisms \citep[e.g.][]{verhamme_2014,Dijkstra_2016}. 
It is now well established that \Lya observables such as the equivalent width or the spectral peak separation correlate with the ionizing escape fraction \citep{Steidel_2018,Izotov_2021}.

While this complex radiative transfer process is a fortunate fact observationally, modeling it is non-trivial and only few analytical solutions exist \citep[e.g.,][]{neufeld90,dijkstra06}. Due to its complexity, typically Monte-Carlo radiative transfer codes are being employed which allows for flexibility in the HI geometry -- but on the other hand show slow convergence. To model the observed \Lya spectra, it is common to use relatively simple geometries, thus reducing the number of free parameters of the model and hence to limit the suite of synthetic spectra to $\lesssim 100,000$. 
These synthetic spectra can reproduce observed ones quite accurately -- which is maybe surprising given their simplicity. 
For instance, \citet{ahn03} introduced the `shell-model' consisting of a moving spherical shell which surrounds a radiation source. 
Since then, the `shell-model' has been often used to fit observed \Lya spectra and learn about the HI distribution and the true systemic redshift of the \Lya emitting source \citep[e.g.,][]{Verhamme_2007,Schaerer2011,Gronke2017}. 
Other models used to systematically fit \Lya spectra include moving slabs \citep{schaerer08}, clumpy multiphase model \citep{Li2020,Li2021} or spherically symmetric halos \citep{Song_2020}.

Between the model parameters multiple degeneracies exist which can lead to a multimodal and non-Gaussian likelihood and makes the fitting process quite computationally expensive. As stated above, in the near future the number of observed \Lya spectra will increase dramatically and a modern, fast pipeline to model them is required.
In this work, we adopt a machine learning algorithm to fit \Lya spectra to obtain this goal. 

While we will focus on the most commonly used model -- the `shell model', this work can easily be extended to include other geometries. Note also that while the physical meaning of the model parameters is frequently discussed in the literature \citep{Gronke2017a,Orlitova_2018} -- and in fact it has been shown that a simple mapping to, e.g., the line-of-sight HI column density is not possible \citep{Vielfaure_2020}, this discussion is not part of this work. 
However, we hope that future studies targeted to this will benefit from our new fitting pipeline.

In this work we present \zelda, an open source Python package to model and fit \lya\ line profiles, as well as to predict \lya\ escape fractions from outflows. An sketch of \zelda's content is displayed in Fig.\ref{fig:zelda_sketch} with the sections where each feature is presented. \zelda\ is based on \flareon\ \citep{GurungLopez_2019b} and \lyart\ \citep{orsi12}. In fact, the computation of the \lya\ escape fractions equivalent to the approach used in \flareon\ and therefore we do not discuss it in this work. For the modeling of \lya\ line profiles and escape fractions, several outflow geometries are included. In particular, \zelda\ includes the procedures for modeling ideal line profiles and mock line profiles that replicate the typical observational limitations that are present in real spectra. The main motivation for modeling ideal line profiles and escape fractions is to populate large simulations with \lya\ emitters \citep[as was, e.g., done in][]{garel12,GurungLopez_2019a,GurungLopez_2020a}. Then, the mock line profiles are useful to understand possible biases, for example in the redshift determination, using simulations \citep{gurung_2020b}. The other main goal of the production of mock spectra is to fit observational data. For the fitting we have included several methodologies, among them, a Monte Carlo Markov Chain approach and a neural network procedure.

\zelda\ is publicly available and ready to use\footnote{\url{https://github.com/sidgurun/Lya_zelda}}. \zelda\ contains all the necessary scripts to reproduce all the results presented in this work. Documentation and several tutorials on how to use \zelda\ are also available\footnote{\url{https://zelda.readthedocs.io/en/latest/}}.

This work is organised as follows: in section \S\ref{s:modeling} we describe the outflow geometry and the computation of the \lya\ line profiles. Then, we describe the architecture of the deep neural network and how we compute the outflow properties and redshift in \S\ref{s:DNN_build}. In \S\ref{s:mcmc} we describe the Monte Carlo Markov Chain implemented in \zelda. Then, we compare the accuracy and computational cost of these methodologies in \S\ref{s:discussion}. In \S\ref{s:observation} we analyse 97 observed \lya\ line profiles with \zelda\ and study the correlations between these. Finally, we make our conclusions in \S\ref{s:conclusions}.

We use rest frame length units to quantify the accuracy in determining the rest frame \lya\ wavelength. However, it is common in the literature to provide this quantity in velocity units \citep[e.g.][]{Verhamme:2018aa,Byrohl_2019}. A wavelength interval $\Delta\lambda$ nearby to the \lya\ wavelength can be express in velocity units as $\Delta v = c \Delta\lambda/\lambda_{\rm Ly\alpha}\sim(247km/s)\times \Delta\lambda / 1$\AA{}, where $c$ is the speed of light and $\lambda_{\rm Ly\alpha}\approx 1215.67$\AA{}.

\section{ Modeling \lya\ line profiles }\label{s:modeling}

        \zelda\ is based on the radiative transfer Monet Carlo code \lyart\ \citep{orsi12}. In summary, \zelda\ computes the \lya\ line profiles from a pre-computed grid of \lyart\ outputs, where the full computation of the radiative transfer of \lya\ photons is made. In this section we detail the outflow and inflow gas geometry set in \lyart (\S\ref{ss:gas_geometry}), the grid specs (\S\ref{ss:grid}), a validation sample that we will be used to test our methodologies (\S\ref{ss:sample}), the accuracy between the line profiles predicted by \zelda\ and those computed by \lyart{} (\S\ref{ss:accuracy_interp}) and how realistic line profiles are generated (\S\ref{ss:mocking}).
    
    \subsection{Outflow gas geometry}\label{ss:gas_geometry}

         \zelda , as \flareon 's successor, includes the three outflow models used in \flareon\ as detailed in \cite{GurungLopez_2019b}. In addition to those, we developed an new \ThinShell\ outflow model for \zelda , in which we focus this paper. The \ThinShell\ model is widely used in the literature \citep[e.g.][]{zheng02,ahn04,verhamme06,orsi12,Gronke2017}. Both, the \ThinShell\ model from \flareon\ and \zelda\ use the same gas distribution, i.e., an isothermal homogeneous spherical thin layer of neutral hydrogen described by an inner and an outer radius $R_{in}$ and $R_{out}$ respectively, with $R_{in}/R_{out}=0.9$. We fix the gas temperature at $T=10,000\,K$.\footnote{While the (effective) temperature has an effect on resonant line transfer, constraining it via spectral profile fitting is difficult and often not possible \citep{Gronke_2015} and we, thus, chose a natural temperature for HI but note that other temperatures could be included in the future.} The neutral hydrogen column density of the gas geometry is \nh. The gas has an homogeneous radial bulk velocity \vexp. Also, the dust optical depth is set to 
            $\tau_a = (1 - A_{\rm Ly\alpha}) \frac{Z}{Z_{\odot}}E_{\odot}N_H$,
        where $E_{\odot} = 1.77 \times 10^{-21}{\rm cm}^{-2}$ is the ratio $\tau_a/N_H$ for solar metallicity, $A_{\rm Ly\alpha}=0.39$ is the albedo at the \lya\ wavelength, $Z_{\odot}=0.02$ \citep{granato00}. 
        
        We have included a new \ThinShell\ model that modifies the intrinsic spectrum injected into the gas cloud with respect the \ThinShell\ model already existing in \flareon. On one hand, we conserved the \ThinShell\ model introduced in \flareon, where the intrinsic spectrum is monochromatic photons exactly at \lya.  On the other hand, in the new \zelda 's \ThinShell\ model we inject a flat spectrum in wavelength units of $f_{\lambda}^{ In}$ with a Gaussian of full width half maximum \w\  centered in \lya\ with equivalent width \ew. From now on, we will refer the new \ThinShell\ model simply as the \ThinShell\ model. 
        
        \zelda\ also incorporates an inflow version of this geometry. In the inflow, the gas distribution is the same, but the bulk velocity is below 0. We describe how the inflow line profiles are computed from the outflow line profiles in Appendix \ref{s:inflow_making}.
        
        \begin{figure*} 
        \includegraphics[width=6.9in]{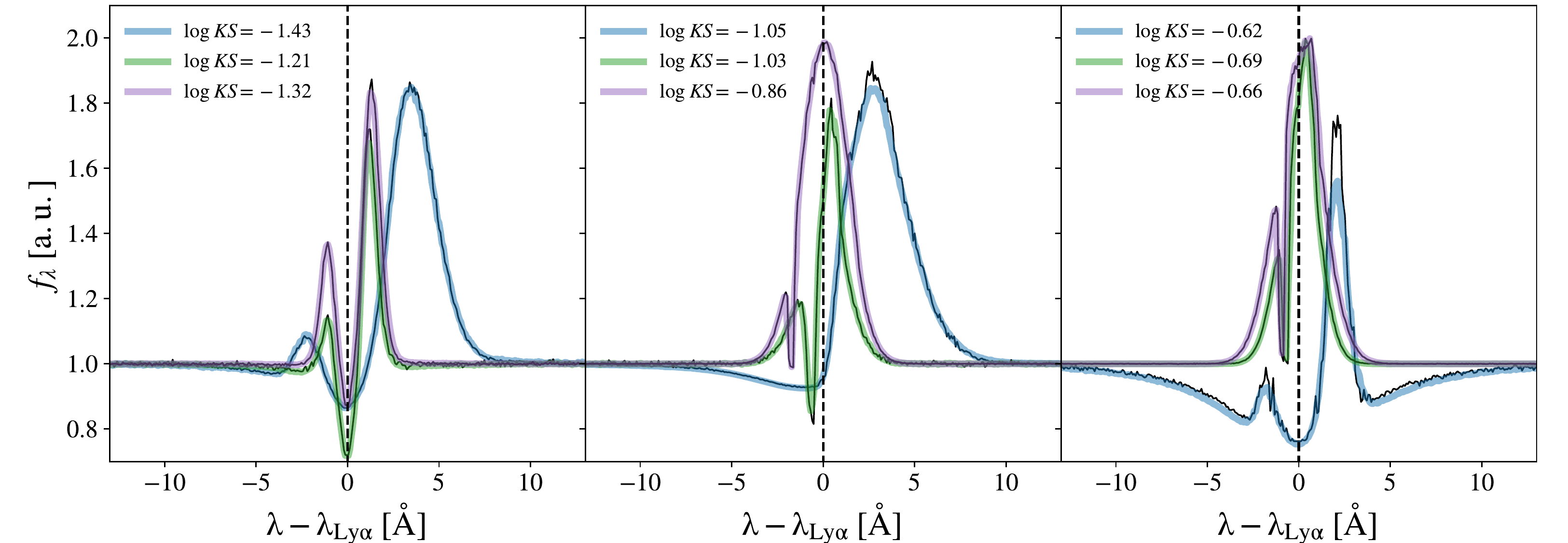}%
        \caption{ Comparison between the \lya\ line profiles computed using \lyart\ (black) and \zelda\ (different colors). These nine examples were randomly chosen from the 2000 random realization of  \lyart . In the left panel we choose the lines which comparison had $-2<\log \; KS < -1.2$, while in the middle panel we imposed $-1.2<\log \; KS < -0.7$ and in the right panel $-0.7<\log \; KS$. The vertical dashed line indicated the \lya\ wavelength. }
        \label{fig:KS_examples}
        \end{figure*}
        
        \subsection{ Grid of \lya\ line profiles }\label{ss:grid}
        
        There are a total of 5 free variables in our \ThinShell\ model: \vexp , \nh , \ta , \ew\ and \w . In practice, we run the RTMC \lyart\ for different values of \vexp , \nh\ and \ta, while the line profiles with different values of \ew\ and \w\ are obtained via post-processing. 
        We run \lyart\ in all the \{\vexp , \nh , \ta\} combinations of
        \begin{flalign}
            V_{\rm exp} [{\rm km\,s^{-1}}] &= [ 0 , 10 , ... , 90 , 100 , 150 , ... , 950 , 1000], && \nonumber  \\
            \log N_H [{\rm cm^{-2}}] &= [ 17.0 , 17.25 , ... , 21.25 , 21.5 ], && \nonumber \\
            \log \tau _a &= [-4.0 , -3.5 , ... -0.5 , 0.0 ] .&&
        \end{flalign}
    
        For each value of \vexp , \nh\ and \ta\, we generate $2\times10^7$ photons with a uniform random distribution of frequency in Doppler units from $x=-1000$ to $x=1000$ with $x = ( \nu - \nu_{\rm Ly\alpha})/\Delta \nu_{D}$ where $\nu$ is the frequency of the photons, $\nu_{\rm Ly\alpha}$ is the \lya\ frequency and $\Delta \nu _{D} = v_{\rm th} \nu_{\rm Ly\alpha} / c$, where $c$ is the speed of light and $v_{\rm th} = \sqrt{2k_B T / m_p}$, where $k_b$ is the Boltzman constant and $m_p$ is the mass of the proton. 
            
        The emerging flux density in Doppler units, $f_x(x , V_{\rm exp}, N_H , \tau_{\rm a})$, can be transformed to flux density in wavelength units through $f_{\lambda} = c \Delta \nu _ D f_x / \lambda^2$, where $\lambda$ is the wavelength. 
        This would give us a tilted $f_{\lambda}$ given the $\lambda$ dependency in the transformation. 
        Then, to mimic a flat input spectrum in wavelength units, we weight each photon by its wavelength $\lambda_p^2$, which give us $f_{\rm \lambda , flat}^{\rm Out}$. 
        
        Once we have the photons for each outflow configuration provided by \lyart , we emulate as a post-process the injection of different intrinsic spectra in the gas geometry. 
        For this goal, we use the recorded input wavelength, $\lambda_{\rm In}$ (uniform in wavelength), and output wavelength $\lambda_{\rm Out}$ of the photons, that contains the RT effects. 
        The \lya\ line profile emerging from an outflow with intrinsic spectrum $f_{\lambda}^{\rm In}$ is computed as the probability distribution function of $\lambda_{\rm Out}$, in which each photon is weighted by $f_{\lambda}^{\rm In} (\lambda_{\rm In}) $.
        
        We do this process for 30 bins of intrinsic line width, \w[\AA{}]$\in$[0.01, 0.05, 0.1, 0.15, 0.2, 0.3, 0.4, 0.5, 0.6, 0.7, 0.8, 0.9 , 1., 1.2, 1.4, 1.6, 1.8, 2., 2.2, 2.4, 2.6, 2.8, 3., 3.25, 3.5, 3.75, 4., 5.25, 5.5, 5.75, 6. ] and for 20 evenly spaced bins from $\log$\ew[\AA{}] = -1.0 to 3.0 . These, in addition to the 29 bins in \vexp, 18 in \nh\ and 10 in \ta , there are in total 3,132,000 grid nodes.
        Finally, in order to compute \lya\ line profiles in arbitrary locations inside the grid volume, we perform linear interpolation between the grid nodes. 
        This is shown in detail in Appendix \ref{s:interpolation}.
        
        \subsection{ Validation sample}\label{ss:sample}
        
        To quantify the performance of the different algorithms implemented in \zelda\ it is necessary to compare with \lya\ line profiles that are not part of the grid used for the interpolation. 
        
        We run additional random 200 combinations of \{\vexp , \nh , \ta \} of \lyart\ with $2\times10^6$ photons for each. 
        These random configurations were chosen with a latin hypercube sampling to homogeneously cover the parameter space \{$\log$\vexp , $\log$\nh , $\log$\ta \}. Then, for each of these configurations, we computed 10 uniformly random combinations of \{$\log$\ew , $\log$\w \}, making a total of 2000 \lya\ line profiles that are independent of \zelda 's line profile grid. 
        Each of \{\vexp, \nh, \ta, \ew, \w \} in this sample covers its full range defined in \S\ref{ss:grid} . 
        For this line profile sample, the full radiative transfer is computed, so they represent the 'real' line profiles, while \zelda 's prediction for these configurations is just an approximation. 
        
        Through this work, we mainly measure the accuracy, focusing on the outflow model, i.e., $V_{\rm exp}>0$. 
        However, as we show in Appendix \ref{s:accuracy_comparisons}, the methodologies that we explore have the same accuracy for the inflow model ($V_{\rm exp}<0$).
        
        \subsection{Accuracy of the interpolation scheme}\label{ss:accuracy_interp}
        
        In this section, we assess the performance of our \lya\ line profile computations that are not included in our grid. In order to quantify the agreement between the \lyart 's and \zelda 's output we compute the Kolmogórov-Smirnov (KS) estimator, which is defined as the maximum separation between two cumulative distributions. 
        
        For illustration, we also show nine random individual examples of the comparison between \lyart\ (black) and \zelda\ (colors) in Fig.~\ref{fig:KS_examples}. In the right panel there are some examples showing the line profiles with the typical highest values of KS, i.e, the cases where \zelda 's predictions are the least accurate. Additionally, in the middle and left panels we show line profiles with intermediate and low KS values, respectively. In these cases, we confirm that the agreement between \lyart 's and \zelda 's outputs is excellent. In some cases it is apparent that the \zelda\ spectra have a higher signal to noise than the \lyart{} spectra. This difference comes from the number of photons used to measure the line profile. For these random runs we injected $2\times10^6$ photons in \lyart\ while in the configurations used for building the grid for \zelda\ we used $2\times10^7$. 
        
\begin{figure} 
        \includegraphics[width=3.4in]{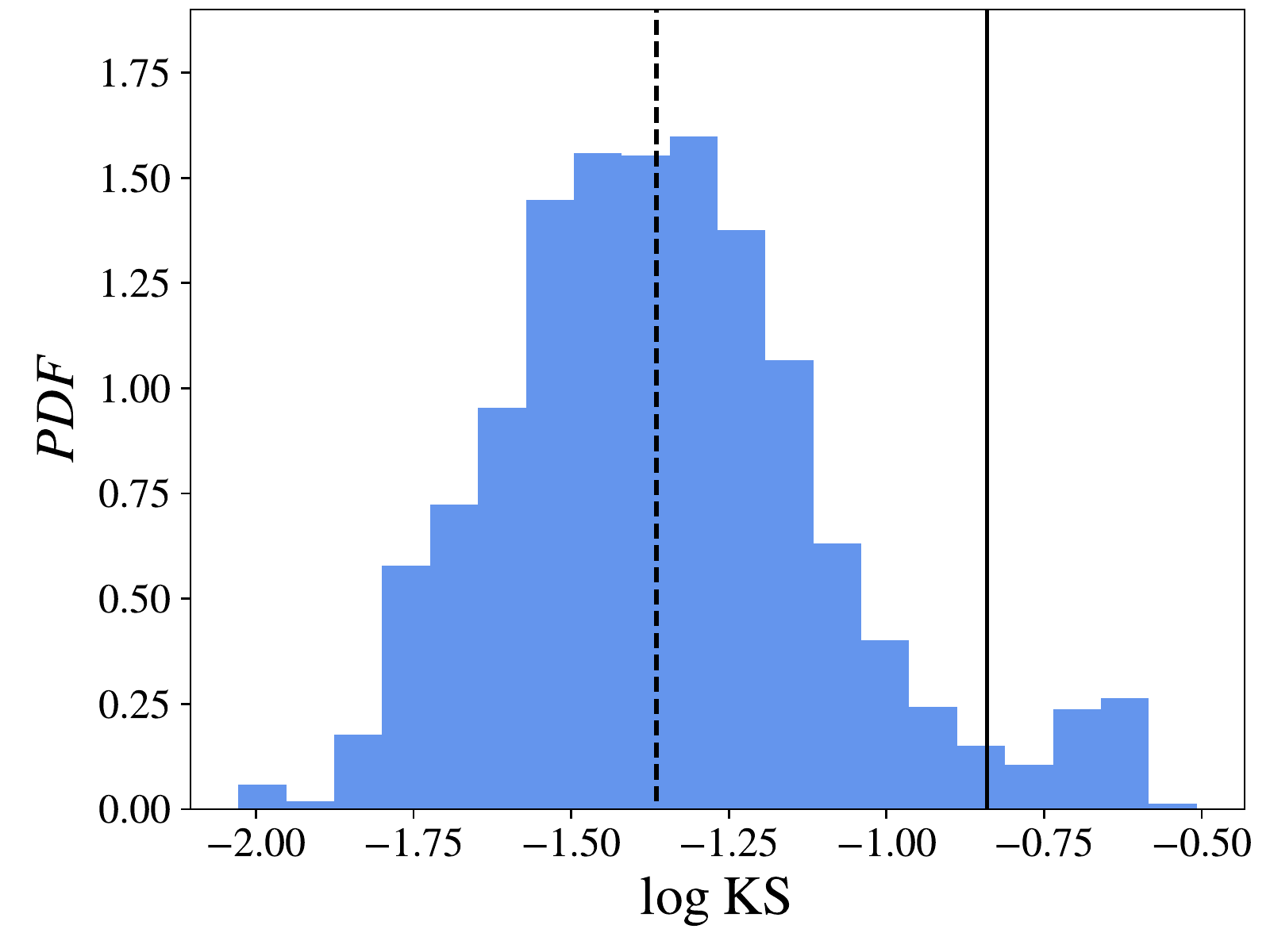}%
        \caption{ Comparison between the \lya\ line profiles of the 2000 random realization of \lyart\ and  those predicted by \zelda . The vertical dashed and solid black lines indicate the 50th and 95th percentiles, respectively.   }
        \label{fig:KS_pdf}
        \end{figure}

        Meanwhile, in Fig.~\ref{fig:KS_pdf} we show the probability distribution of the KS estimator. Overall, \zelda 's estimation is sufficiently accurate in comparison with the full RT computation of \lyart . The median KS value is $\sim 10^{-1.4}$ while the 95\% of the studied cases exhibit KS$<10^{-0.8}$. 
        
        \begin{figure*} 
        \includegraphics[width=6.9in]{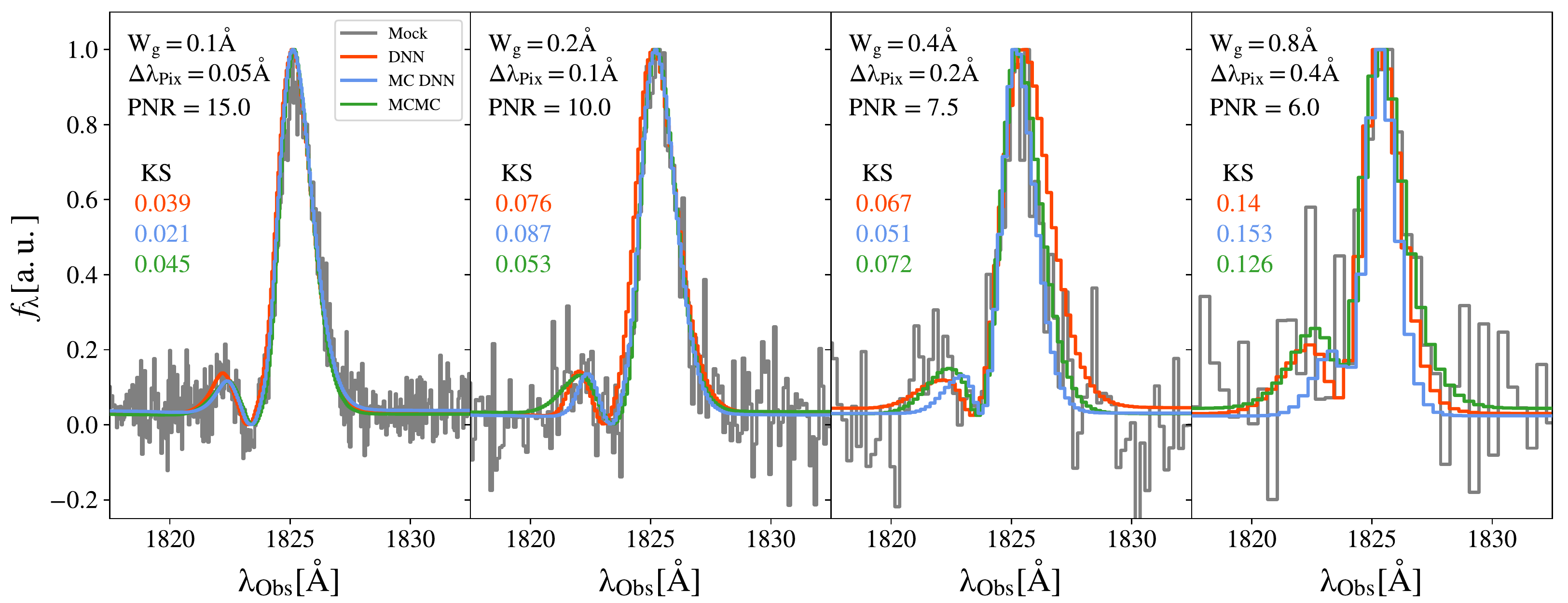}%
        \caption{ Illustration the spectral quality as a function of \wg, \dl\ and \sn. In grey we show a mock line profile with \vexp=50$km/s$, \nh=$10^{20}cm^{-2}$, \ta=0.01, \ew=$10^{1.5}$\AA{} and \w=0.5\AA{} at $z=0.5$. From left to right the spectral quality decreases, with parameters \{\wg[\AA{}], \dl[\AA{}] ,\sn\}, \{0.1,0.05,15\}, \{0.2,0.4,10\}, \{0.4,0.2,7.5\} and \{0.8,0.4,6\} respectively. The colored lines show the best fitting line profile for the different fitting schemes. In particular, red marks deep learning, blue the Monte Carlo deep learning and green the Monte Carlo Markov Chain methodology. On the left of each panel we display the KS estimator values for each fit, from top to bottom the DNN, MC DNN and MCMC methodologies. }
        \label{fig:spectral_quality}
        \end{figure*}
        
    \subsection{Generating realistic \lya\ line profiles}\label{ss:mocking}
        
        As described later, in section \S\ref{s:observation}, we compare \zelda 's predictions with real observations of \lya\ line profiles. 
        To make a fair comparison, we put \zelda 's spectrum to the same quality level than that with which we are making the comparison. 
        This is also necessary for the training of the deep neural network, as we explain later in \S\ref{s:DNN_build}.
        
        In general, there are three main variables to characterize the quality of a spectrum. First, the spectral resolution $R=\lambda/{\rm W_g}$ that effectively dilutes the spectrum with a Gaussian kernel of width \wg . 
        Second, the pixelization of the spectrum, i.e., the binning in wavelength used for the sampling the flux, \dl . And third, the level of signal compare to the noise of the spectrum. To characterize the third variable, we use the ratio 
        
        \begin{equation}
            S/N_p = \frac{f_{\lambda}^{\rm Ly\alpha} ( \lambda _ {\rm max } ) }{\Delta f_{\lambda}^{\rm Ly\alpha}(\lambda _ {\rm max }) },
        \end{equation}
        where $\lambda _ {\rm max }$ is the wavelength of the maximum of the \lya\ line profile and $\Delta f_{\lambda}^{\rm Ly\alpha}$ is the uncertainty in $f_{\lambda}^{\rm Ly\alpha}$.
        
        The line profiles computed by \lyart\ and predicted by \zelda\ are ideal. They exhibit an excellent quality with almost infinite resolution, small pixel size $\Delta \lambda = 0.08$\AA{} and a  high value of $S/N_p$. 
        With the current instruments, it is unrealistic to expect observed \lya\ line profiles to present a similar quality. 
        To reduce the quality of the spectrum predicted by \zelda\ we follow the process described in \cite{gurung_2020b}. 
        In summary:
        
        \begin{enumerate}
            \item We dilute the flux density by convolving it with a Gaussian kernel of with $W_g$.
            \item We re-bin the spectrum and it is evaluated in $\lambda_{\rm pix}$ as
            \begin{equation}
            \label{eq:pixelization}
            \displaystyle
            f_{\lambda , \rm  pix }^{\rm Ly\alpha}(\lambda_{\rm pix})  =  
            {
            \displaystyle
            {\int ^{\lambda_{\rm pix}+\Delta \lambda_{\rm pix}/2} 
            _{\lambda_{\rm pix}-\Delta \lambda_{\rm pix}/2}
            {f_{\lambda}^{\rm Ly\alpha} (\lambda) \;  d\lambda }} 
            \over 
            {\Delta \lambda_{\rm pix} } 
            }
            .
            \end{equation} 
             
            \item We compute the maximum of the line profile and we add white noise for a given value of \sn .
        \end{enumerate}

    When dealing with \lya\ line profiles at redshift $z>0$, first, the line is redshifted and then we apply the process described above to reduce the spectrum quality. Note that \wg\ and \dl\ are defined in the observed frame.

    In Fig.\ref{fig:spectral_quality} we illustrate the spectral quality as a function of \wg, \dl\ and \sn. In grey we show a mock line profile at $z=0.5$ with  \vexp=50$km/s$, \nh=$10^{20}cm^{-2}$, \ta=0.01, \ew=$10^{1.5}$\AA{} and \w=0.5\AA{}. The combination of parameters \{\wg[\AA{}], \dl[\AA{}] ,\sn\} changes from left to right as \{0.1,0.05,15\}, \{0.2,0.4,10\}, \{0.4,0.2,7.5\} and \{0.8,0.4,6\} respectively. The spectral quality decreases greatly from left to right, as the number of independent wavelength bins is reduced and also their signal to noise ratio.
    
    In the following sections, we will generate mock \lya\ line profiles encapsulating most of the actually observed spectra. In particular, we will cover \wg\ from 0.1\AA{} ($\sim$24 \kms) to 2\AA{} ($\sim$500 \kms), \dl\ from 0.05\AA{} to 1.0\AA{} and \sn\ from 5.0 to 15.0.
 
\renewcommand{\arraystretch}{1.4}
\begin{table*}
\caption{ Parameters associated with the line profiles displayed in Fig.\ref{fig:spectral_quality}.}
\begin{tabular}{ccccccccc}
 Parameter & Unit    & True & \multicolumn{3}{c}{\wg=0.1\AA{} \dl=0.05\AA{}} & \multicolumn{3}{c}{\wg=0.2\AA{} \dl=0.1\AA{}} \\
               &            &      & \multicolumn{3}{c}{\sn=15.0}                   & \multicolumn{3}{c}{\sn=10.0}           \\ \cmidrule(r){1-2} \cmidrule(r){3-3}  \cmidrule(r){4-6} \cmidrule(l){7-9}
               &            &      & DNN & MC DNN & MCMC & DNN  & MC DNN & MCMC  \\ \cmidrule(r){1-2} \cmidrule(r){3-3}  \cmidrule(r){4-6} \cmidrule(l){7-9}
z              &  & 0.5 & $0.5$ & $0.5001^{+4.5e-04}_{-1.6e-04}$ & $0.5002^{+2.8e-04}_{-1.9e-04}$ & $0.4998$ & $0.5^{+5.7e-04}_{-2.3e-04}$ & $0.5002^{+2.4e-05}_{-2.0e-05}$      \\
\vexp          & [$km\; s^{-1}$] & 50.0 & $50.2$ & $57.5^{+7.1e+01}_{-1.8e+01}$ & $62.9^{+6.0e+01}_{-2.3e+01}$ & $48.1$ & $50.0^{+8.3e+01}_{-2.2e+01}$ & $71.1^{+2.7e+00}_{-4.3e+00}$      \\
$\rm \log$\nh  & [$cm^{-2}$] & 20.0 & $19.9$ & $19.9^{+2.2e-01}_{-6.7e-01}$ & $19.7^{+3.7e-01}_{-3.2e-01}$ & $20.2$ & $20.0^{+3.2e-01}_{-7.0e-01}$ & $19.8^{+3.5e-02}_{-3.1e-02}$      \\
\ta            &  & 0.01 & $0.036$ & $0.058^{+2.2e-01}_{-5.4e-02}$ & $0.102^{+6.3e-01}_{-4.4e-02}$ & $0.0$ & $0.011^{+2.1e-01}_{-1.1e-02}$ & $0.15^{+2.5e-02}_{-2.9e-02}$      \\
$\rm \log$\ew  & [\AA{}] & 1.5 & $1.65$ & $1.64^{+2.3e-01}_{-1.6e-01}$ & $1.79^{+7.3e-02}_{-1.1e-01}$ & $1.65$ & $1.72^{+3.6e-01}_{-1.5e-01}$ & $1.8^{+3.3e-02}_{-4.0e-02}$      \\
\w             & [\AA{}] & 0.5 & $0.4$ & $0.35^{+1.6e-01}_{-1.2e-01}$ & $0.44^{+9.7e-02}_{-1.7e-01}$ & $0.2$ & $0.29^{+2.3e-01}_{-1.2e-01}$ & $0.64^{+2.3e-02}_{-3.5e-02}$
\\
\\
 Parameter & Unit   & True & \multicolumn{3}{c}{\wg=0.4\AA{} \dl=0.2\AA{}} & \multicolumn{3}{c}{\wg=0.8\AA{} \dl=0.4\AA{}} \\
                  &         &      & \multicolumn{3}{c}{\sn=7.5}                   & \multicolumn{3}{c}{\sn=6.0}           \\ \cmidrule(r){1-2} \cmidrule(r){3-3}  \cmidrule(r){4-6} \cmidrule(l){7-9}
               &            &      & DNN & MC DNN & MCMC & DNN  & MC DNN & MCMC  \\ \cmidrule(r){1-2} \cmidrule(r){3-3}  \cmidrule(r){4-6} \cmidrule(l){7-9}
z              &  & 0.5 & $0.5004$ & $0.5006^{+7.3e-04}_{-7.1e-04}$ & $0.5005^{+3.2e-05}_{-6.2e-04}$ & $0.5003$ & $0.5007^{+6.1e-04}_{-6.5e-04}$ & $0.5008^{+4.6e-05}_{-7.9e-05}$      \\
\vexp          & [$km\; s^{-1}$] & 50.0 & $137.4$ & $91.1^{+1.6e+02}_{-5.7e+01}$ & $104.2^{+1.2e+01}_{-7.9e+01}$ & $49.6$ & $57.6^{+1.2e+02}_{-3.4e+01}$ & $124.9^{+7.5e+00}_{-1.2e+01}$      \\
$\rm \log$\nh  & [$cm^{-2}$] & 20.0 & $19.5$ & $19.2^{+9.2e-01}_{-1.4e+00}$ & $19.4^{+1.0e+00}_{-1.1e-01}$ & $19.7$ & $19.1^{+8.9e-01}_{-1.1e+00}$ & $18.8^{+1.4e-01}_{-9.4e-02}$      \\
\ta            &  & 0.01 & $0.001$ & $0.006^{+8.1e-02}_{-5.7e-03}$ & $0.032^{+4.9e-01}_{-3.0e-02}$ & $0.003$ & $0.003^{+5.6e-02}_{-2.3e-03}$ & $0.002^{+1.5e-02}_{-2.0e-03}$      \\
$\rm \log$\ew  & [\AA{}] & 1.5 & $1.56$ & $1.59^{+2.1e-01}_{-1.7e-01}$ & $1.72^{+1.3e-01}_{-1.2e-01}$ & $1.72$ & $1.71^{+2.5e-01}_{-2.3e-01}$ & $1.6^{+3.0e-02}_{-3.2e-02}$      \\
\w             & [\AA{}] & 0.5 & $0.94$ & $0.47^{+3.5e-01}_{-2.0e-01}$ & $0.81^{+1.3e-01}_{-3.9e-01}$ & $1.0$ & $0.44^{+3.8e-01}_{-1.9e-01}$ & $1.19^{+7.7e-02}_{-1.1e-01}$
\end{tabular}
\label{tab:quality_params}
\end{table*}

  \begin{figure*} 
    \includegraphics[width=2.4in]{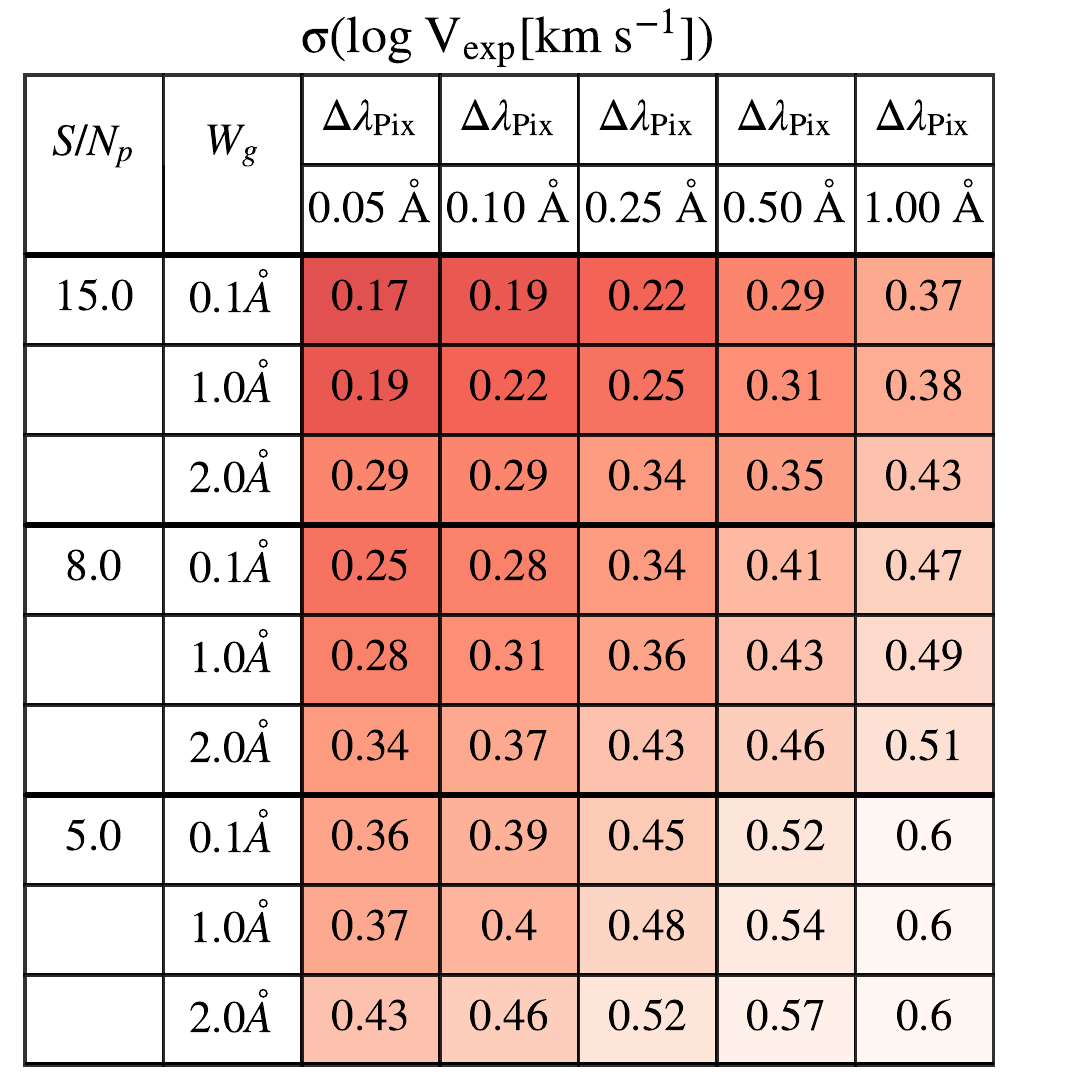}%
    \includegraphics[width=2.4in]{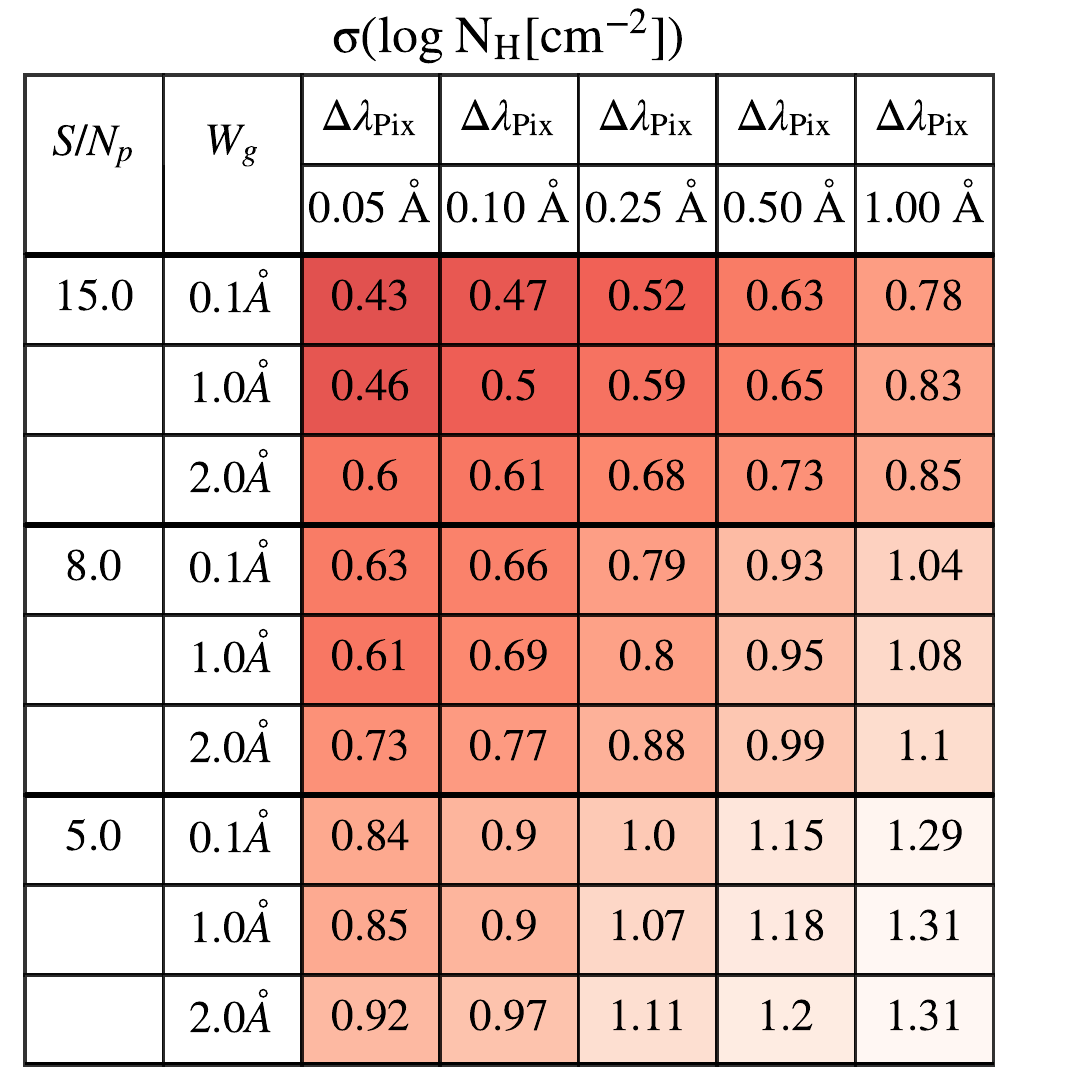}%
    \includegraphics[width=2.4in]{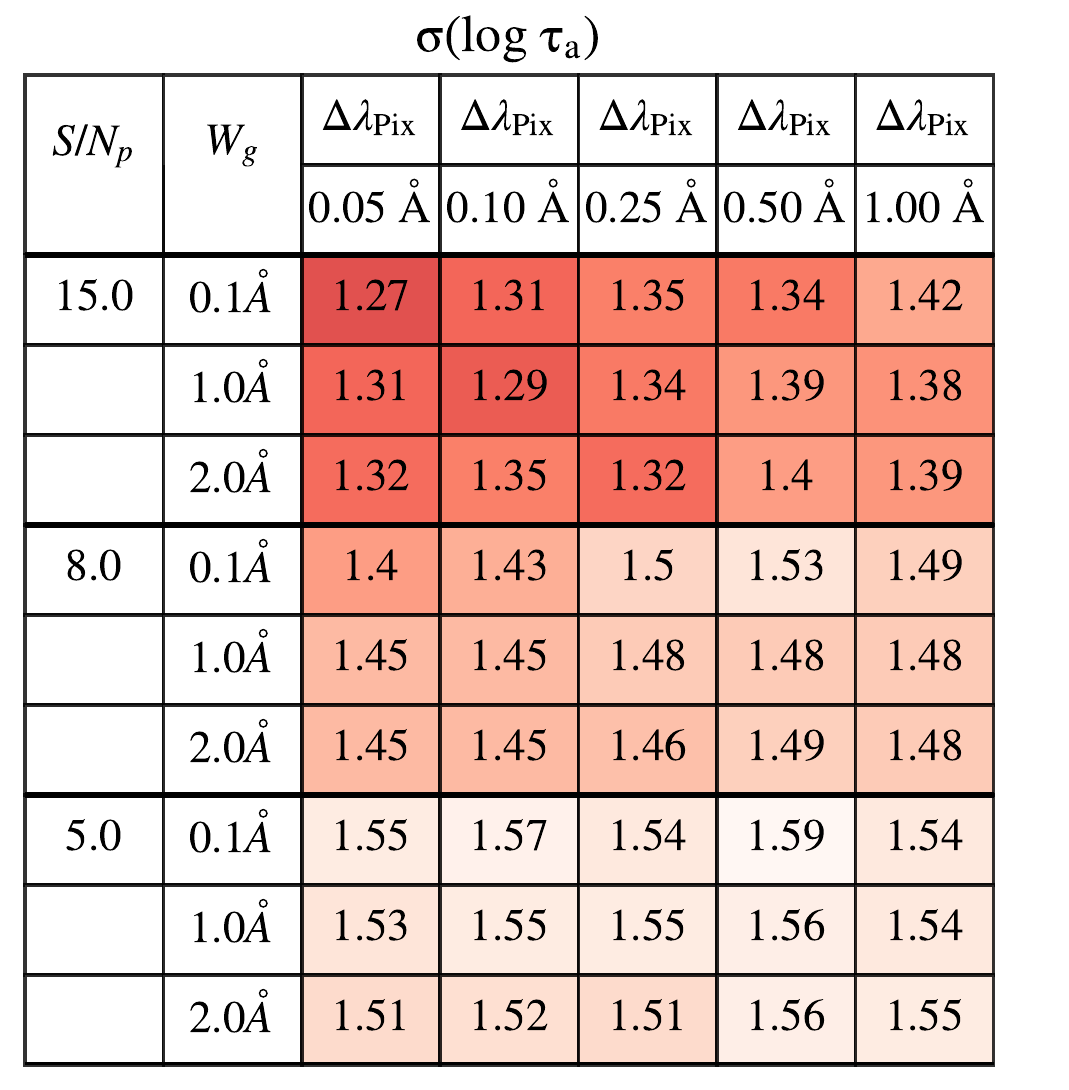}
    
    \vspace{0.2in}
    
    \includegraphics[width=2.4in]{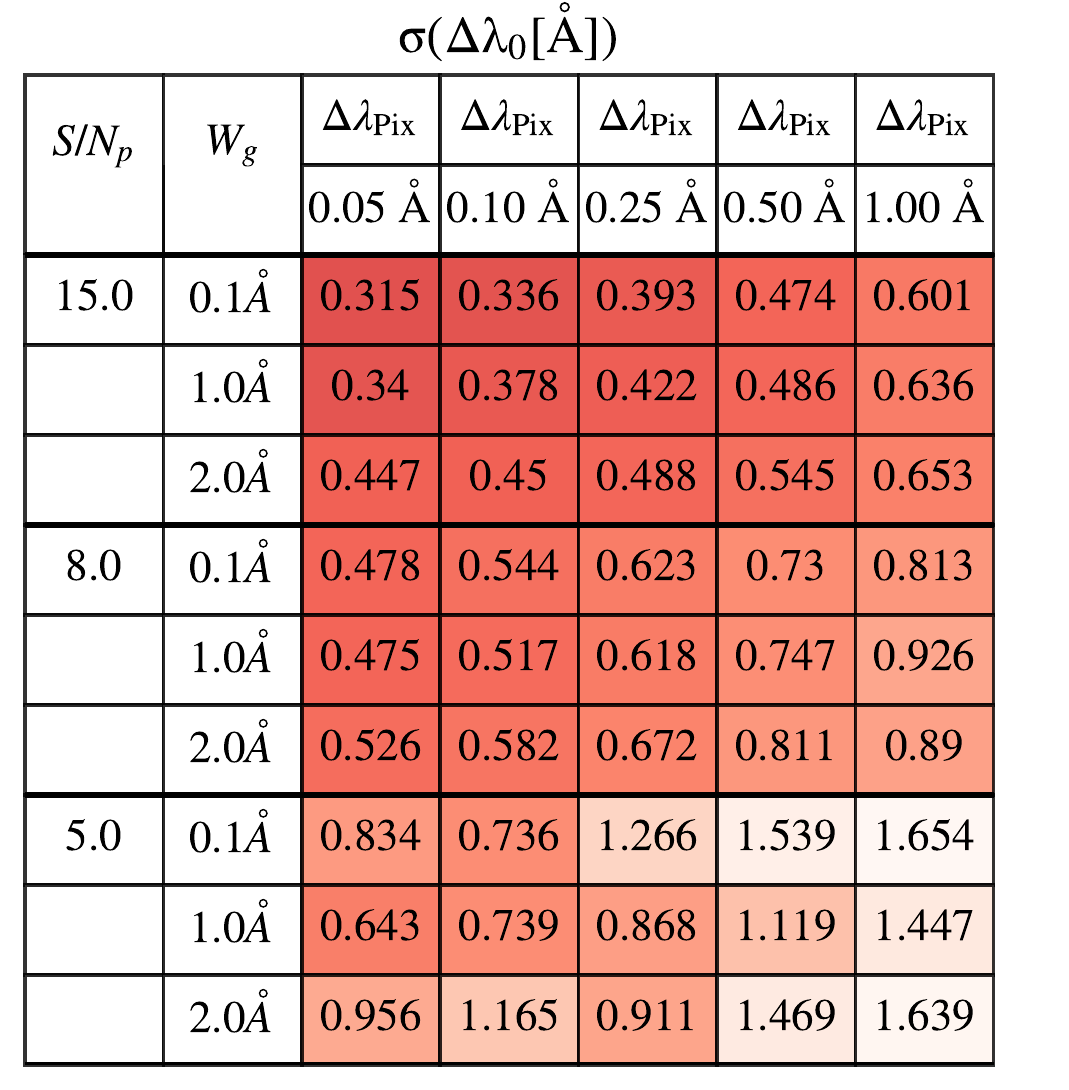}%
    \includegraphics[width=2.4in]{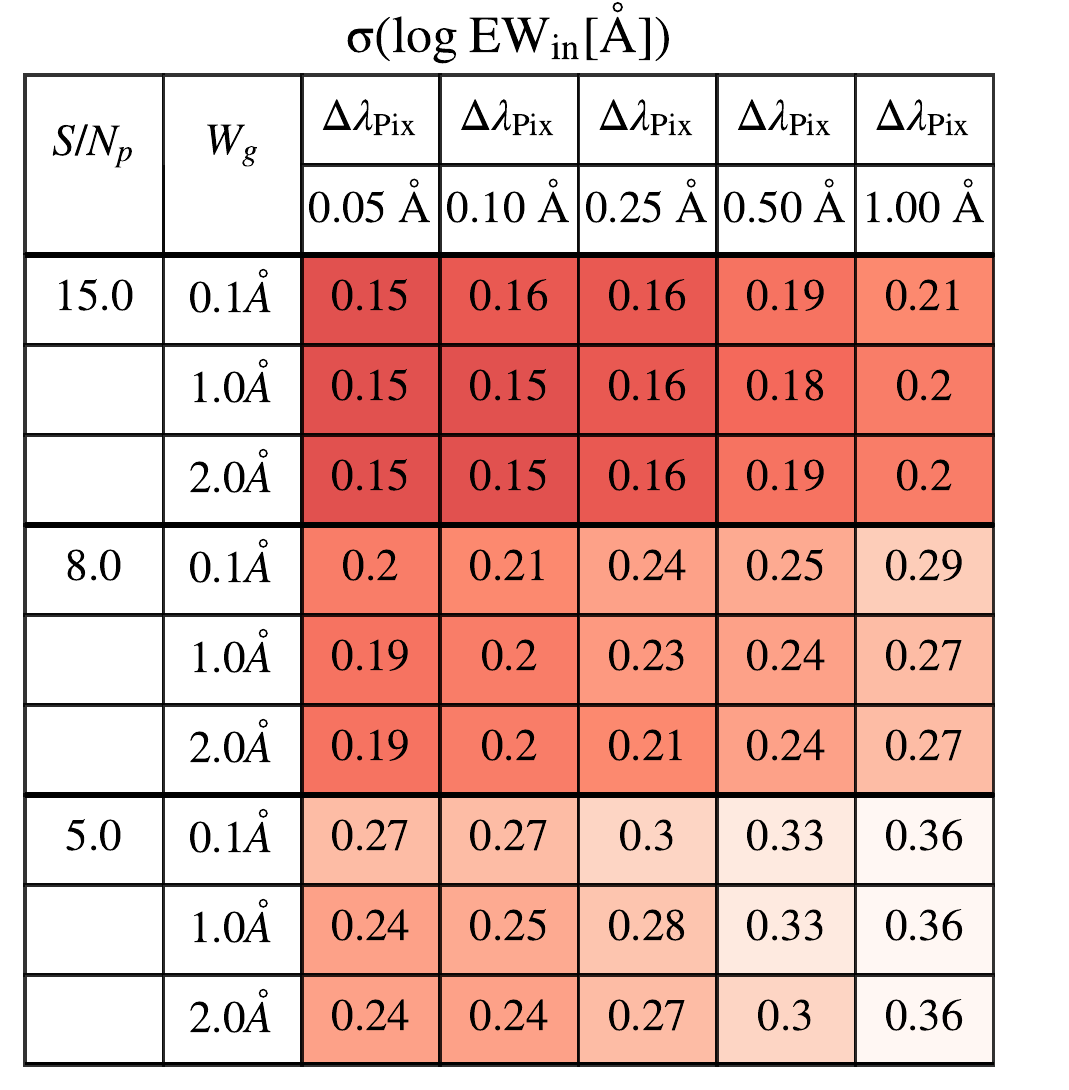}%
    \includegraphics[width=2.4in]{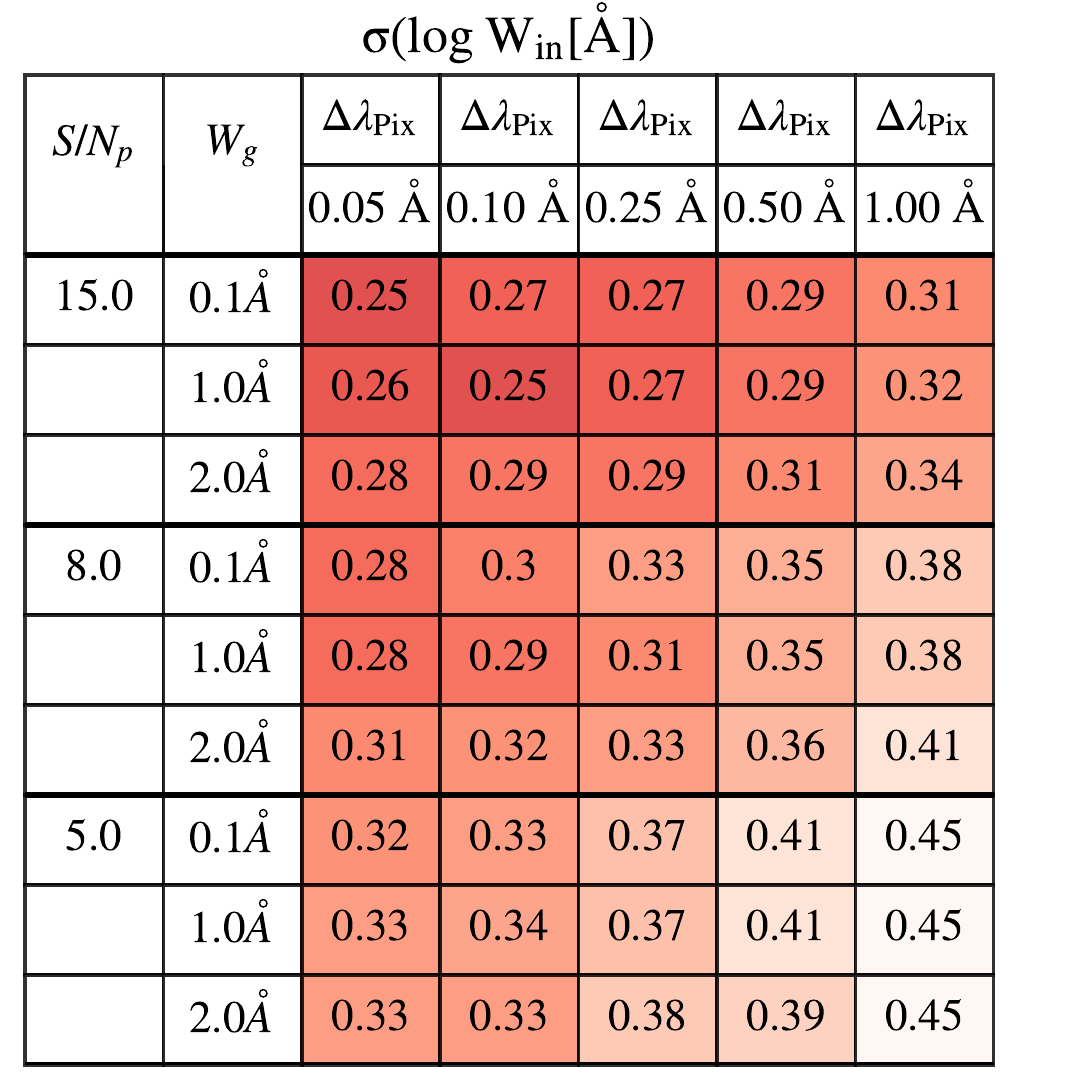}%
    \caption{ Standard deviation of the difference between the true and the predicted inflow/outflow parameters in the direct DNN methodology. In the row, \vexp, \nh\ and \ta\ from left to right. In the bottom row, \dlt , \ew\ and \w\  from left to right. Cells are colored by their value and darker means lower (better).     }
    \label{fig:props_1_iter}
    \end{figure*}
    
\section{ Fitting \lya\ line profiles with deep learning }\label{s:DNN_build}

    A novel feature of \zelda\ is that it incorporates a fitting procedure on the basis of deep neural networks (DNN): namely, \zelda\ can find a best-fitting parameter set given a \lya\ spectrum. In this section, we describe our DNN scheme and how the fitting is performed. In \S\ref{ss:network_arc} we describe the DNN architecture and training set. Then, \S\ref{s:DNN_application} describes how the DNN is used to estimate the outflow properties, while in \S\ref{ss:feature_importance} we make a feature importance analysis in order to understand which parts of the \lya\ line profile contain relevant information about the outflow parameters. 

    \subsection{ Deep neural network architecture and training}\label{ss:network_arc}

    Here, we describe our DNN scheme. First, in \S\ref{sss:input_output} we describe the input and output of the neural network. Then, in \S\ref{ss:training} we describe the sample used for the training set.

    \subsubsection{ Input and output}\label{sss:input_output}
        
    The input of our DNN scheme is an array of 1003 variables. 1000 of these contain the information about line profile, and the other 3 about its quality and a proxy for the redshift of the source. In terms of the line profile, we apply the following methodology:

    \begin{enumerate}
        \item Find the wavelength global maximum of the line profile $\lambda_{\rm max}$. We use this wavelength as a proxy for the true \lya\ wavelength of the line $\lambda_{\rm True}$. In general $\lambda_{\rm True} \neq \lambda_{\rm max}$. Therefore, the proxy redshift is $z_{\rm max}= \lambda_{\rm max} / \lambda_{\rm Ly\alpha}-1$.
        
        \item Convert the line profile to the proxy rest frame, $f_{\lambda, \rm max }^{\rm Ly\alpha}$. Specifically, we convert the array where $f_{\lambda}^{\rm Ly\alpha}$ is evaluated in the observed frame, $\lambda^{\rm Obs}_{\rm Arr}$ to the rest frame wavelength as if $\lambda_{\rm True} = \lambda_{\rm max}$, i.e., $\lambda^{\rm 0}_{\rm Arr} = \lambda^{\rm Obs}_{\rm Arr}/(1+z_{\rm max}) $. 
        
        \item Normalize our line profile by its maximum $f_{\lambda}^{\rm Ly\alpha}(\lambda_{\rm max})$. We do this so that all the line profiles have a similar dynamical range in density fluxes. This increases the accuracy in the predictions of the DNN.
        
        \item We re-bin $f_{\lambda, \rm max }^{\rm Ly\alpha}$ into 1000 bins from $\lambda_{\rm Ly\alpha}-18.5$\AA{} to $\lambda_{\rm Ly\alpha}+18.5$\AA{} (corresponding to $v\in [\pm 2719]\,{\rm km}\,{\rm s}^{-1}$) using linear interpolation between the values of $f_{\lambda, \rm max }^{\rm Ly\alpha}$ evaluated in $\lambda^{\rm 0}_{\rm Arr}$. We chose this range because it covers all the \lya\ line profile features. Additionally the value of 1000 bins in wavelength is arbitrary as long as it is big enough to sample properly this wavelength range. We checked that increasing the resolution in the binning does not lead to better results. 
    \end{enumerate}
    Finally, the other 3 variables in our input arrays are  $z_{\rm max}$,  \wg\ and  \dl .

    \zelda\ contains two deep neural networks (DNN), which are trained to predict the outflow and inflow properties associated with a \lya\ line profile. Each of the two DNN that predict the inflow/outflow properties uses a regression algorithm for six variables: $\log | V_{\rm exp}|$, $\log N_H$, $\log \tau_a$, $\log EW_{\rm in}$, $\log W_{\rm in}$ and the displacement between the wavelength set as \lya\ and the true \lya\ wavelength in the proxy rest frame, $\Delta \lambda_{\rm True}$. For this last quantity, the true \lya\ wavelength in the observed rest frame, $\lambda^{\rm Obs}_{\rm True}$, can be reconstructed as
    
    \begin{equation}
        \Delta \lambda_{\rm True} = \lambda_{\rm True}^0 - \lambda _{\rm Ly\alpha} = \lambda _{\rm Ly\alpha} \left( \frac{\lambda^{\rm Obs}_{\rm True}}{\lambda_{\rm max}} - 1 \right),
    \end{equation}
    where we have used that the true \lya\ wavelength in the proxy frame is $\lambda^{\rm Obs}_{\rm True}/(z_{\rm max}+1)$. Once  $\lambda^{\rm Obs}_{\rm True}$ is computed, the redshift of the source is set simple as $z=\lambda^{\rm Obs}_{\rm True}/\lambda_{\rm Ly\alpha}-1$.
    
    \subsubsection{Deep neural network training }\label{ss:training}
    
    Our training sets are composed by \lya\ line profiles spawning the whole range of \vexp , \nh , \ta , \ew\ and \w\ covered by \zelda\ and a wide range of \wg , \dl\ and \sn . In particular, our default deep neural network (DNN) to predict the outflow properties spawns uniformly $\rm \log W_g[$\AA{}$] \in$[ -1 , 0.3 ] , $\rm \log \Delta \lambda_{Pix} $[\AA{}]$ \in$[ -1.3 , 0.3 ], $\log S/N_p \in$[ 0.7 , 1.6 ] and $z\in$[0.001 , 4.0]. 
    We decided to use these dynamical ranges as they cover most of the spectroscopic \lya\ lines in the literature. 
    Furthermore, sources at larger redshift are prone to exhibit a \lya\ line profile clearly affected by IGM effects \citep{laursen11,Byrohl_2019,GurungLopez_2020a}, which are not included in our model. 
    For this reason, we train only up to redshift 4 \citep[note, however, that even for $z\lesssim 3$ the IGM can affect the \lya\ line shape][]{GurungLopez_2020a,Byrohl2020}. 
    Note that \zelda\ includes all tools to build the training sets for the neural networks described here. 
    Users may choose their own dynamical ranges for all properties described here if they require a custom DNN. 
    
    Both the inflow and outflow DNNs are trained in the same outflow/inflow parameter hyper-volume:
    
    \begin{itemize}
        \item $\log \; | V_{\rm exp} | [{\rm km\,s^{-1}}]$ $\in [  1.0   ,  3.0   ]$,

        \item  $\log N_{\rm H} [{\rm cm^{-2}}]$ $\in [ 17.0   , 21.5   ]$,

        \item  $\log \tau_{a}$ $ \in [ -4.0 , 0.0  ]$,

        \item  $\log EW_{\rm in} $[\AA{}] $ \in  [ 0.7 , 2.3  ]$,

        \item  $\log W_{\rm in} $[\AA{}] $ \in  [ -2.0 , 0.7  ]$,

    \end{itemize}
    
    \noindent where for inflows $V_{\rm exp}<0$ and for outflows $V_{\rm exp}>0$. Also, in principle \zelda\ is designed to estimate the properties of \lya\ line profiles, therefore we decided to train only $EW_{\rm in}$ greater than $10^{0.7}$\AA{}. 
    As we will show in \S\ref{s:observation} all the observed spectrum that we analyze is well above this threshold.
    
    To determine the best DNN configuration for our problem, we tested several configurations with different number of hidden layers and different hidden layer sizes . 
    In summary, we found that a DNN with 5 hidden layers, each with 256 nodes is optimal to predict the inflow and outflow properties given our input for the DNN.  
     
    Note that in \zelda, we have incorporated all tools to produce the training sets. In this way, if users desired to have their own custom DNN, with a particular redshift range, \dl , \wg , etc, or even change the hidden layer configuration, they would be able to train their own deep neural network.
        
    \begin{figure*} 
    \includegraphics[width=6.9in]{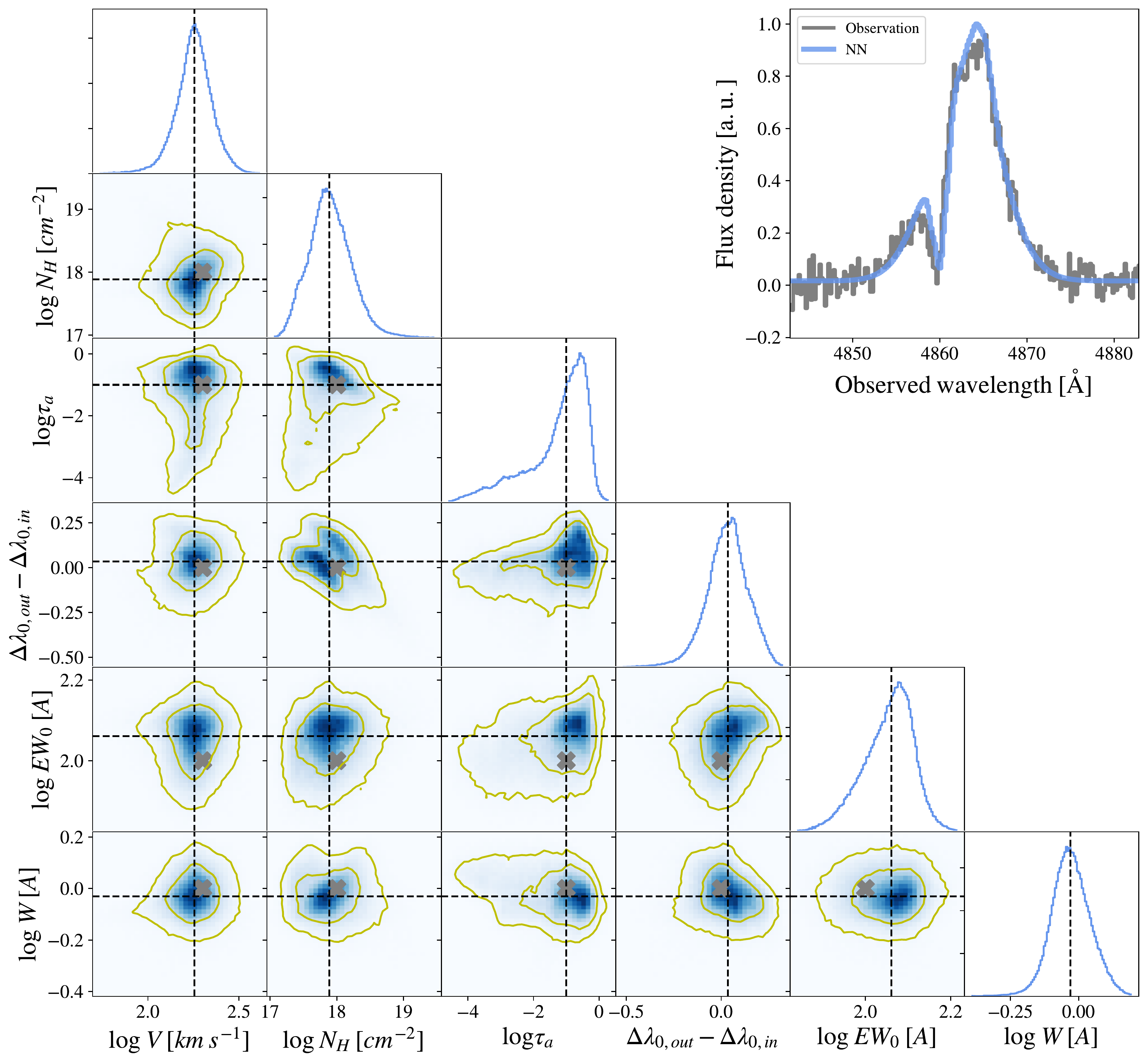}
    \caption{ Illustration of the Monte Carlo methodology using our DNN for a mock \lya\ line profile, at redshift 3.0, \vexp =200km/s , \nh=$\rm 10^{18}cm^{-2}$, \ta=0.1, \w=0.1\AA{} and \ew=100\AA{} and quality \wg=0.5\AA{} , \dl=0.1\AA{} , \sn = 10. The grey cross is the true outflow parameters, the black dashed lines indicate the percentile 50 of the distributions of the outflow parameters predicted by the DNN after perturbing the input \lya\ line profile consecutively by its uncertainty. In the top right panel we display the input Lya line profile in grey and in blue, \zelda's prediction, which corresponds to the percentile 50 of the outflow parameters distributions.  The yellow curves indicate the 1$\sigma$ and 2$\sigma$ contours of the 2D distributions.  }
    \label{fig:outflow_dnn_ilustration}
    \end{figure*} 
    
    \begin{figure*} 
    \includegraphics[width=6.9in]{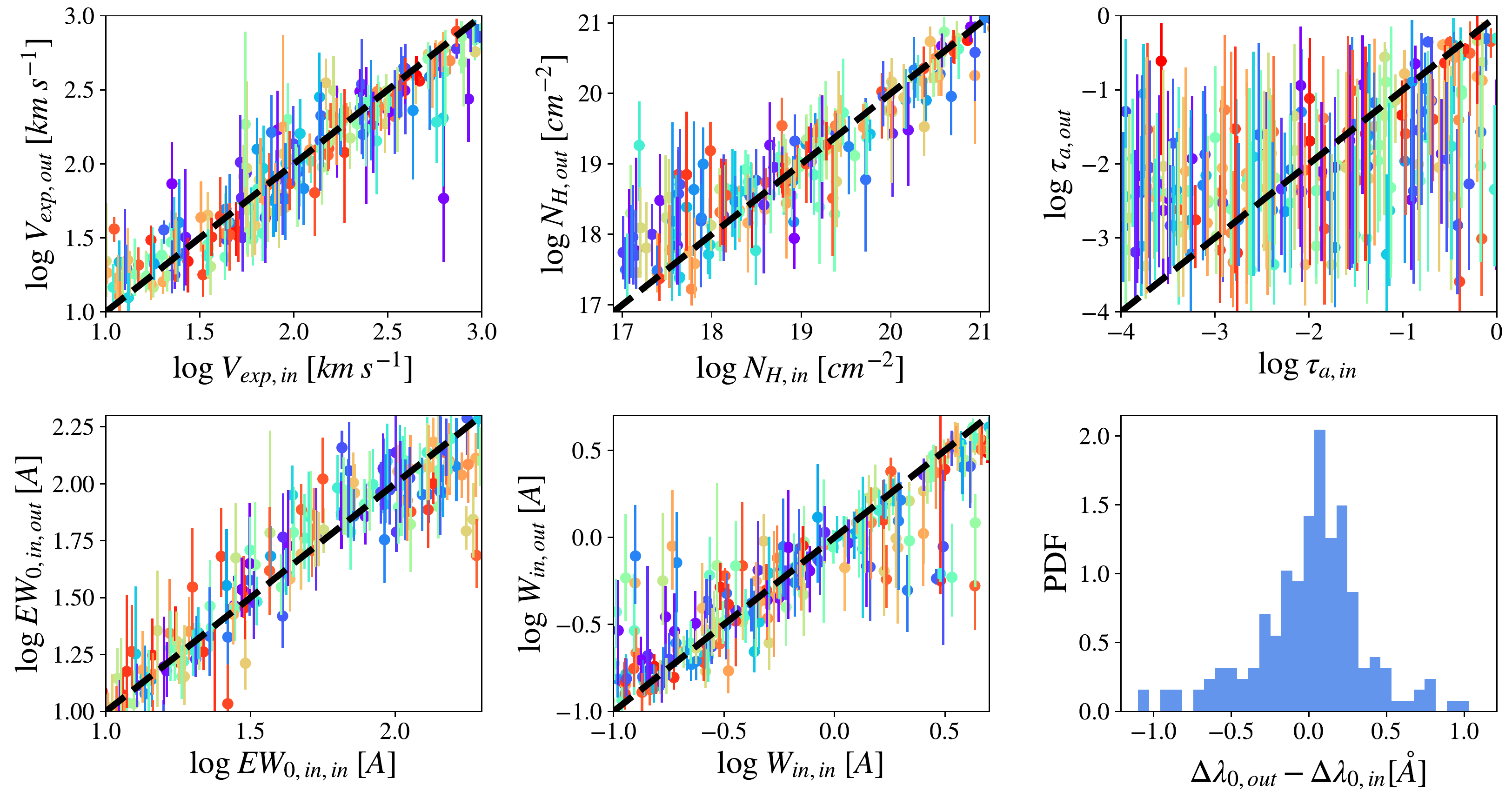}%
    \caption{ Accuracy of DNN predictions. 200 mock line profiles computed making interpolation in the grid with random outflow parameters and from redshift 0.0001 to 4.0 with quality \wg=0.5\AA{} , \dl=0.1\AA{} , \sn = 10 .}
    \label{fig:outflow_dnn_ilustration_acc}
    \end{figure*} 
    
    \begin{figure*} 
    \includegraphics[width=2.31in]{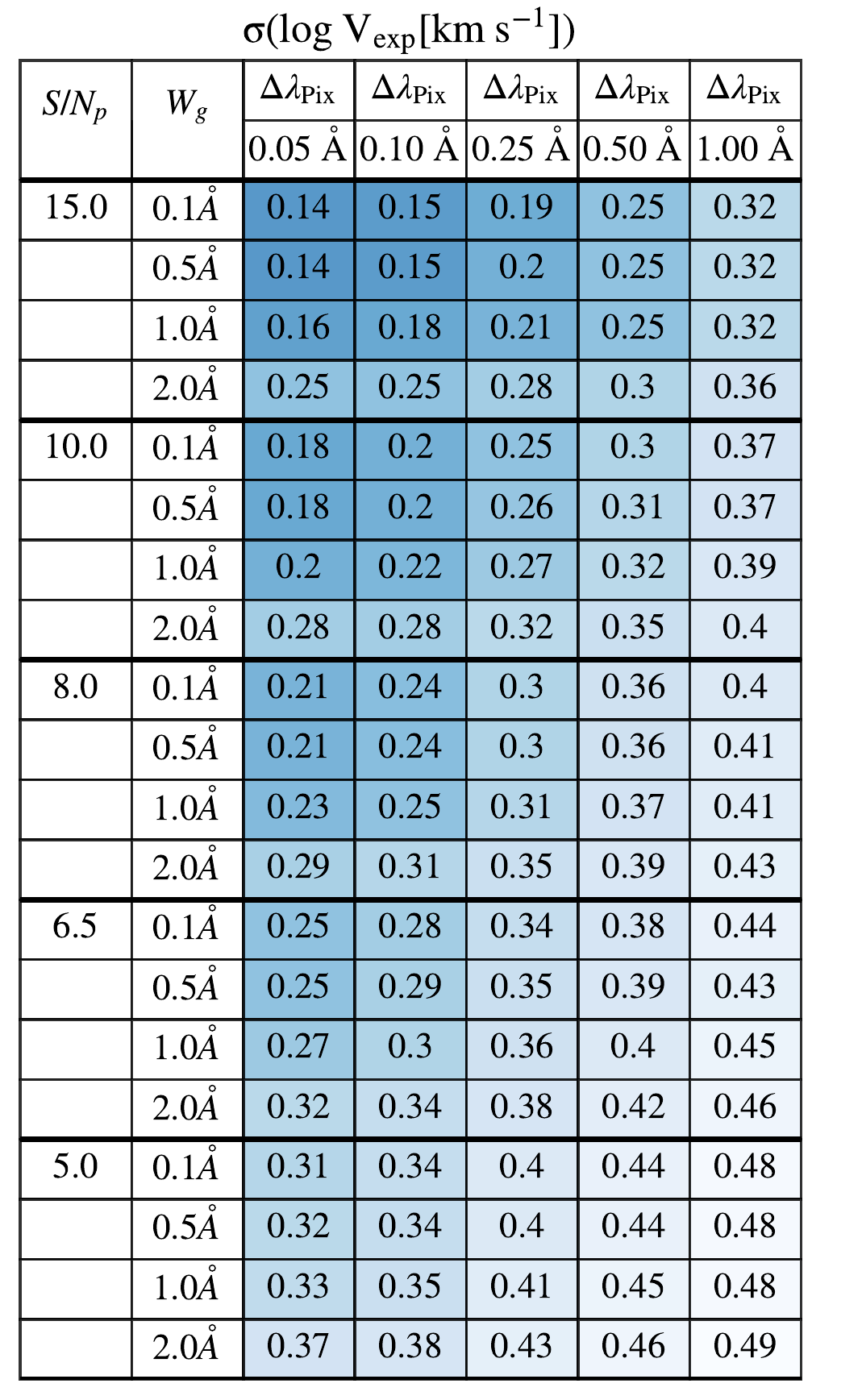}%
    \includegraphics[width=2.31in]{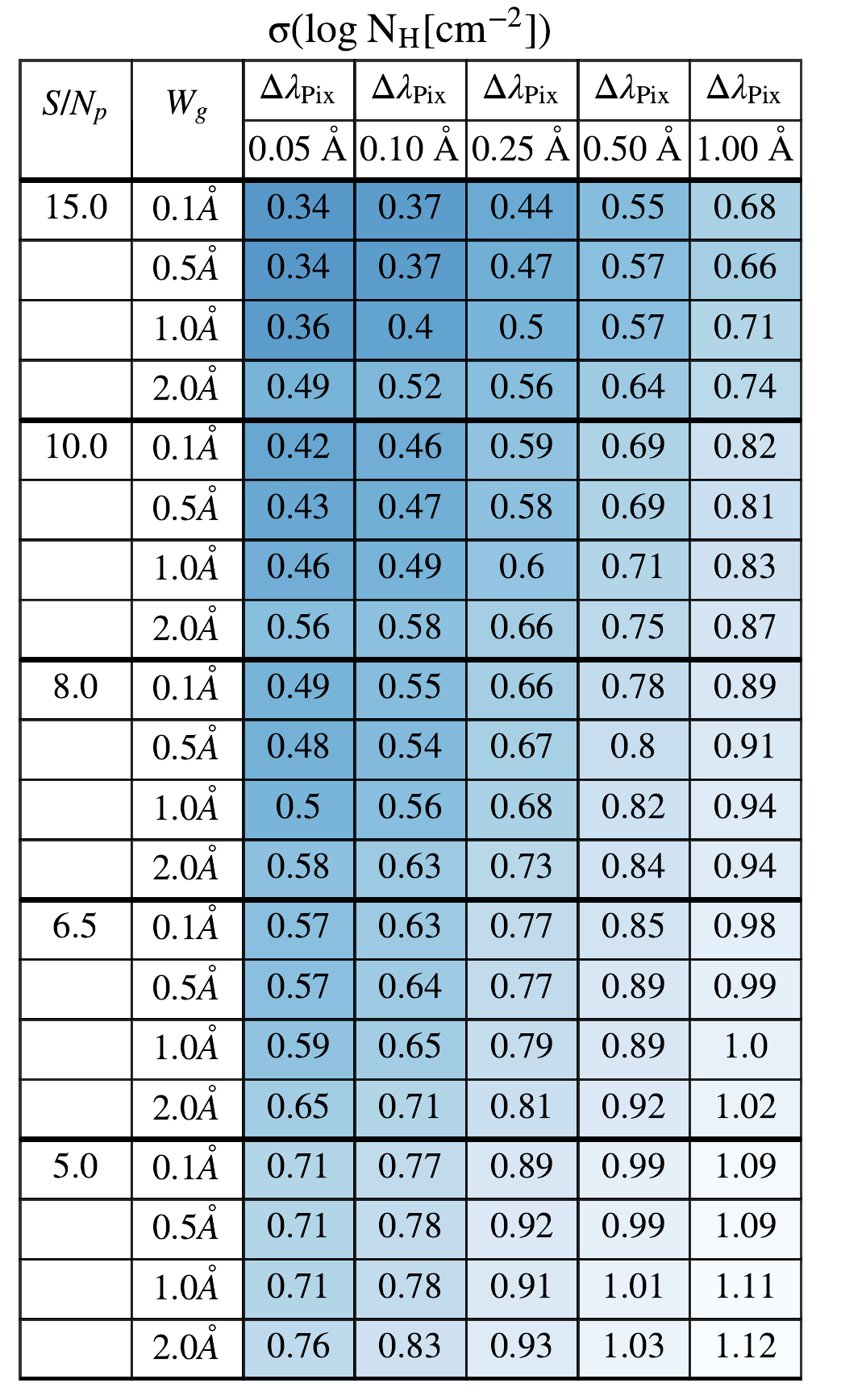}%
    \includegraphics[width=2.31in]{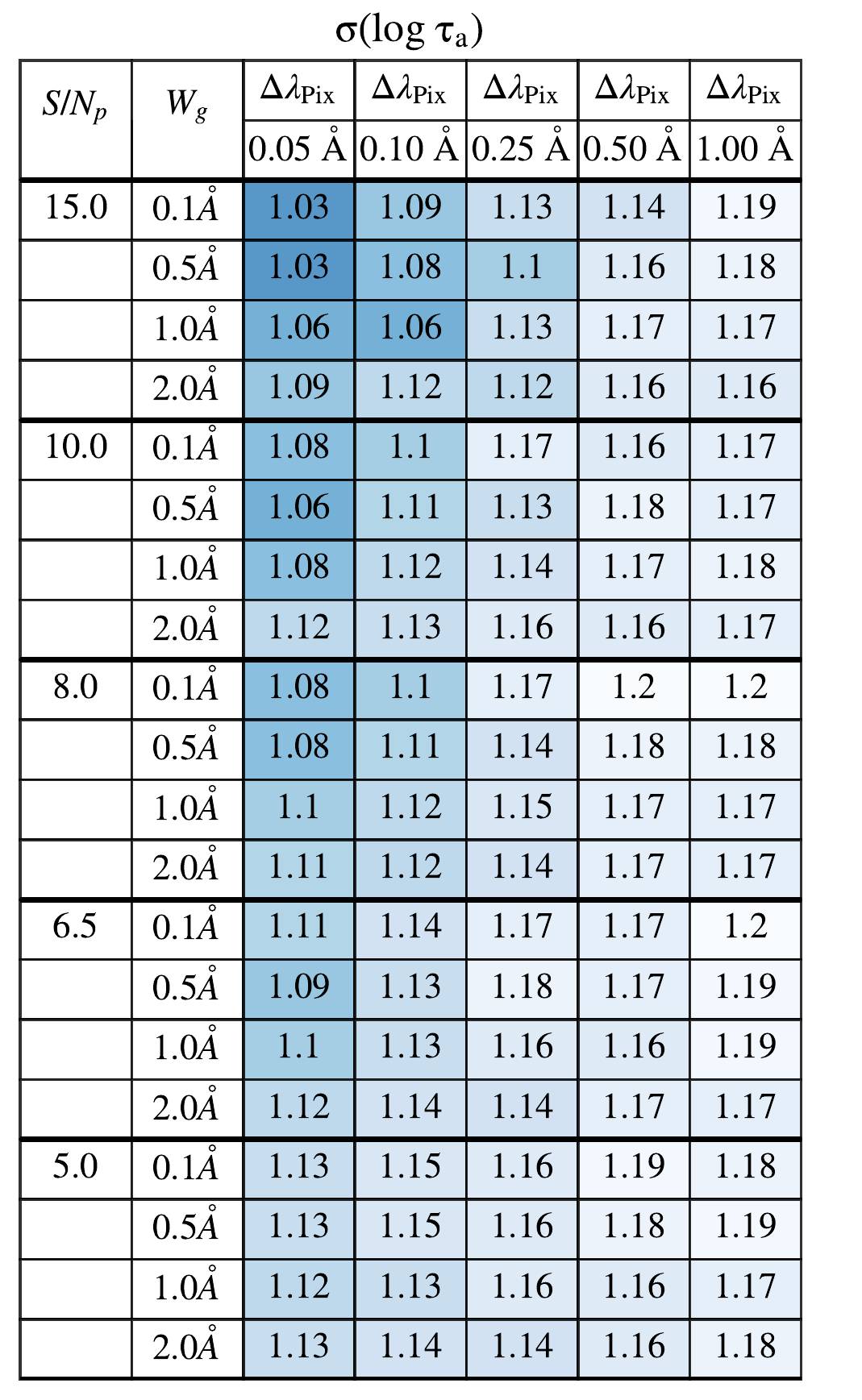}
    
    \vspace{0.20in}
    
    \includegraphics[width=2.31in]{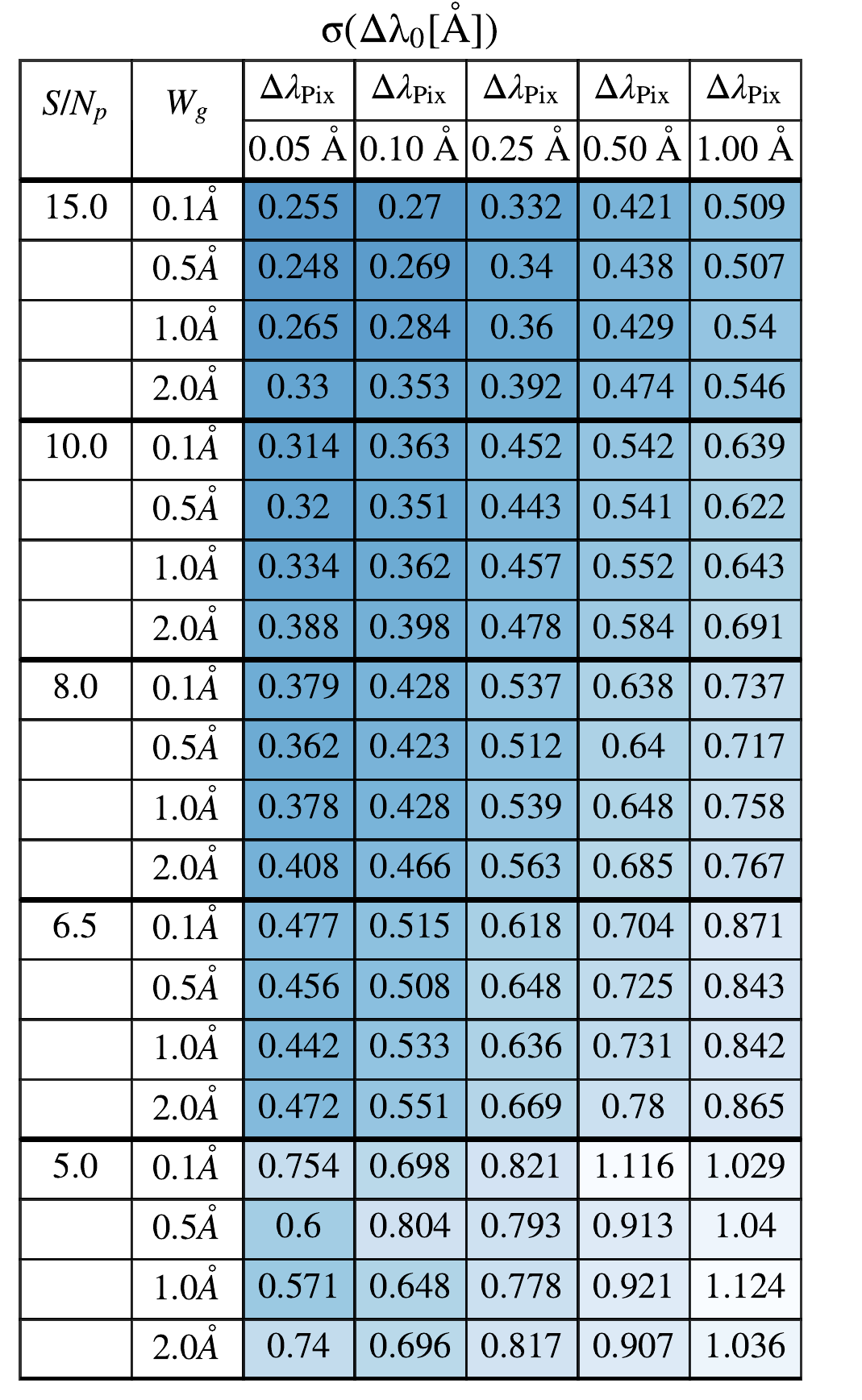}%
    \includegraphics[width=2.31in]{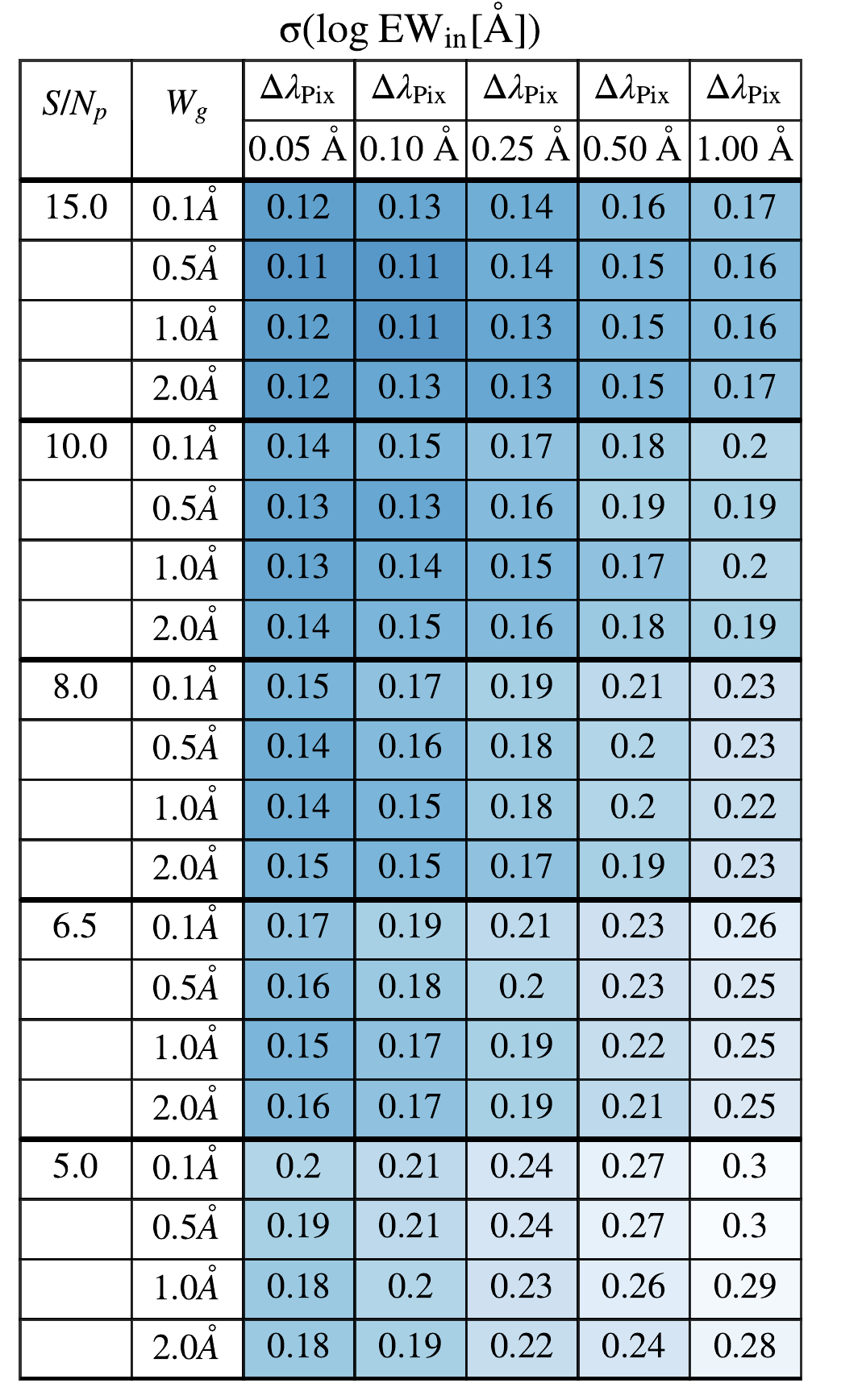}%
    \includegraphics[width=2.31in]{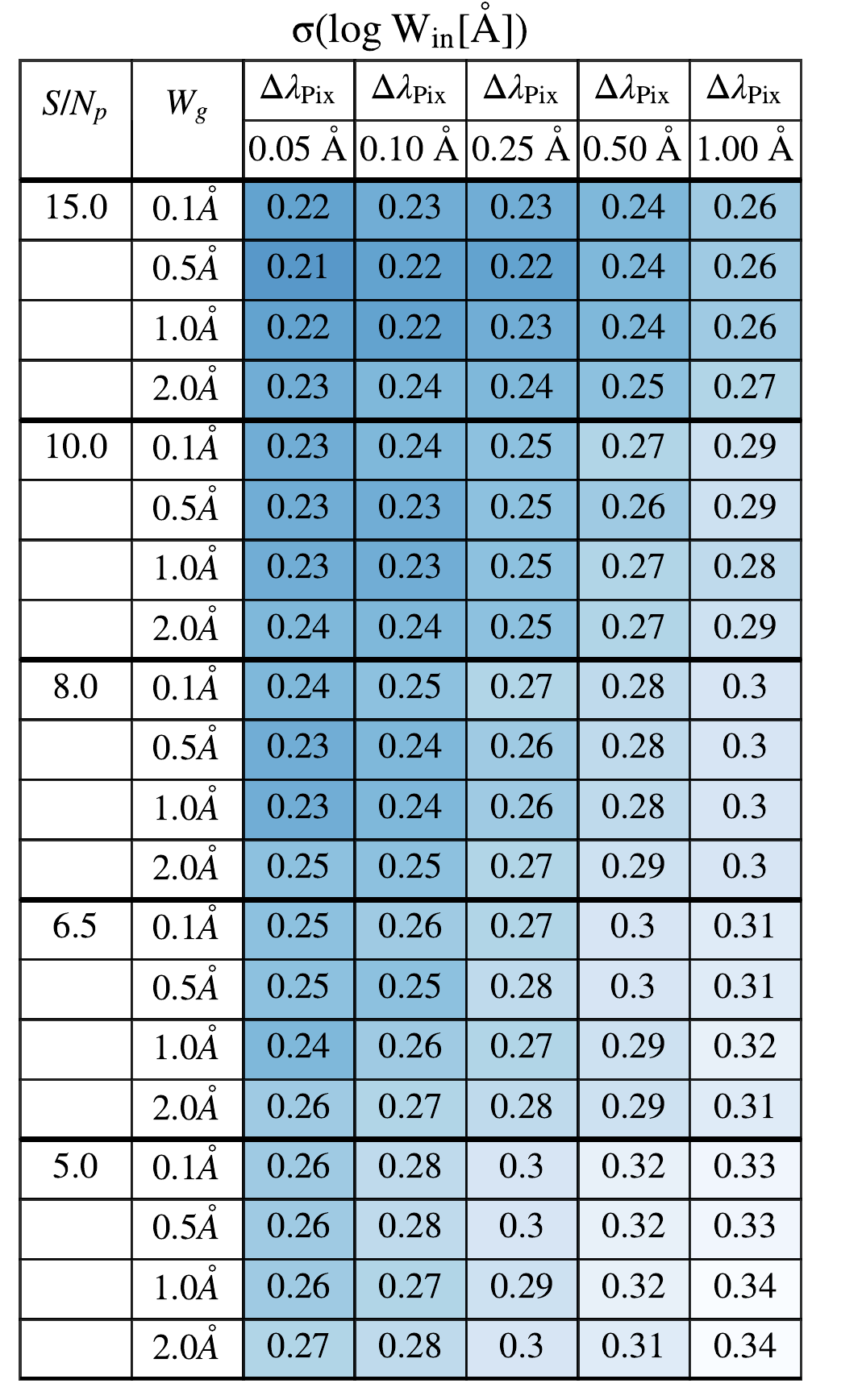}%
    \caption{ Standard deviation of the difference between the true and the predicted inflow/outflow parameters for the MC DNN methodology. In the upper row, \vexp, \nh\ and \ta\ from left to right. In the bottom row, \dlt , \ew\ and \w\  from left to right. Cells are colored by their value and darker means lower (better).  }
    \label{fig:props_1_acc}
    \end{figure*}
    
    \begin{figure*} 
    \includegraphics[width=6.9in]{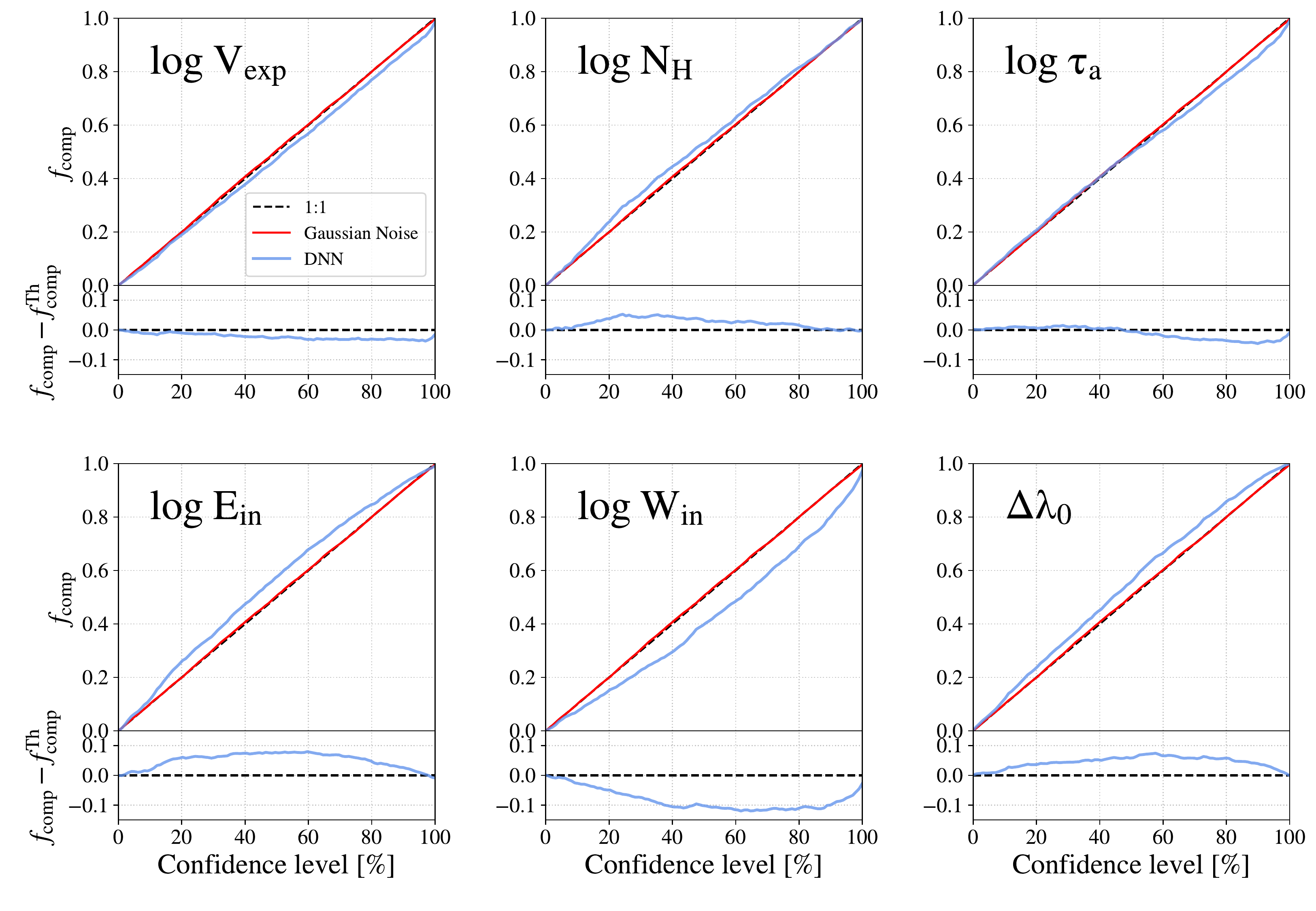}%
    \caption{ Comparison between the fraction of cases that a measurement is compatible with the intrinsic true value as a function of the confidence level. This particular figure shows the results for a sample of ~2000 line profiles with quality \wg=0.5\AA{} , \dl=0.1\AA{} and \sn = 10. }
    \label{fig:NN_errorbars}
    \end{figure*} 
    
    \begin{figure*} 
    \includegraphics[width=6.9in]{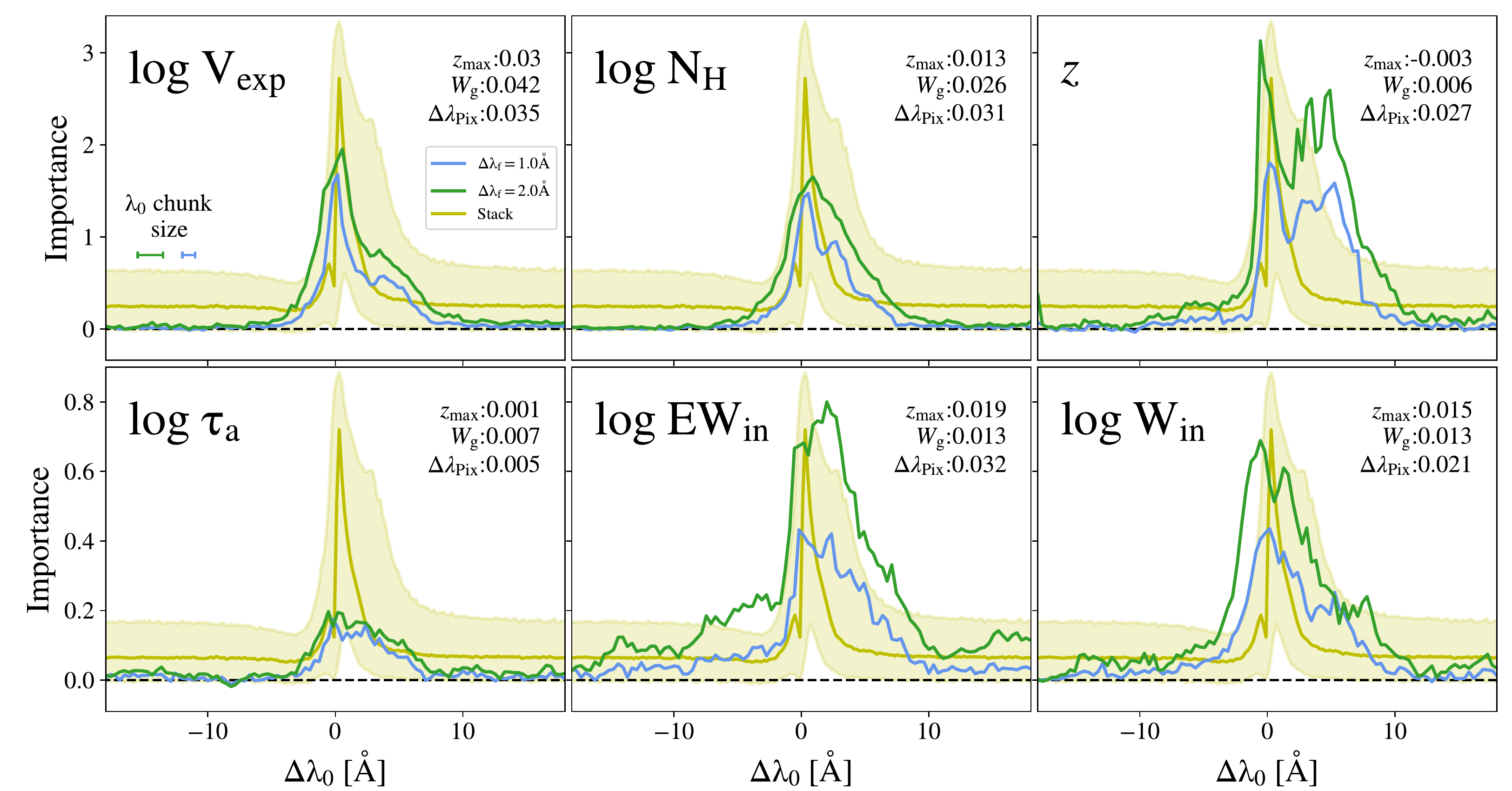}%
    \caption{Feature importance analysis for the quality configuration \wg=0.5\AA{}, \dl=0.25\AA{} and \sn=15.0. Top panels show the importance in predicting \vexp , \nh\ and $z$ from left to right respectively. Bottom panels show the importance in finding \ta , \ew\ and \w from left to right respectively. The blue lines show the importance of each wavelength chunk around the corresponding rest frame wavelength. In the top right corner of each panel we display the importance of \wg, \dl\ and  $z_{\rm max}$ determining the value of the DNN output variables. The horizontal dashed line indicates zero importance. Finally, the size bar in the top left panel illustrated the size of the wavelength chunks used in the analysis. In solid yellow we display the stack of the line profiles used, computed as the 50th percentile. The yellow shade indicates the 16th and 84th percentiles of the dristribution of line profiles.}
    \label{fig:feature}
    \end{figure*} 

    \subsection{ Inflow and outflow properties estimation in the DNN  }\label{s:DNN_application}

    We have incorporated two different methodologies in \zelda\ to estimate the inflow and outflow properties for a given \lya\ line profile using the trained DNN. These methodologies are applied indistinctly to the inflow DNN and to the outflow DNN. Basically, in the first methodology, the DNN is applied once to the \lya\ line profile (\S\ref{sss:1_iter}). Meanwhile, in the second methodology we perform a Monte Carlo perturbation of the line profile (\S\ref{sss:N_iter}), which gives an estimate of the uncertainty of the measurement.
   
    \subsubsection{Direct output from the DNN}\label{sss:1_iter}
    
    The first and most basic methodology for measuring the inflow/outflow properties of a \lya\ line profile is just to use the line profile {\it as it is}.
    
    To estimate the accuracy of this methodology, we generate several samples of mock \lya\ line profiles. We build these samples starting from the extra \lyart\ sample described in \S\ref{ss:sample}. This \lya\ profile set covers a wider range in \ew\ than the sample used for training the neural network described in \S\ref{ss:training}. Therefore, for each \{\vexp,\nh,\ta\} combination, we generate 10 new line profiles with randomly uniform \ew\ values across the range used in the DNN training, resulting in 2000 \lya\ line profiles. These line profiles are ideal in terms of signal to noise, resolution and wavelength sampling. We produce  samples with realistic quality by conducting the procedures described in \S\ref{ss:mocking} to the sample of \lya\ line profiles just described. Then, for each combination of \{\sn,\dl,\wg\}, we measure the accuracy of each property as the standard deviation of the difference between the true and the predicted values of the six output variables.
    
    In Fig.\ref{fig:spectral_quality} we show examples of the best fitting line profile produced by the DNN methodology (red) to mock line profiles with different spectral quality. The accuracy of the fit depends on the spectral quality of the line. This is quantified in Tab.\ref{tab:quality_params}, where the parameters of the best fitting spectrum are listed. In general, for the best quality configuration, the predicted parameters are close to the intrinsic values. Then, as we decrease the spectrum quality, the predicted parameters become less accurate. 
    
    In Fig.~\ref{fig:props_1_iter} we list the accuracy for the six variables that the DNN predicts, i.e., \vexp , \nh\ and \ta\ in the top row from left to right and \dlt, \ew\ and \w\ in the bottom row from left to right. In particular, we did this analysis for the sample described in \S\ref{ss:sample} for each combination of the quality variables \sn=[5.0, 8.0, 15.0], \wg=[0.1,1.0,2.0]\AA{} and \dl=[0.05,0.1,0.25,0.5,1.0]\AA{}. The line profiles were homogeneously distributed from $z$=0.001 to $z$=4. In general, the accuracy of this methodology depends on the quality of the line profile, achieving better results as the quality improves, i.e., larger values of \sn\ and lower of \wg\ and \dl. We find that this methodology predicts with a great accuracy \vexp, \nh, \ew, \dlt\ and \w. Specially for $\log$\vexp, $\log$\ew and $\log$\w\ which are recovered with a 0.6, 0.36 and 0.45 uncertainty respectfully even in the most challenging cases with the worst quality. Then, for \nh\ the accuracy is below one order of magnitude for the majority of the quality configurations explored here. For samples with good quality, the uncertainties in $\log$\vexp, $\log$\nh, $\log$\ew and $\log$\w are as low as 0.17 , 0.43, 0.15 and 0.25.  We also find that the uncertainty in determining \dlt\ is $\sim 0.31$\AA{} ($\sim76.5\;km/s$) for the best quality configurations and  $\sim 1.6$\AA{} ($\sim395.2\;km/s$) for the most challenging. This is an improved accuracy compared to other redshift estimation methods used in the literature (e.g., \citealp{Verhamme:2018aa,Byrohl_2019,Muzahid_2019,Runnholm_2020}), see \cite{gurung_2020b} for a comparison. 
    
    Regarding \ta\ we find uncertainties larger than one order of magnitude over the full quality range considered. This might be caused by the fact that the dust in some cases does not impact significantly the line profile shape. Other works in the literature that extract outflow properties from line profiles have found the same challenge estimating \ta\ \citep[e.g.][]{Gronke2017}.  
     
    \subsubsection{Monte Carlo iterations of the DNN}\label{sss:N_iter}
    
    The second methodology to extract the inflow/outflow parameters from a given \lya\ line profile  consists in perturbing the flux density $f_\lambda^{\rm Ly\alpha}(\lambda)$ by its uncertainty $\Delta f_\lambda^{\rm Ly\alpha}(\lambda)  $ and using the result line profile as input for the DNN iteratively. We will refer to this procedure as the Monte Carlo methodology.  For each random perturbation of the line profile the DNN predicts a different set of \{$\log$|\vexp|, $\log$\nh, $\log$\ta, \dlt, $\log$\ew,  $\log$\w\}. Then, the inflow/outflow parameters predicted by this methodology are defined as the percentile 50th of the distribution of them. Also, their uncertainty is computed using the corresponding percentiles of the distributions. For example, the top and bottom 1-$\sigma$ uncertainty would be the 16th and the 84th percentiles. 
    
    In Fig.~\ref{fig:outflow_dnn_ilustration} we illustrate the Monte Carlo methodology for a mock \lya\ line profile (shown as a grey cross in the corner plots and in grey in the top-right panel) at redshift 3.0 and with quality  \wg=0.5\AA{} , \dl=0.1\AA{} , \sn = 10. The outflow parameters used to generate the line profile are \vexp =200$km/s$ , \nh=$\rm 10^{18}cm^{-2}$, \ta=0.1, \w=0.1\AA{} and \ew=100\AA{}. The line profile predicted by out Monte Carlo methodology is displayed in blue, as well as the 1D and 2D distribution of the posteriors. The values of the percentiles 50th are displayed as the dashed black lines. 

    Additionally, we also illustrate the accuracy of the the MC methodology in Fig.~\ref{fig:outflow_dnn_ilustration_acc}, where we compare the intrinsic (horizontal axis) and predicted (vertical axis) outflow properties for a sub set of 200 line profiles from the 2000 used in \S\ref{sss:1_iter}. In particular we make this analysis for a quality configuration of \wg=0.5\AA{} , \dl=0.1\AA{} and \sn = 10. For this quality we find a tight correlation between the intrinsic and predicted parameters in general. Also the error in estimating \dlt\ is almost always below 1\AA{}.  

    In Fig.~\ref{fig:spectral_quality} we show four particular examples of the MC DNN performance for different spectral quality configurations. As in the DNN methodology, by looking at Tab.~\ref{tab:quality_params}, we find that spectrum with less quality have lower accuracy in the estimated parameters. In these cases, for the good quality line profiles, the shape of the spectrum and the outflow parameters are well recovered. 
    
    Here we characterize the accuracy in the outflow/inflow parameters as a function of the \lya\ line profile quality. For this, we repeat the exercise done for the direct DNN methodology, but for the MC methodology. In the top row of Fig.\ref{fig:props_1_acc} we list the standard deviation of the difference between the intrinsic and predicted parameters ($\log$|\vexp|, $\log$\nh, $\log$\ta\ from left to right) as well as in bottom panel for \dlt, $\log$\ew,  $\log$\w\ from left to right. In general, we find that, as well as in the case of the direct DNN measurement, the better the quality, the higher is the accuracy. 
    The accuracy of the MC methodology is also excellent, with uncertainties as low as 0.14 for $\log$|\vexp|, 0.34 for $\log$\nh, 0.25\AA{} ($\sim61.7\; km/s$) for \dlt, 0.12 for $\log$\ew\ and 0.22 for $\log$\w\ . Meanwhile, the uncertainty in \ta\ is about one order of magnitude. 

    A direct comparison between the accuracy of both methodologies (Fig.~\ref{fig:props_1_iter}, Fig.~\ref{fig:props_1_acc}) shows that the  MC methodology is slightly more precise than the direct measurement through the explored range of quality. 
    For example, this becomes apparent in the accuracy of \dlt\ for the best configuration considered (\wg=0.1\AA{} , \dl=0.05\AA{} and \sn = 15). Meanwhile for the direct DNN output $\sigma$(\dlt)=0.31\AA{} ($\sim76.5\; km/s$), for the MC methodology $\sigma$(\dlt)=0.25\AA{} ($\sim61.7\; km/s$). This is the case also for the other outflow/inflow properties, but \dlt\ is the one with the most prominent differences in accuracy. 

    We have explored the converge of the outflow/inflow parameters as a function of the number of iterations in the MC analysis. We have found that for 1000 iteration the percentiles converge and that increasing the number of realization does not lead to different results. Therefore, unless otherwise stated, we have used 1000 iterations in all the shown MC DNN measurements.

    \subsubsection{ Uncertainty accuracy in the Monte Carlo DNN methodology }

    In this section we test the accuracy of the uncertainties computed by the Monte Carlo deep neural network methodology. In general, it is challenging to have a good estimation of the uncertainty of the quantities predicted by deep neural networks \citep[see][]{Kuleshov2018}. For example, it might be the case that for a 90\% confidence level, more (less) than the 90\% of the times the true solution is compatible with the measurement, which would mean that the uncertainty is overestimated (underestimated). 

    To assess the accuracy of the MC DNN methodology, we compare the fraction of times that a measurement is compatible with the true observable as a function of the confidence level. Here we study the same samples used previously to quantify the accuracy of the MC methodology as a function of the line profile quality.  For example, for a confidence level of a 10\% we compute the fraction of cases $f_{\rm comp}$ in which the true quantity is between the 45th and 55th percentiles of the posterior of each outflow/inflow property returned by the MC DNN approach. 
    
    We performed this analysis for all the quality configurations explored in Fig.~\ref{fig:props_1_acc}, but here we arbitrarily focus on the sample with \wg=0.5\AA{}, \dl=0.1\AA{} and \sn = 10, as we find similar results in the other samples. Of course the accuracy of the uncertainties depends on the number of MC realization performed in the analysis. We find that for 1000 iterations the uncertainties have already converged. As illustrated in Fig.~\ref{fig:NN_errorbars}, in general the biases in the uncertainties derived using the MC DNN methodology are below the 10\% level (blue). For comparison, we also display the case of purely Gaussian measurements (red). In some cases the outflow/inflow parameter uncertainties are overestimated while others are underestimated. In this particular case, the uncertainty of \vexp, \ta\ and \w\ are underestimated. Meanwhile, the uncertainty of \nh , \ew\ and \dlt\ are overestimated. Although, this analysis probes that the uncertainty calculation is not perfect, it also shows that its bias is below the 10\%, and we regard it as a good enough estimation that we will improve it in future version of the code. 
   
    \subsection{Feature importance analysis}\label{ss:feature_importance}

In this section we perform a feature importance analysis to our DNN in order to analyze which DNN input variables are more determinant to predict the line profile properties.

There are several methodologies to estimate the impact of a DNN feature into the DNN output. One of the most common technique is {\it Feature Perturbation} \citep{Lundberg_2017,Hooker_2018}. In this methodology, the importance of a given feature in a sample  is computed by shuffling the values of a that particular feature among all the sample objects and passing the new shuffled sample to the DNN. Then, the comparison between the accuracy obtained using the original sample and the new sample with the shuffled feature determines the importance of that feature. In general, the accuracy of the DNN using the new shuffled sample will be worst, as the information contained by the target feature is removed. For example, the importance of a given feature $F$ in determining a given output variable $T$ could be define as ${\rm I}_{\rm F}^{\rm T} =  \sigma^{\rm T}_{\rm F} / \sigma^{\rm T} - 1$, where $\sigma^{\rm T}$ is the standard deviation of the difference between values of the true output variable and predicted output for the original sample. Also, $\sigma^{\rm T}_{\rm F}$ is the analogous of  $\sigma^{\rm T}$ but in the sample in which we have shuffled the feature $F$. In this way, if the accuracy heavily decreases after shuffling a given feature ($\sigma^{\rm T}_{\rm F}$ rises), it would suggest that the shuffled feature has a significant importance (${\rm I}_{\rm F}^{\rm T}$ increases). In the same way, ${\rm I}_{\rm F}^{\rm T}=0$ indicates that the accuracy is not affected by removing the studied feature. Note that since the quality of the estimate is decreased, we have generally $\sigma_{\rm F}^{\rm T} \ge \sigma^{\rm T}$ and, hence, $\rm I_{ F}^{T} \ge 0$.

We analyze the sample of $\sim 2000$ line profiles described in \S\ref{sss:1_iter} with quality \wg=0.5\AA{}, \dl=0.25\AA{} and \sn=15.0. We tested that we find similar trends for the other quality configurations.

First, we focus on the non-spectral features in the input of the DNN.
In order to study the importance of \wg, \dl\ and  $z_{\rm max}$ we apply a  similar procedure as the explained above. However, since all the line profiles have exactly the same values of  \wg=0.5\AA{}, \dl=0.25\AA{}, shuffling these features across the sample would not change the result. Instead, we assign each line profile random values of \wg\ and \dl\ homogeneously distributed in the range in which the DNN was trained (see \S\ref{ss:training}). Meanwhile, $z_{\rm max}$ is shuffled across the population, which is homogeneously distributed from $z=0.001$ to $4$. For each line profile we perform 1000 perturbations. 

We show the results in the top left corner of each panel of Fig.~\ref{fig:feature}. Each panel corresponds to a different output variable, \vexp, \nh\ and $z$ from left to right of the top row and \ta , \ew\ and \w\ from left to right in the bottom row. In general we find that the importance of these three features is different on predicting each of the six output variables. Also, these features exhibit a small impact  ($\rm I<0.05$) on the output of the DNN. In the most extreme case, the importance of \wg\ determining \vexp\ is 0.042, which means that the accuracy drops a 4.2\% when \wg\ is removed from the analysis. Additionally, we find that \wg, \dl\ and  $z_{\rm max}$ have a impact greater than 1\%  in predicting \vexp, \nh, \ew\ and \w. Meanwhile, only \dl\ has an impact greater than 1\% in determining the redshift of source. Finally, none of the three variables seem to be important to determine \ta , as they affect less than 1\% the accuracy.   
 
Regarding the importance of the spectral features, we find it interesting to study the importance of the rest frame wavelength into determining the output outflow parameters and redshift. However, note that, the flux density that we use as input to the DNN is in the proxy rest frame that assumes that the maximum of the line profile is \lya, as explained in \S\ref{sss:input_output}. Therefore, we need to make some adjustments to the procedure described above, but keeping the same philosophy. So, instead of perturbing the features corresponding to the line profile, we are going to perturb the observed line profiles directly. 

Another challenge in determining the importance of a given rest frame wavelength is that the flux density bins are not independent among them. This implies that even if we remove the information of only one spectral bin, the neighbour bins would still contain information about the removed bin. This can affect to the determination of the importance of a given line profile bin, as its information is still fed to the DNN through nearby bins. In order to overcome this challenge, instead of removing the information of individual density flux bins, we remove the information of flux density chunks of width 1\AA{} (blue) and 2\AA{} (green) in the rest frame of the source. 

To sum up, the procedure to study the importance of a given density flux chunk centered at $\rm \lambda_0^{F}$ with width $\rm \Delta\lambda_{f}$ is the following. First, we convert from observed frame to rest frame all the line profile sample. Second, we measure the PDF of all the density flux bins in the rest frame wavelength window ranging from $\rm \lambda_0^{F}-\Delta\lambda_{f}/2$ to  $\rm \lambda_0^{F}+\Delta\lambda_{f}/2$. Third, we substitute the density flux values in the same wavelength chunk by random values of flux density following the same PDF. In this way, we remove all spectral the information  within $\rm \lambda_0^{F}\pm\Delta\lambda_{f}/2$. Finally, we apply the DNN to the new sample with the removed information and measure the accuracy in the outflow parameters. We make this analysis for 100 rest frame wavelength chunks and 1000 perturbations each. 

We present the results of the feature importance analysis of the rest frame wavelength in Fig.\ref{fig:feature}. We show the result of this analysis for two chunk sizes, $\rm \Delta\lambda_{f}=1.0$\AA{} (blue) and $\rm \Delta\lambda_{f}=2.0$\AA{} (green). Overall, both wavelength chunk sizes exhibit similar trends. In particular, the importance of the wavelength chunks of size $\rm \Delta\lambda_{f}=2.0$\AA{} is greater than those of size $\rm \Delta\lambda_{f}=1.0$\AA{}. This is caused by the fact that the larger the chunk size, the more information is removed.

In general, we find that each rest frame wavelength chuck contributes differently to each DNN output variable. A common pattern is that the regions $\pm 5$\AA{} of \lya\ contribute the most to determining the output. Also, the importance tends to extend more towards redder wavelengths than to bluer. This might be due to the fact that in the \ThinShell\ outflow, in the explored parameter volume, there is more variance in the shape of the red peaks than in blue peaks. For example, for large values of \vexp\ the line profiles tend to exhibit only an asymmetric wide red peak with a red tail. Meanwhile, no combination of parameters produce a blue peak with a blue tail as extended as in the red case. It is noticeable that some wavelengths chunks play an important role in determining \vexp, \nh\ and $z$, as the importance goes up to 1.5, i.e., the accuracy drops a 150\% when these chunks are removed. In comparison, the chunks in \ta, \ew\ and \w\ have less importance as they go up to $\sim 0.4$ (notice the different scale in the Y-axis). 

From this analysis it is clear how some of the output variables are estimated in the DNN. Overall, we find that that naturally the typical spectral width is imprinted in the importance curve. However, there are some additional interesting features. The most apparent  is \ew, where the range $\Delta\lambda_0<-5$\AA{} has a significant importance. In contrast, these chunks exhibit little importance (<0.05) in the other output variables. One possible interpretation is that the DNN uses this wavelength range to estimate the continuum, while it uses the chunks close to \lya\ to determine the injected flux. The combination of these two properties would give \ew.  

Another interesting feature appears in the importance of determining the redshift, which exhibits a small bump at $\Delta\lambda_0\sim-5$\AA{}. As pointed out by \cite{Verhamme:2018aa}, the position of the blue and red peaks in a \lya\ line profile can give a good estimation for the redshift. The small bump in the importance could be seen as if the DNN was using these wavelength to estimate the redshift in some cases. 

\begin{figure*} 
    \includegraphics[width=2.31in]{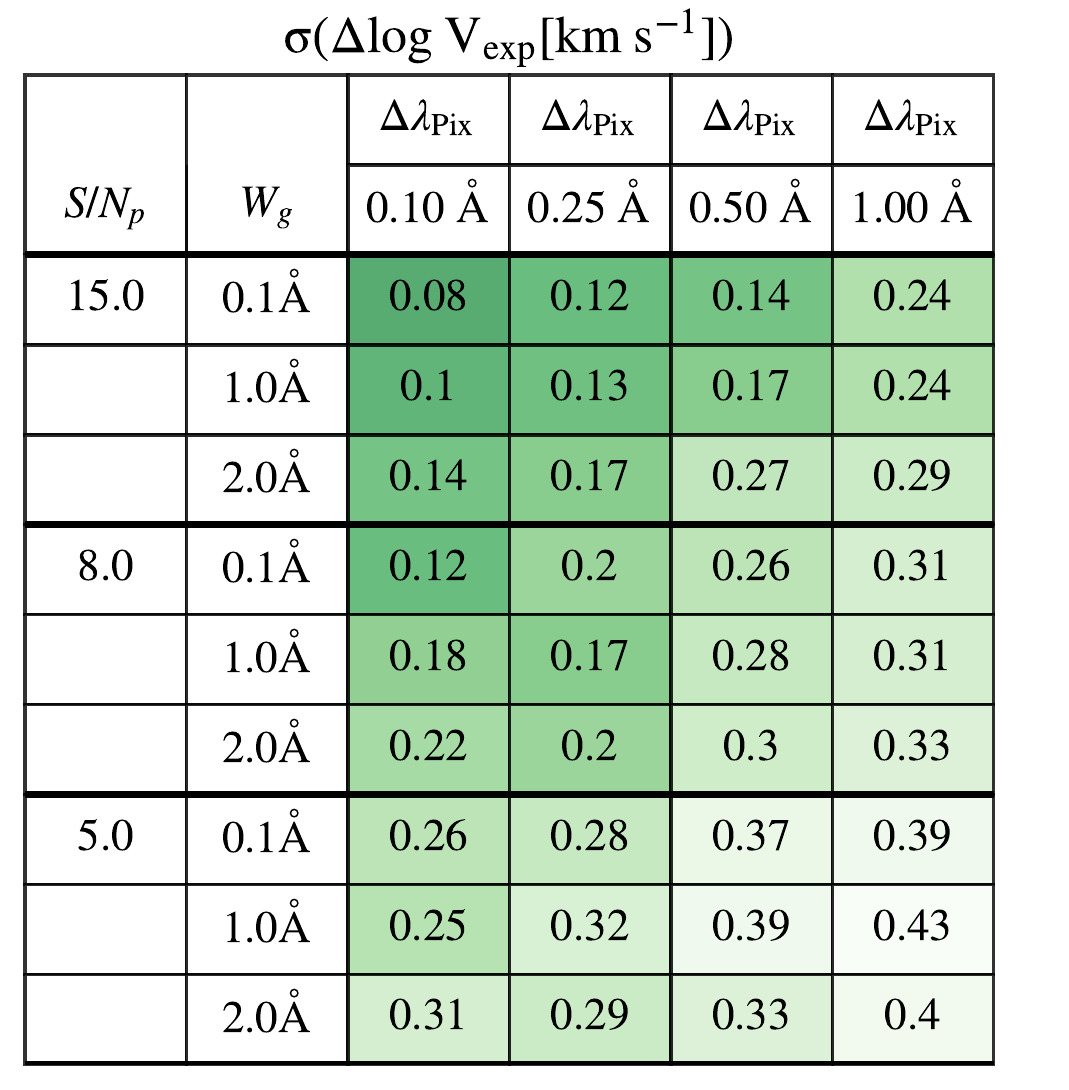}%
    \includegraphics[width=2.31in]{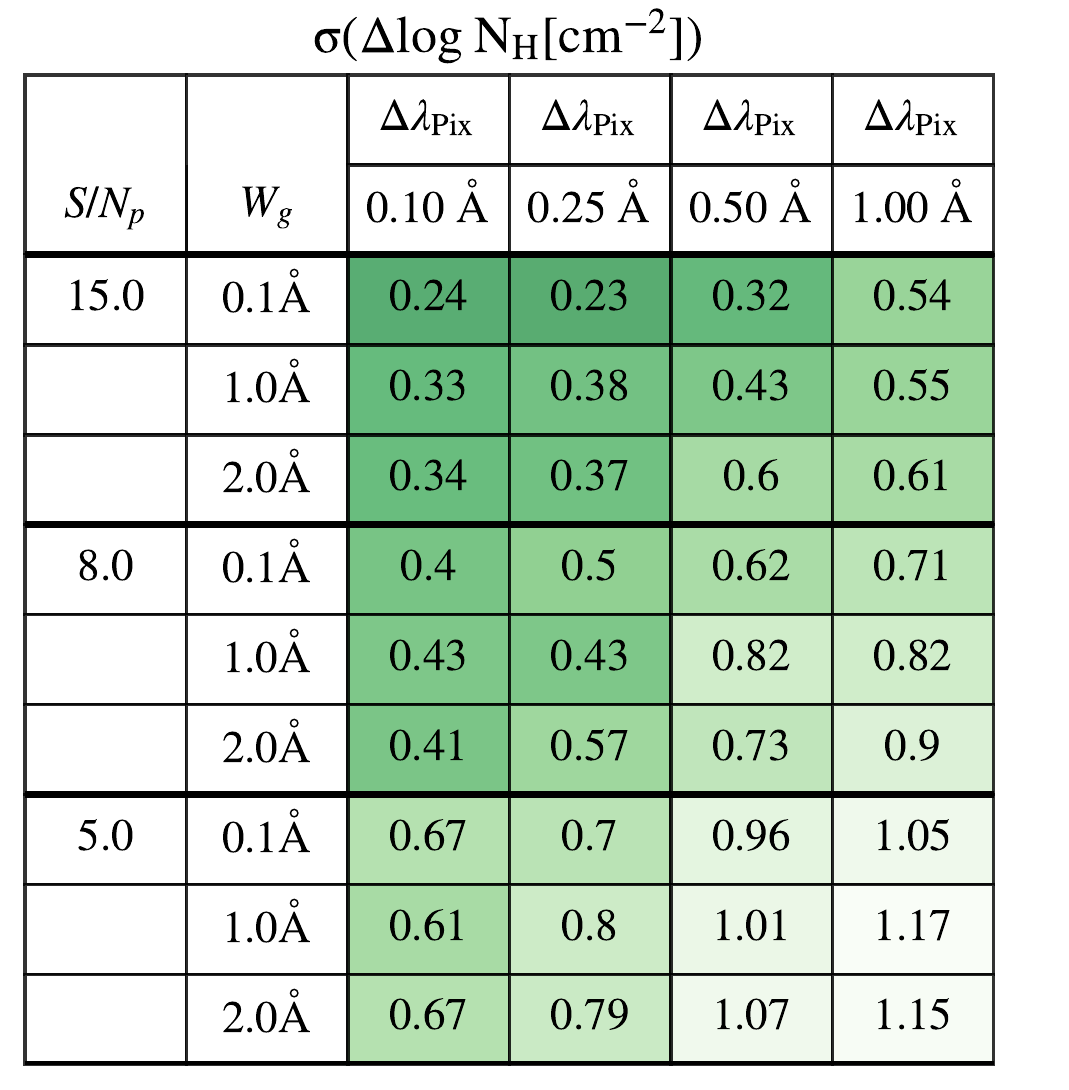}%
    \includegraphics[width=2.31in]{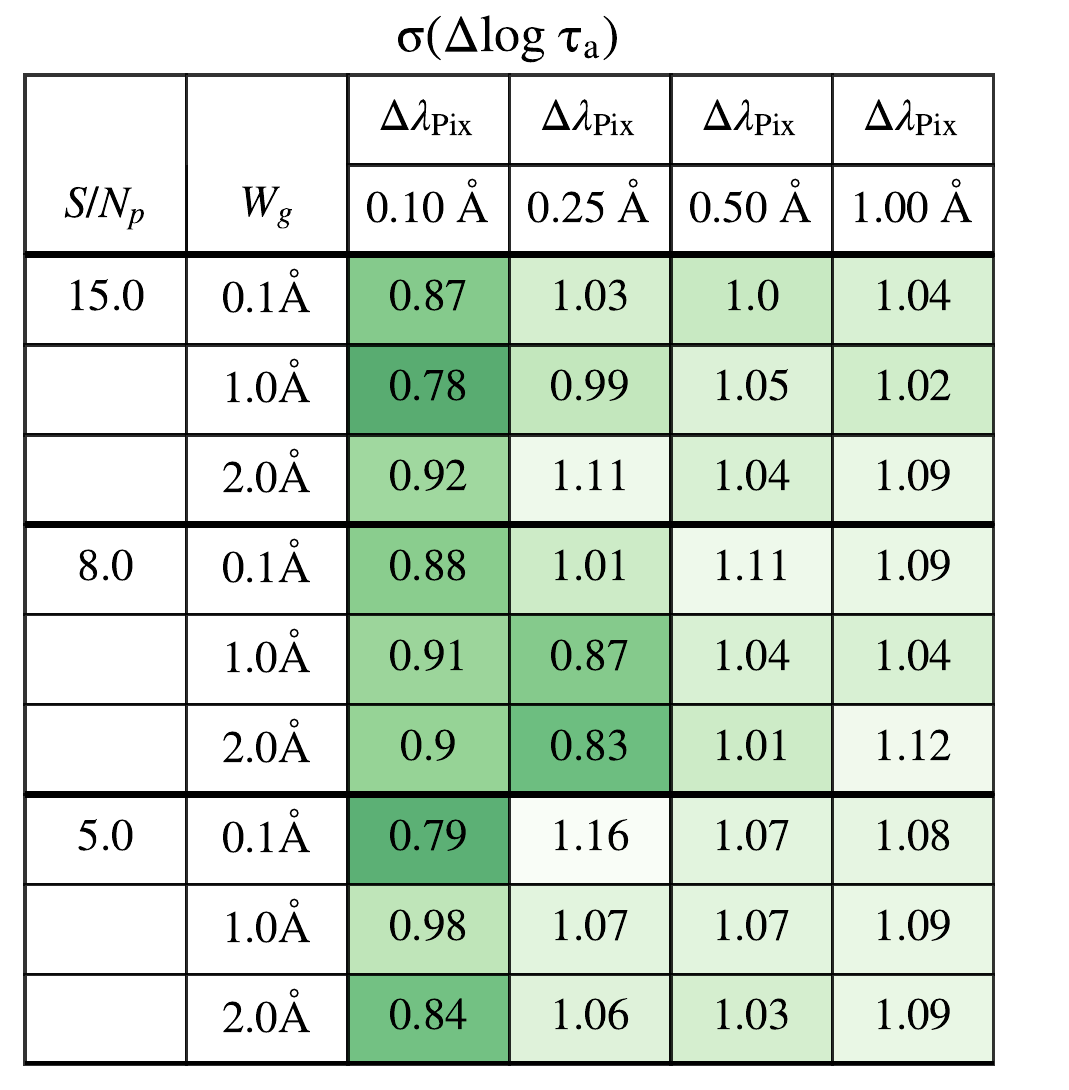}
    
    \vspace{0.2in}
    
    \includegraphics[width=2.31in]{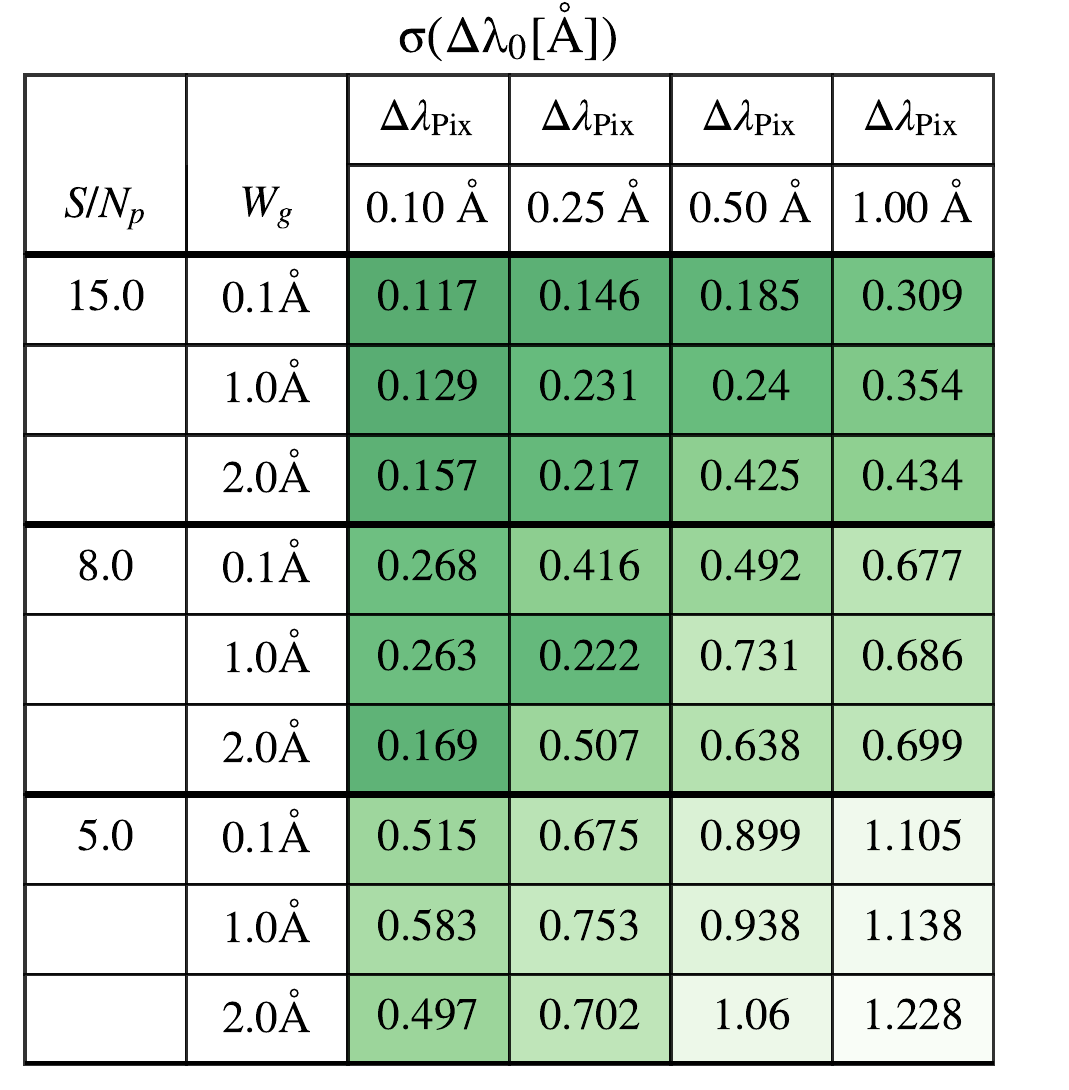}%
    \includegraphics[width=2.31in]{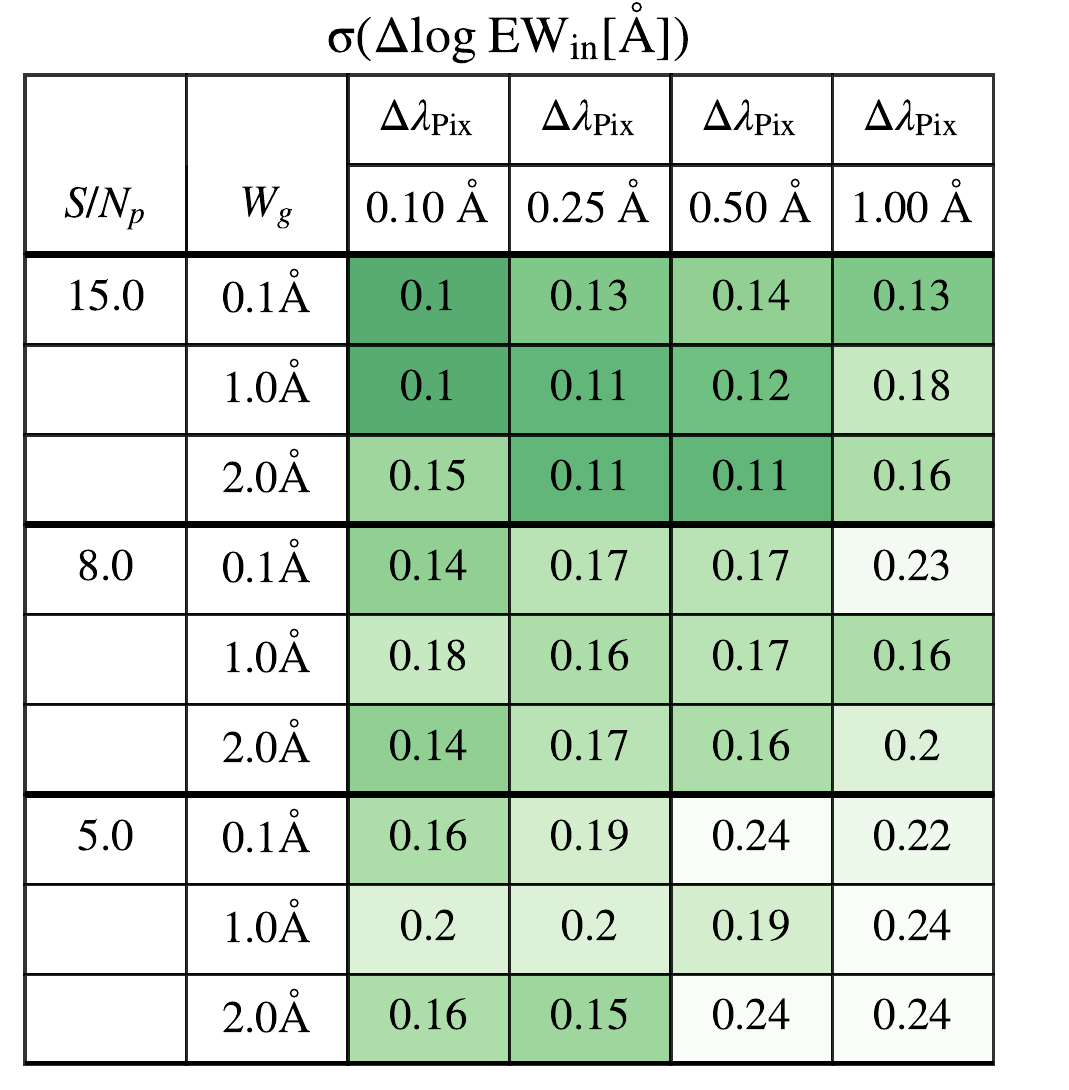}%
    \includegraphics[width=2.31in]{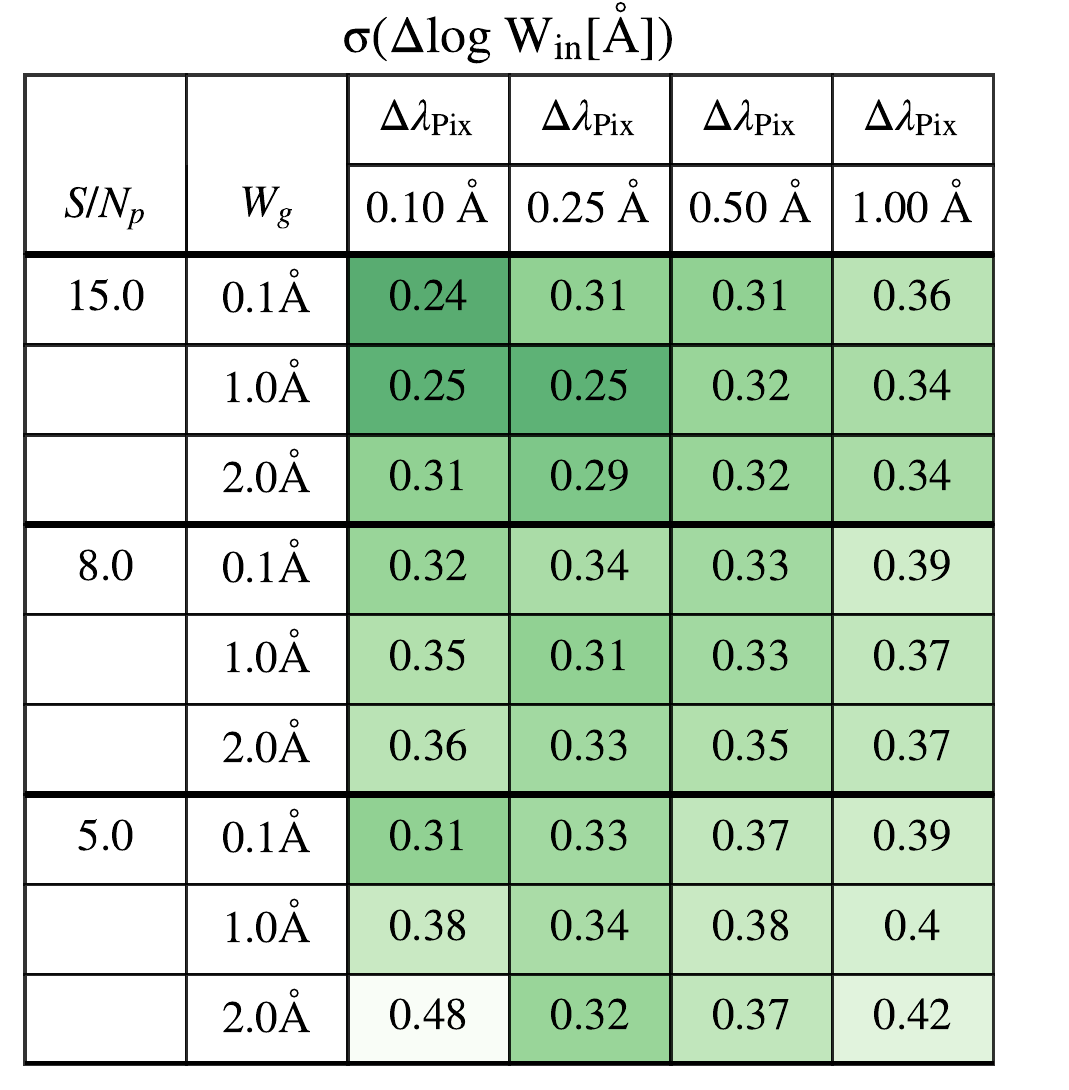}%
    \caption{ Standard deviation of the difference between the true and the predicted inflow/outflow parameters in the MCMC methodology. In the row, \vexp, \nh\ and \ta\ from left to right. In the bottom row, \dlt , \ew\ and \w\  from left to right. Cells are colored by their value and darker means lower (better).  }
    \label{fig:props_mcmc}
    \end{figure*}

    \section{ Fitting \lya\ line profiles with Monte Carlo Markov Chain }\label{s:mcmc}
    
    In addition to the deep neural network approaches, \zelda\ also includes a Monte Carlo Markov Chain methodology to infer the outflow/inflow properties from an observed \lya\ line profile. This technique has already been explored in the literature, giving reasonable results for fitting observed \lya\ line profiles \citep[e.g.][]{Gronke_2015}. In \S\ref{ss:methodology_MCMC} we describe the MCMC methodology, while in \S\ref{ss:performance_MCMC} we analyze its performance on mock spectrum. 
    
    \subsection{Methodology}\label{ss:methodology_MCMC}
    
    In our MCMC scheme the fitting is done in the observed frame using \texttt{emcee} \citep{Foreman_Mackey_2013}. There are six variables in our MCMC approach: \{\vexp, \nh, \ta , \ew  , \w , $z$\}. For each step of each walker we compute a \lya\ line profile with the same \wg\ and \dl\ as the observed spectrum, as described in \S\ref{s:modeling}. In particular, given the observed density flux of a \lya\ line profile $f_{\rm \lambda}(\lambda)$ and its uncertainty $\sigma_{\rm \lambda}(\lambda)$ with $n$ wavelength bins evaluated in $\lambda_n$, we minimize the logarithmic of the likelihood, i.e.,
    
    \begin{equation}
        \log \mathcal{L} = \frac{1}{2} \displaystyle\sum\limits_{n}\left[ 
        \frac{
        (f_{\rm \lambda}(\lambda_n)
        - 
        f_{\rm \lambda}^m(\lambda_n) )^2
        }
        {\sigma^2_{\rm \lambda}(\lambda_n)} - \ln\sigma^{2}_{\rm \lambda}(\lambda_n) \right],
    \end{equation}\label{eq:mcmc}
    where $f_{\rm \lambda}^m(\lambda_n)$ is a mock \lya\ line profile computed with the same \wg\ and \dl\ as the observation and \{\vexp, \nh, \ta , \ew  , \w , $z$\} of that walker in that step. 

    The procedure followed in the MCMC scheme is:
    
    \begin{enumerate}
        \item{ We define a redshift range where to run the MCMC. For this, first we find the wavelength global maximum $\lambda_{\rm max}$ of the line and compute the redshift $z_{\rm max}$ assuming that $\lambda_{\rm max}$ corresponds to the observed \lya\ wavelength. Then the redshift interval is $z_{\rm max}\pm0.002$.}
        
        \item{ In order to initialise the model parameters we perform a fast Particle Swarm Optimization (PSO) using \texttt{PySwarms} \citep{Miranda_2018} minimizing 
        
        \begin{equation}
        \chi^2 =  \displaystyle\sum\limits_{n}\left[ 
        \frac{
        (f_{\rm \lambda}(\lambda_n)
        - 
        f_{\rm \lambda}^m(\lambda_n) )^2
        }
        {\sigma^2_{\rm \lambda}(\lambda_n)}  \right].
        \end{equation}
        This analysis is a first attempt to fit the observed line profile and, although even if the fit is not perfect, it narrows down the six dimensional volume. The initial position of the walkers is then set around the PSO solution. 
        
        Additionally, in \zelda\ there are other two methods for initialising the walkers for the MCMC. i) covering homogeneously the 6D volume and ii) using the output of the MC DNN analysis. However, in the results shown in this work we always make the walkers initialization with the PSO algorithm.}
        
        \item{ We perform a first MCMC iteration with  500 walkers, with a 'burn-in' phase of 200 steps and a consecutive run of 1000 steps. In general, the this leads to a complex and multi modal likelihood distribution \citep[see][]{Gronke_2015}. This process yields the parameters with maximum likelihood. }
        
        \item{ Then we perform a second iteration of the MCMC, running with the same settings but using for initializing the walkers around the peak positions computed in the previous step. }
        
        \item{ We repeat this process until the solutions of the actual MCMC run and the previous iteration are compatible. Normally, only two iterations are required. }
        
    \end{enumerate}

\begin{figure*} 
    \includegraphics[width=6.9in]{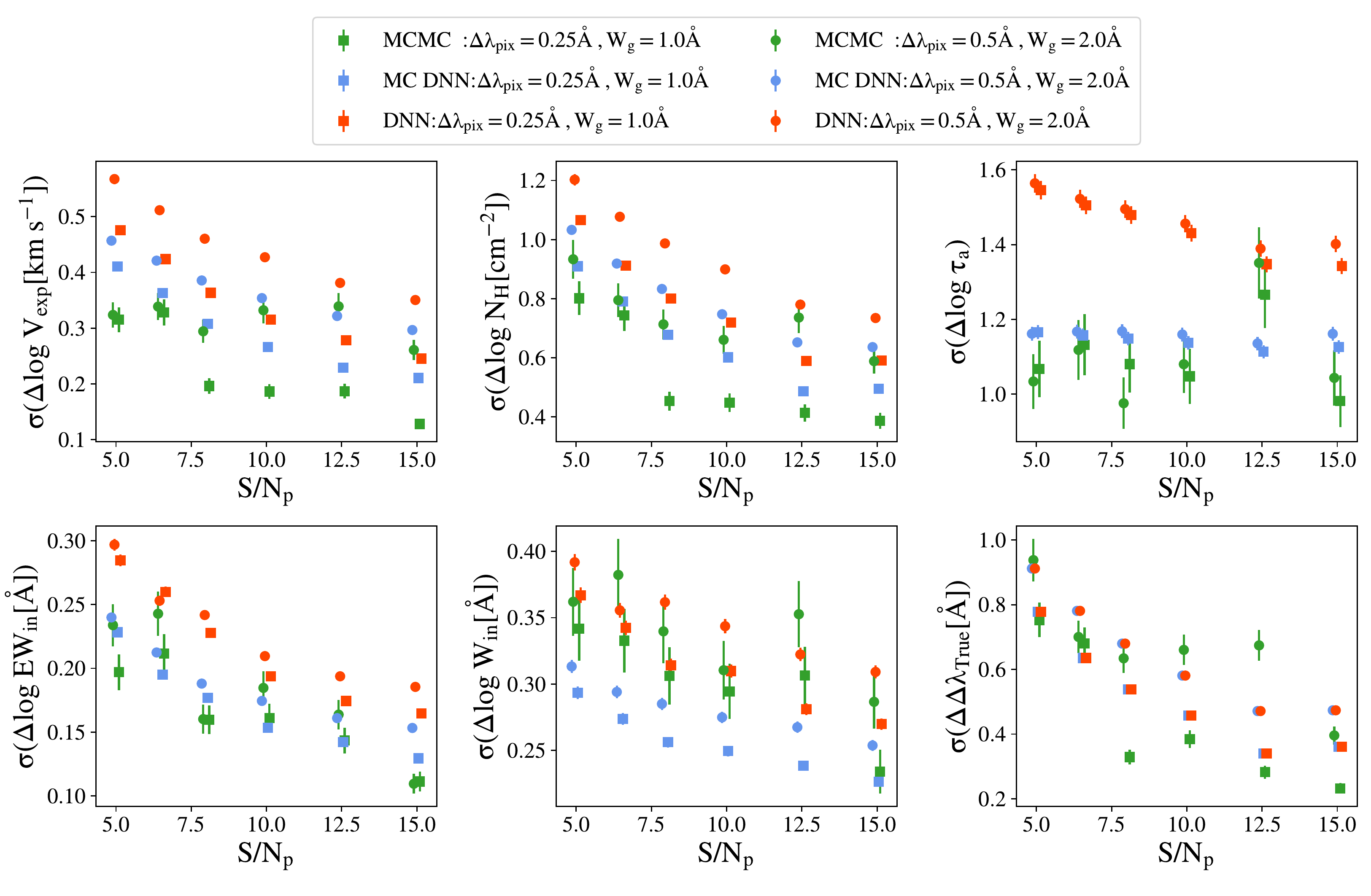}
    \caption{  Standard deviation between the input and predicted quantities for the DNN (red), MC DNN (blue) and the MCMC (green) methodologies as a function of \sn. The shown outflow properties are \vexp, \nh, \ta\ (top from left to right), \ew , \w\ and $\rm \Delta\lambda_{True}$ (bottom from left to right). Here we show the results for two spectrum qualities: \wg=1.0\AA{}, \dl=0.25\AA{} (squares) and \wg=2.0\AA{}, \dl=0.5\AA{} (dots). }
    \label{fig:comparison_acc}
    \end{figure*}
    
\subsection{Performance}\label{ss:performance_MCMC}

In Fig.~\ref{fig:spectral_quality} we show the four examples of line profiles of different qualities fitted with the MCMC methodology (green), while the outflow parameters are listed in Tab.\ref{tab:quality_params}. The MCMC methodology predicts outflow parameters close to those intrinsic. As the spectral quality is decreased, the parameters become less accurate progressively but are still consistent with the true parameters.

In Fig.~\ref{fig:props_mcmc} we list the accuracy of the MCMC methodology as a function of the quality of the line profile. For these analysis we used 50 randomly chosen mock line profiles from the samples described in \S\ref{sss:1_iter}. For the 50 mock line profiles, the redshift and outflow properties \{\vexp,\nh,\ta,\ew,\w\} are the same across all the quality configurations. We made this analysis with only 50 line profiles per quality configuration due to the heavy computational cost of the MCMC analysis (see \S\ref{s:discussion}).

Overall, as for the DNN and MC DNN schemes, the performance of the MCMC methodology improves as the quality of the spectra increases. We find that for the best quality configurations considered, \sn=15, \dl=0.1\AA{} and \w=0.1\AA{}, the \vexp\  accuracy is as good as $10^{0.08}km/s$. For this quality configuration, the accuracy in the determination of the \lya\ wavelength is also good with $\rm \sigma(\Delta\lambda_0)=\sim0.12$\AA{}. 

\section{ Comparison between methodologies }\label{s:discussion}

In this section we compare directly the different methodologies available in \zelda. As discussed, \zelda\ contains three methodologies to fit \lya\ line profiles that could be split into two categories; those using the deep neural network (DNN and MC DNN) and that using the Monte Carlo Markov Chain algorithm (MCMC). Each of these categories have a different philosophy in the fitting procedure. While the MCMC fits the shape of line profile to get the outflow parameters, the deep neural network procedures fit the outflow parameters and as a 'by product' they reproduce the shape of the line.  These methodologies also have different accuracy and different computational cost, being the MCMC procedure much more expensive than the deep neural network approaches (see below).

On one hand, the MCMC samples the parameter space looking for the region in which the mock line profile fits better the shape of the the target line profile through Eq.\ref{eq:mcmc}. We have tested the accuracy of the MCMC procedure in \S\ref{ss:performance_MCMC} on mock line profiles, finding that, by fitting the shape of the line profile with the MCMC, the output outflow parameters match the input parameters (with some uncertainty). On the other hand, DNN and MC DNN make use of a deep neural network in which the input is the line profile and its quality and the output are the outflow parameters (see \S\ref{sss:input_output}). In this sense, the deep neural network is trained to fit the relation between the line profile and the outflow parameters. In \S\ref{s:DNN_application} we tested that the outflow parameters are well recovered (also with some uncertainty) using the DNN and MC DNN methodologies. Then, as consequence of the accurate estimation of the outflow parameters, the line profile shapes are also relatively well recovered (KS$\sim10^{-1.2}$, see \S\ref{sss:shapes}.)

\begin{figure*} 
    \includegraphics[width=3.2in]{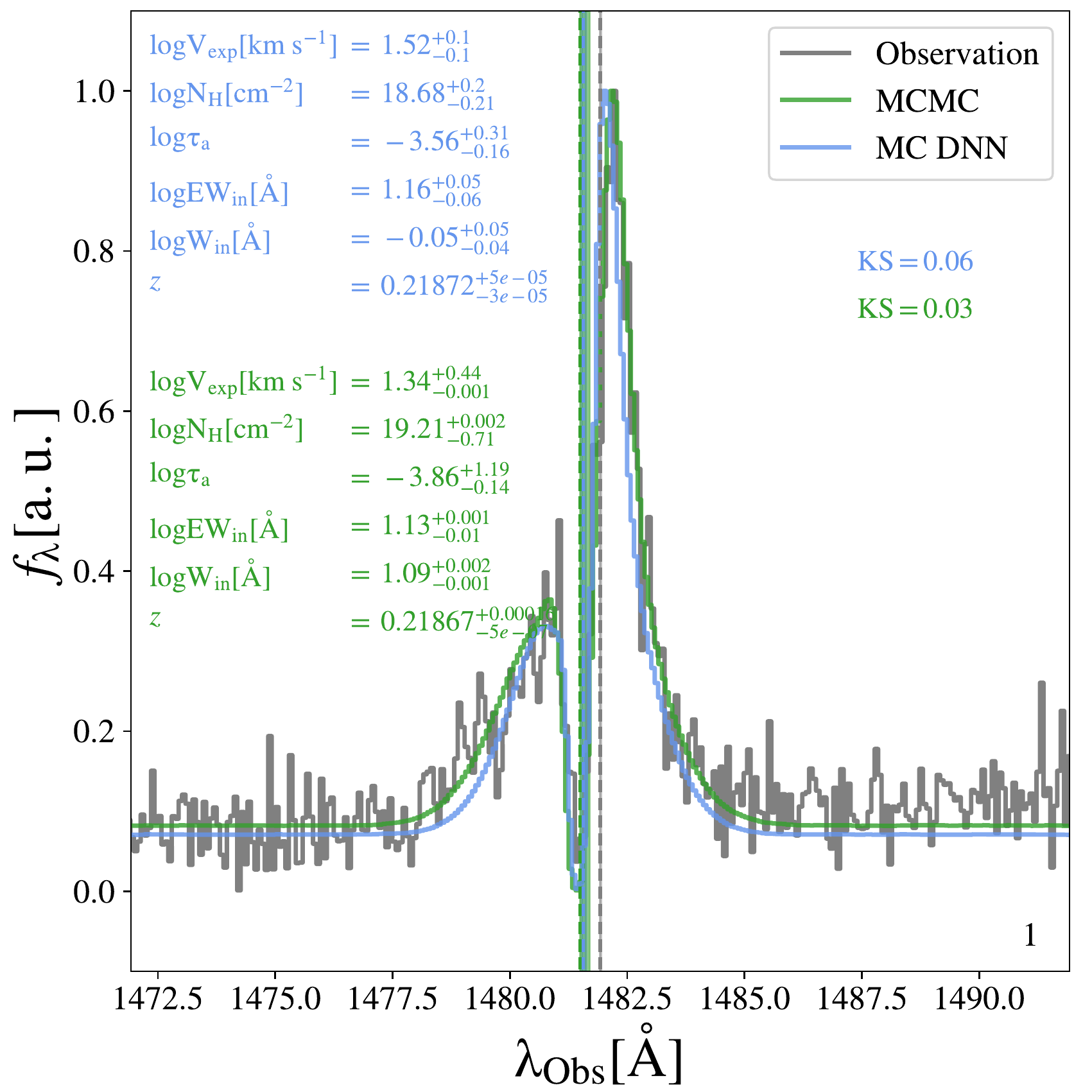} %
    \includegraphics[width=3.2in]{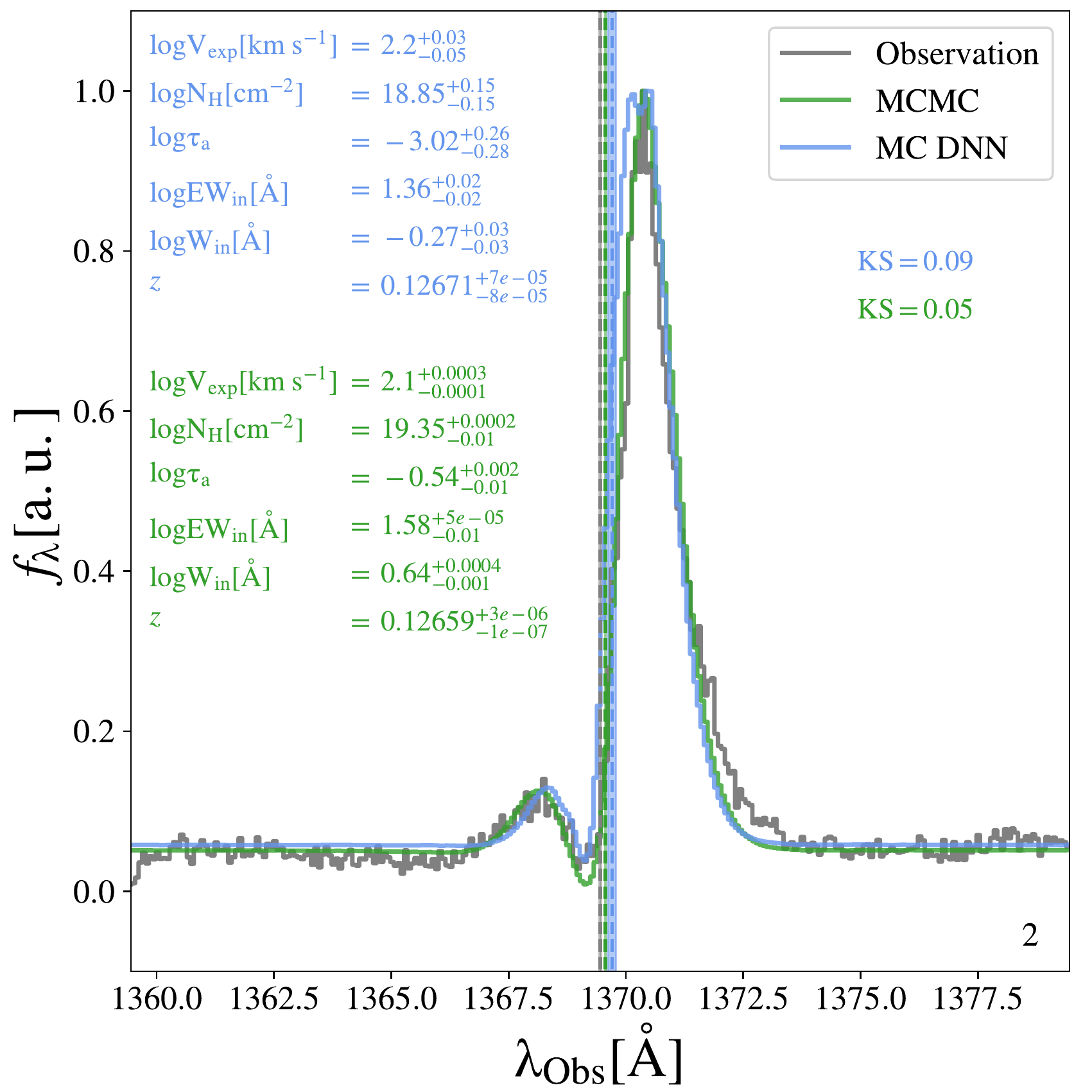}
    \caption{ Two examples of the performance of the MC DNN and MCMC approaches to extract the outflow properties of real observed \lya\ line profiles. The model lines given by the MC DNN (MCMC) methodology are shown in blue (green) and the value of the outflow parameters and the 1-$\sigma$ uncertainties are indicated in the same color in the top left (middle left). The vertical dashed lines and shaded regions mark the value and the 1-$\sigma$ uncertainty of the true \lya\ wavelength of the observation (grey) and the \lya\ wavelength predicted by the MC DNN (blue) and MCMC (green). The KS estimator is given on the right of each panel. On top, for the MC DNN and on the bottom for the MCMC approach. }
    \label{fig:examples_observed}
    \end{figure*}
    
    \begin{figure*} 
    \includegraphics[width=6.8in]{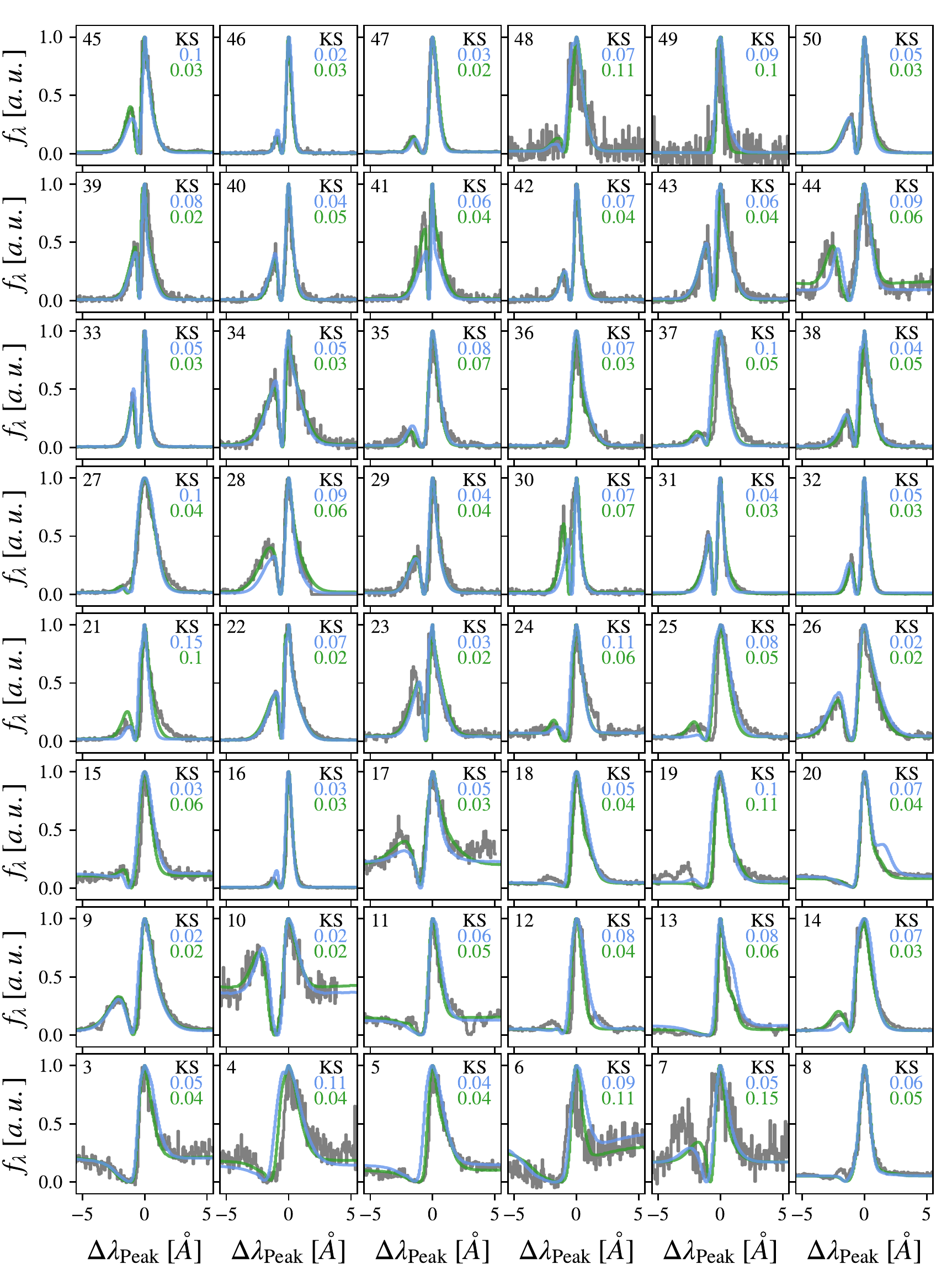}
    \caption{ Comparison between the observed line profiles (grey) and the predicted line profiles by the MC DNN (blue) and MCMC (green) methodologies. In order to make a better comparison of the shape, the line profiles are shown assuming that the maximum of the line is the true \lya\ wavelength. In the top right corner we display the KS estimator values for the MC DNN (top, blue) and the MCMC (bottom, green) methodologies. }
    \label{fig:comparison_NN_MC_lines_0}
    \end{figure*}
    
    \begin{figure*} 
    \includegraphics[width=6.8in]{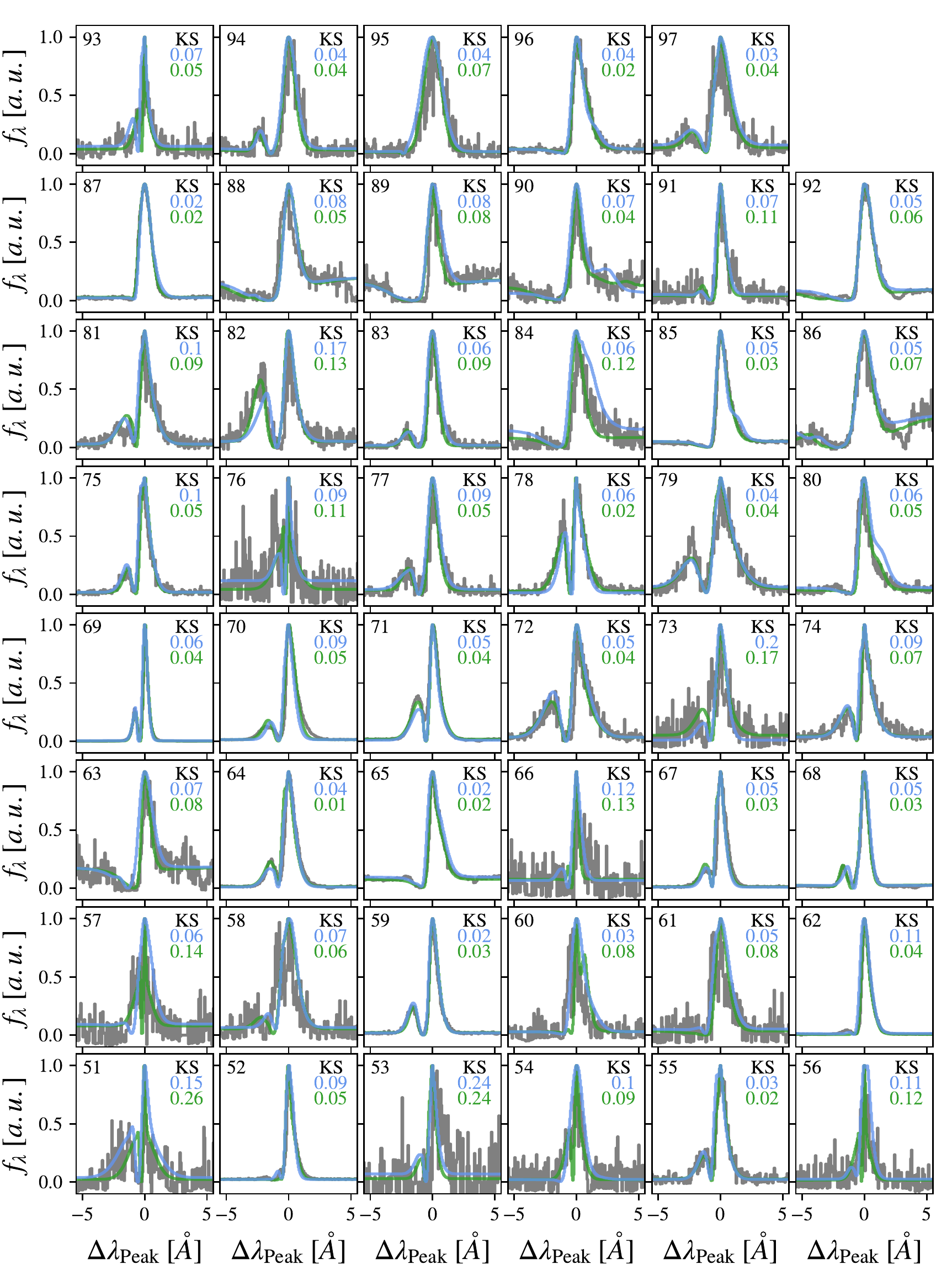}
    \caption{ Continuation of Fig.~\ref{fig:comparison_NN_MC_lines_0}.    }
    \label{fig:comparison_NN_MC_lines_1}
    \end{figure*}
    
    \begin{figure*} 
    \hspace*{\fill}%
    \includegraphics[width=3.0in]{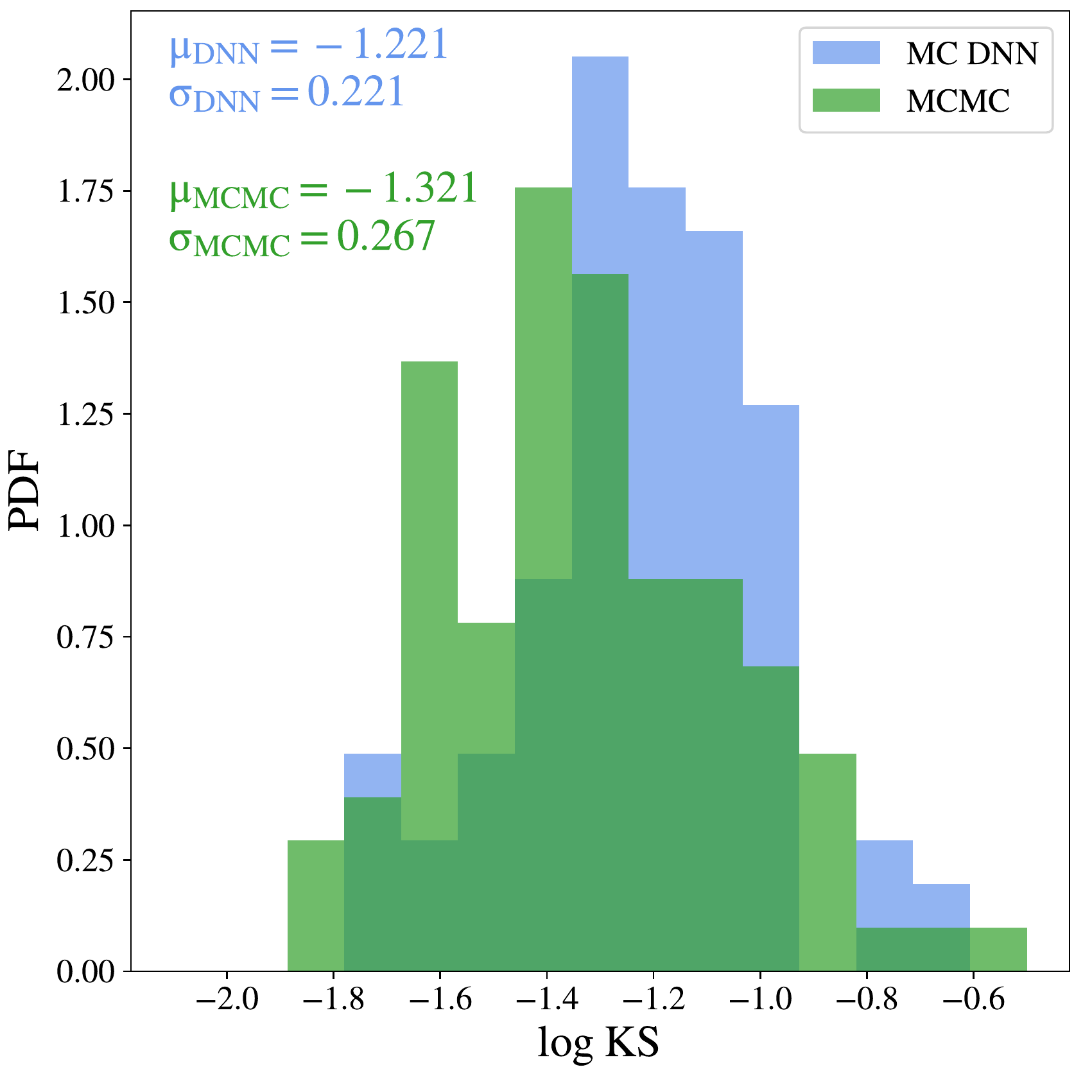}\hfill%
    \includegraphics[width=3.0in]{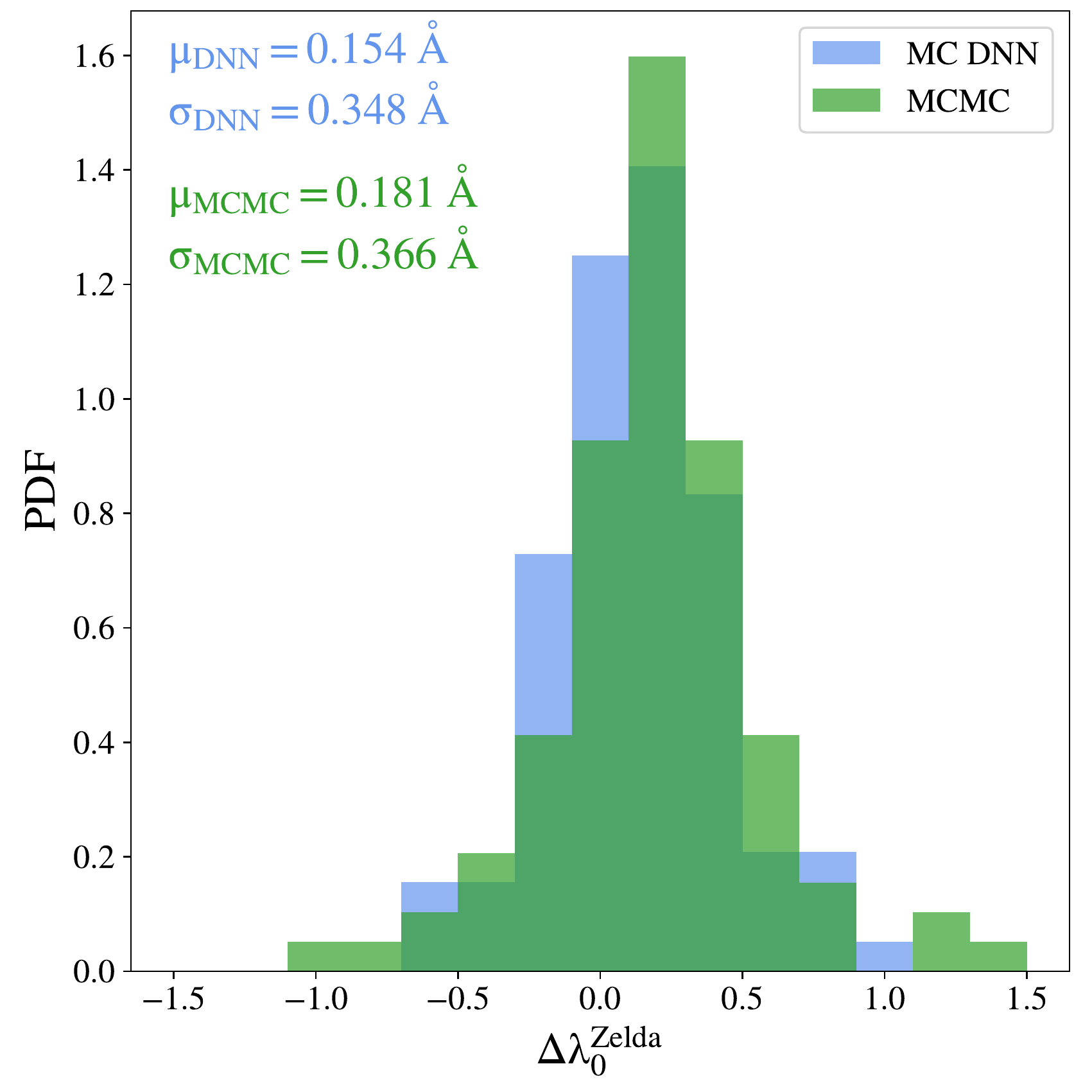}%
    \hspace*{\fill}%
    \caption{ {\bf Left}: Distribution of the logarithm of the KS estimator for the observed \lya\ line profiles by the MC DNN (blue) and the MCMC (green) methodologies. {\bf Right}: Distribution of the difference between the true \lya\ wavelength (given by the systemic redshift) and the wavelength assigned as \lya\ in the rest frame of the galaxy. For both panels, the mean ($\mu$) and the standard deviation ($\sigma$) of the distributions are indicated in the top left (MC DNN) and in the middle left (MCMC). }
    \label{fig:redshift_accuracy_obs}
    \end{figure*}  
    
The MCMC and the methodologies using the deep neural network also exhibit different accuracy. In Fig.\ref{fig:comparison_acc} we compare the accuracy of the DNN (red), MC DNN (blue) and MCMC (green) methodology as a function of the \sn\ for two quality configurations: \wg=1.0\AA{}, \dl=0.25\AA{} (squares) and \wg=2.0\AA{}, \dl=0.5\AA{} (dots). For the DNN and MC DNN methodologies we used the samples of ~2000 mock line profiles also used in \S\ref{sss:1_iter}. Thus, the accuracy values presented here are the same those in Fig.\ref{fig:props_1_iter} for the DNN and in Fig.\ref{fig:props_1_acc} for the MC DNN. Meanwhile, due to the computational cost of the MCMC procedure we used a subsample of 100 line profiles. The error bars indicate the uncertainty in the standard deviation of the difference between the input and output parameters. Overall we find that as we improve the spectral quality, the accuracy in determining the parameters improve. In general, we find that the accuracy of the MCMC is the best, although it is closely follower by the MC DNN and then by the DNN. The MCMC methodology exhibits the best accuracy in determining \vexp, \nh , \ta\ and \dl -- albeit at a much higher computational cost as discussed below. Then, the accuracy in \ew\ and  is similar between the MC DNN and the MCMC methodologies. Finally, for \w , the MC DNN procedure seems the most accurate. 

Finally, in terms of computational cost, the MCMC methodology is considerably more expensive than the MC DNN and the DNN. The MC DNN approach is computationally cheap as it basically only needs to load the trained neural network and perform matrix operations. The typical executions time for the MC DNN methodology is $\sim10s$ (1000 iterations, one core) per line profile  and needs 12MB of {\it Random Access Memory} (RAM). Meanwhile the MCMC is expensive, as in each step of each walker a mock line profile has to be computed from the \lyart 's grid and compared with the observations. The MCMC analysis takes about $1.5h$ (one core) per line profile  and it needs $\sim$10GB of RAM memory, as it has to keep in RAM \lyart 's grids. The cheap cost of the MC DNN methodology allow for a wider variety of analysis. For example, in \S\ref{ss:feature_importance} we performed a feature importance analysis, showing which regions of the line profile contain information about the outflow parameters. In principle, the same analysis can be done using the MCMC, however, due to the computational cost, performing a feature importance analysis with the MCMC method is challenging. For the analysis presented in  \S\ref{ss:feature_importance} we predicted the outflow parameters using the MC DNN for a total of $8\times10^{6}$ line profiles. For the time expenses of the MC DNN, this was feasible.

For the current number of available \lya\ line profiles (a few hundreds in \lasd) the MCMC is a helpful tool and in a few hundreds of hours of computational time the full analysis can be made. However, future surveys will provide thousands of \lya\ spectra and the use of deep neutral networks such as \zelda\ will become a crucial tool to analyse them.

\section{ \zelda\ performance on observational data. }\label{s:observation}

    So far we have demonstrated the accuracy of \zelda 's methodologies to extract the inflow/outflow parameters from mock \lya\ line profiles. In this section we apply our methodologies to the observed \lya\ line profiles available in the literature. We describe the used observation data in \S\ref{ss:data_props}, while in \S\ref{ss:results_obs} we present our results.
    
    The systemic redshift is an observable that can be measured with other means, such as finding other emission and/or absorption lines unaffected by the complex \lya\ RT, therefore, in principle,  more reliable. Then, we can compare the systemic redshift obtained from the \lya\ line profile and the other methodologies to quantify \zelda 's accuracy. 

    \subsection{Observational data description}\label{ss:data_props}
    
    The observational data used in this work was obtained through the {\it Lyman alpha Spectral Database}  \citep[\texttt{LASD}][]{Runnholm_2020}\footnote{\url{http://lasd.lyman-alpha.com}}, that contains to date more than 300 \lya\ emission lines at between redshift 0 and 6.6. As we want to test the redshift accuracy of \zelda\ we focus on the local \lasd\ sample with systemic redshifts estimated using other emission or absorption lines than \lya . \lasd\ contains a total of 107 \lya\ line profiles with systemic redshift. 
    From the original 107 line profiles we remove i) line profiles with a very steep continuum around \lya,   ii) spectra with low \sn\ and iii) spectra that contain apparently the \lya\ emission line of several sources. The excluded spectrum is discussed in Appendix \ref{s:excluded}.  This leaves a total of 97 \lya\ line profiles from redshift 0 to 0.44 for which the line profiles are shown in grey in Figs.\ref{fig:examples_observed}, \ref{fig:comparison_NN_MC_lines_0} and \ref{fig:comparison_NN_MC_lines_1}. 
        
    The \lya\ line profiles used in this work were obtained  by the Cosmic Origins Spectrograph \citep[{\it COS}][]{Green_2012} on board the {\it Hubble Space Telescope} ({\it HST}) in the General Observers (GO): GO 11522 and 12027 \citep[PI: Green,][]{salzer_2001,Wofford_2013}, GO11727 and 13017 \citep[PI: Heckman,][]{Heckman_2011,Heckman_2015}, GO 12269 \citep[PI: Scarlata,][]{Songaila_2018}, GO 12583  \citep[PI: Hayes][]{Hayes_2014,Rivera-Thorsen_2015}, GO12928  \citep[PI: Henry][]{Henry_2015}, GO 13293 and 14080 \citep[PI: Jaskot][]{Jaskot_2014,Jaskot_2017}, GO 14201 \citep[PI: Malhotra][]{Yang_2017} and GO 13744 \citep[PI: Thuan][]{Izotov_2016,Izotov_2018,Izotov_2020}.
    
    These \lya\ line profiles have an excellent spectral quality. In particular, for these line profiles \dl\ might take two individual values; 0.0598\AA{} and 0.0735\AA{}, which results in an excellent wavelength sampling. This is given by the two the medium resolution gratings G130M and G160M used to observed the sources. 
    The spectral resolution is also high, and ranges from \wg=0.073\AA{} to 0.10\AA{} with median 0.085\AA{}. Finally, the signal to noise ratio of the maximum of the line spawns a wide range from \sn=6.5 to $\sim 400$ with a median of $\sim$38 and only a $\sim11\%$ of sample exhibit \sn\ values below 15. 
    Considering this, most of the line profiles studied here have similar quality to the best case studied in the previous sections. 

    The name, systemic redshift, \lya\ luminosity and observed equivalent width for each of the observed line profiles is listed in Tables~\ref{tab:real_mcmc_LUM_0} and \ref{tab:real_mcmc_LUM_1}. 

    \subsection{Results}\label{ss:results_obs}
    
    Here we analyse the 97 outflow parameters and redshift estimated for the 97 observed \lya\ line profiles using the MC DNN and the MCMC methodologies. For these 97 line profiles we only used the outflow DNN, as none of them show inflow characteristics, such as a more prominent blue peak than a red peak. All the outflow parameters given by the MC DNN and MCMC methodologies are listed in Tables~\ref{tab:real_mcmc_0}, \ref{tab:real_mcmc_1} and \ref{tab:real_mcmc_2}. In \S\ref{sss:shapes} we focus on the line profile shapes recovery, in \S\ref{sss:z_accuracy} on the redshift accuracy, while on \S\ref{sss:comparison_metho} we compare the properties obtained with the MC DNN and the MCMC methodologies. Finally, we study the possible correlation between the outflow parameters between themselves on \ref{sss:obs_properties}. 
    
    For illustration, we display two detailed examples in Fig.~\ref{fig:examples_observed} in the observed frame. The KS estimator values for each methodology are shown in the left of each panel. We find that both, the MCMC (green) and the MC DNN (blue) approaches manage to fit the observed spectra (shown in grey) well (KS<0.1). For both line profiles, the two approaches agree, as the continuum, the red and the blue peaks are well reproduced. Also, the line profiles predicted by the MCMC and the MC DNN methodologies are similar with only minor differences. 
    
    Additionally, the systemic \lya\ wavelength given by auxiliary lines is marked in a vertical grey dashed line. 
    For these two cases we find a good agreement between the measured and the estimated systemic redshift. In particular, for the line profile in the left, the difference between the true and the predicted \lya\ wavelengths is $\sim -0.35$\AA{} in the rest frame ($\sim -86\kms$). For the line profile in the right, this difference is $\sim0.1$\AA{} in the rest frame ($\sim -24\kms$).
    
    In terms of the outflow parameters, both methodologies predict relatively similar values. These are shown on the top left in blue for the MC DNN and in the middle left in green for the MCMC with their respective $\pm 1-\sigma$ uncertainties.  In particular, in both cases, the predicted \vexp , \nh\ and \ew values are close. Then, the methodologies predict different values for \w\ in both cases. In the first one (left) there is more than one order of magnitude of difference between the predicted \w, while in the second (right) there is half and order of magnitude of difference. This might indicate that in this particular configurations \w\ is not important to shape the line profiles, as the agreement between both methodologies and the observed data is quite good (KS<0.1). This is also the case for \ta ,  that might differ more than two orders of magnitude depending on the methodology. 
    
    \subsubsection{ Performance reproducing the observed \lya\ line profile shape }\label{sss:shapes}    
    
    In Fig.~\ref{fig:comparison_NN_MC_lines_0} and \ref{fig:comparison_NN_MC_lines_1} we display the other 96 observed \lya\ line profiles (grey) and the prediction given by the MC DNN (blue) and MCMC (green) methodologies. Each panel is labeled with a number in the top left. This number matches the labels of Tables.\ref{tab:real_mcmc_0}, \ref{tab:real_mcmc_1} and \ref{tab:real_mcmc_2}, where all the outflow properties are listed. In this comparison we focus on the shape of the line profiles. Therefore, the line profiles are shown in the the proxy rest frame, as described in \S\ref{sss:input_output}, in which we assume that $\lambda_{\rm max}$ is the true \lya\ wavelength. Then, in order to quantify the goodness of the fit we display in the top left corner of each panel the KS estimator value for the MC DNN (top blue) and MCMC (bottom green) computed in the observed frame. Meanwhile, the KS estimator distribution is shown in the left panel of Fig.~\ref{fig:redshift_accuracy_obs}.
    
    We find, in general, that both, the MCMC and the MC DNN produce an  agreement of KS<0.1 between the predictions and the observations. Also, in general, \zelda\ fits better (lower KS values) the line profiles with a better spectral quality. In particular,  we find that the mean KS estimator value is $10^{-1.221}$ for the MC DNN while it is $10^{-1.321}$ for the MCMC methodology. For example, for the line profiles with a good signal to noise, (e.g. 3, 5, 8, 9, 16, 22, 26, 29, 31, 32, 33, 38, 46. 55, 59, 64, 65, 67. 68 or 85) we find an excellent agreement with typical values of KS$\sim 0.05$. Also, for other line profile with a lower signal to noise, such as the cases 5, 10, 24, 34, 43, 48, 54, 61, 79 or 95, the agreement is also good (KS$\sim 0.09$). 
    
    Additionally there are some cases in which the MCMC reproduces well the observed line profile, while with the MC DNN there are small differences between the observation and the prediction. Some examples are cases 4, 20, 30, 44, 80 and 82. We find that in the cases in which the quality of the fit is lower for both methodologies, the main peak is still well reproduced while the secondary peak might differ. For example, this happens in cases 4, 7, 19 and 23 . This analysis shows that the shape of the line profiles are well (KS<0.1) recovered by both methodologies in general. 

    \begin{figure*} 
    \includegraphics[width=6.9in]{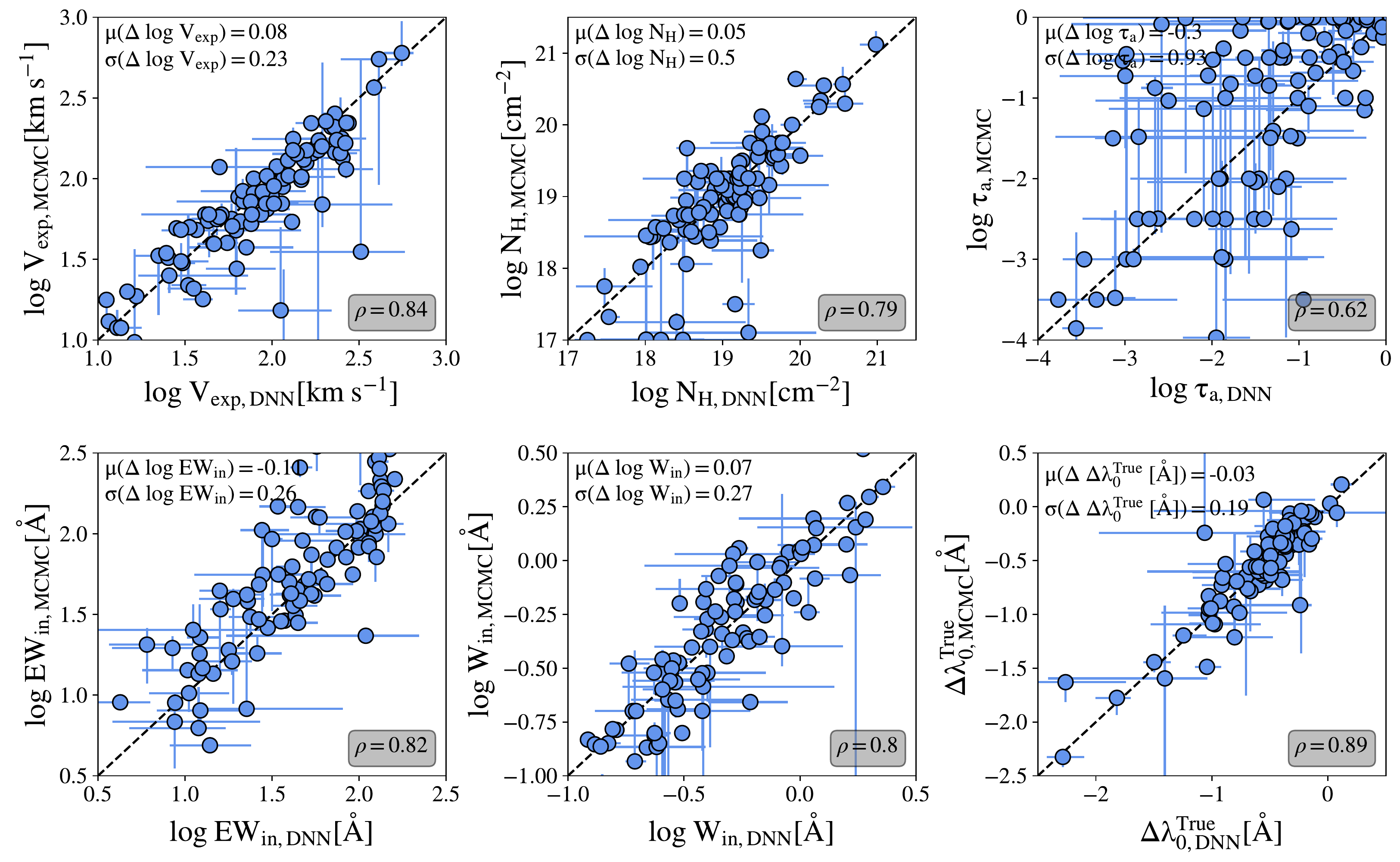}
    \caption{ Comparison between the line profile properties predicted by the MC DNN (horizontal axis) and the MCMC (vertical axis) procedures. \vexp , \nh\ and \ta\ are shown in the top row from left to right. Also, \ew, \w\ and \dlt\ are displayed in the bottom row from left to right. The filled circles show the percentile 50th of the PDF of each property. Meanwhile, the error bars indicate the percentiles 16th and 84th. The one-to-one relation is marked as the black dashed line. The mean $\mu$ and the standard deviation $\sigma$ of the difference between the properties predicted by MC DNN and MCMC are shown in the top left of each panel.  The Pearson correlation coefficient between the MCMC and the MC DNN prediction is displayed in the bottom left of each panel. }
    \label{fig:comparison_NN_MC_props}
    \end{figure*}
    
    \begin{figure*} 
    \includegraphics[width=6.9in]{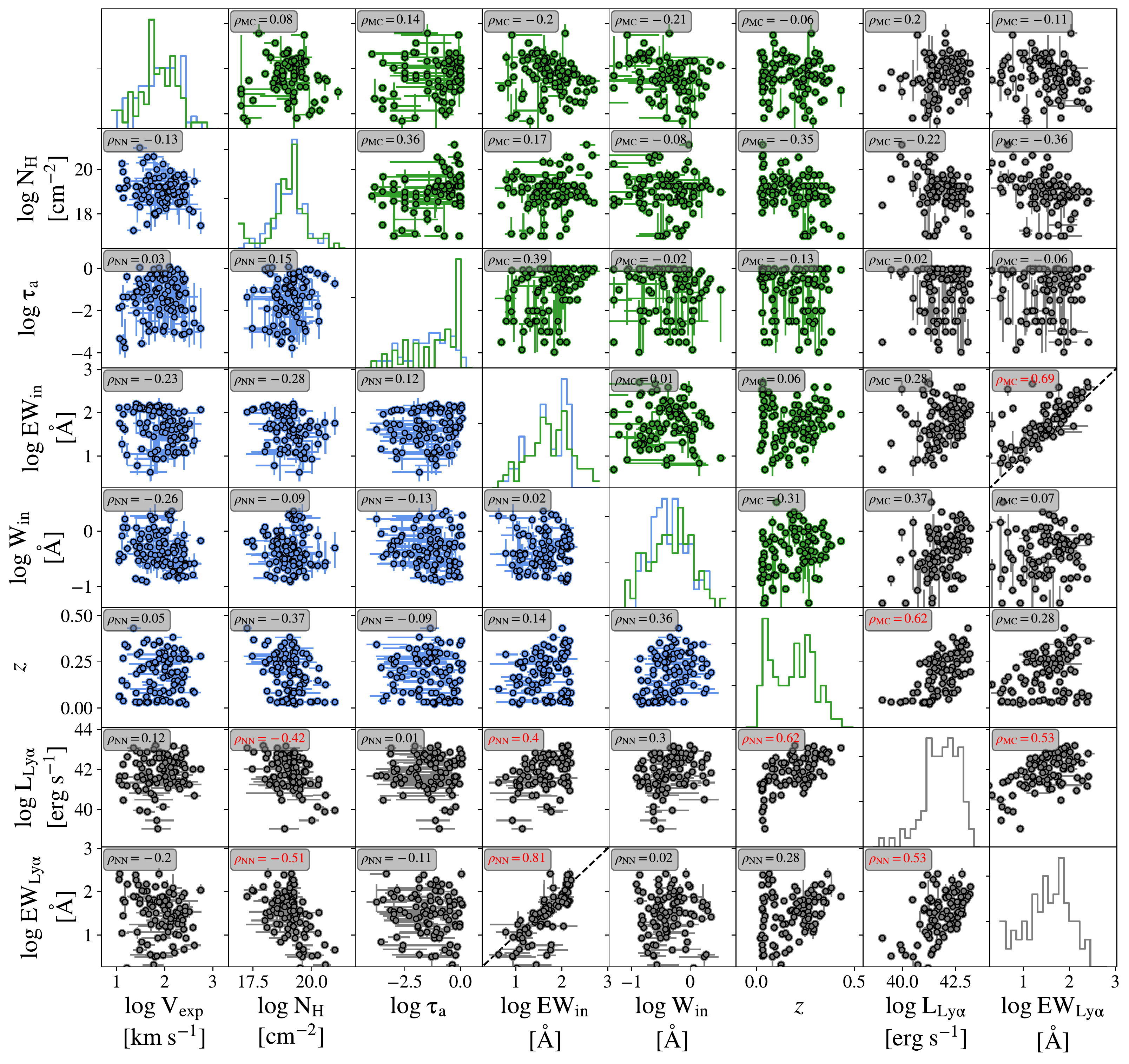}
    \caption{ Correlation figure of the properties derived by \zelda\ and also the \lya\ luminosity and equivalent with of the line computed by \lasd. The diagonal (line of panels from the top left to the bottom right) shows the PDF of the properties. The figure is divided in two parts. The part above the diagonal makes use of the properties predicted by the MCMC (green), while the part below the diagonal makes use of the MC DNN predictions (blue). Meanwhile, if properties computed by \lasd\ are present, they displayed en grey. \vexp, \nh, \ta, \ew , \w , $z$, \llya\ and \ewl\ are shown from left to right and from top to bottom respectively. Also, the Pearson coefficient is displayed in the top left corners. }
    \label{fig:properties}
    \end{figure*}
    
    \begin{figure*} 
    \includegraphics[width=5.5in]{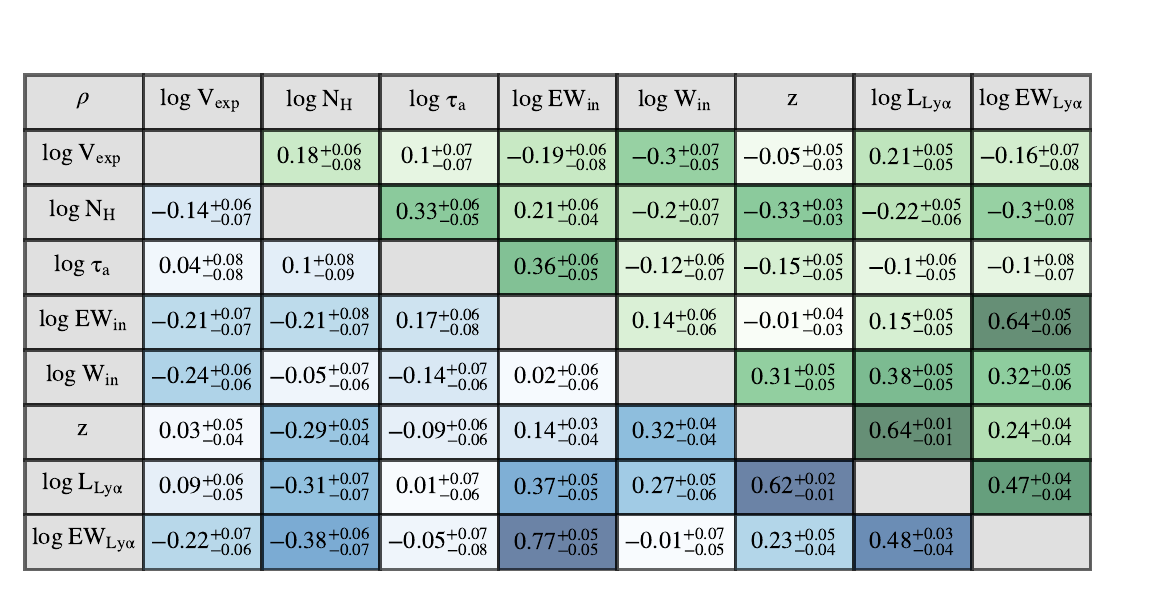}
    \caption{ Pearson correlation coefficients with their $\pm 1 \sigma$ (corresponding to the 84th and 16th percentiles) uncertainty for the 1000 parameters perturbations. In the bottom left from the diagonal (grey) we show the (anti)correlation coefficient using the MC DNN output (blue). In the top right corner we display the same but for the MCMC output (green).  }
    \label{fig:pearson_perturbed}
    \end{figure*}
    
    \subsubsection{ Accuracy in systemic redshift estimation }\label{sss:z_accuracy}
        
    In order to quantify the shift between the redshift given by the MC DNN and the MCMC methodologies and the systemic redshift, we use the difference in rest frame wavelength, i.e.,   
    \begin{equation}
        \Delta \lambda ^{\rm Zelda}_0 = \lambda _ {\rm Ly\alpha} \frac{z^{\rm Zelda}-z^{\rm Sys}}{ 1+z^{\rm Sys}},
    \end{equation}
    where $z^{Sys}$ is the true systemic redshift given by a \lya\ independent measurement and $z^{\rm Zelda}$ in the redshift predicted by our various methodologies.\footnote{Note that $\Delta \lambda ^{\rm Zelda}_0$ is not exactly identical to \dlt, as \dlt is calculated assuming that the maximum of the line profile is the \lya\ wavelength. However, these two quantities differ less than 0.001\AA{}.}
    
    In the right panel of Fig.~\ref{fig:redshift_accuracy_obs}, we show the performance of recovering the systemic redshift by the MC DNN (blue) and the MCMC (green) methodologies. We find that both the MC DNN and the MCMC algorithms exhibit similar performances with the MC DNN showing slight more accurate results. In particular, the dispersion around the true \lya\ wavelength in rest frame in the MC DNN approach is 0.348\AA{} ($\sim$85\kms) and in the MCMC approach is  0.366\AA{} ($\sim$89\kms).  This accuracy in the redshift estimation would lead to accurate clustering measurements down to $\rm\sim2cMpc/h$ in the monopole and $\sim4{\rm cMpc/h}$  in the quadrupole of the two point correlation function \citep{gurung_2020b}. We also find that both methodologies slightly overestimates the systemic wavelength in the rest frame by $\sim0.154$\AA{} ($\sim37km/s$) in the MC DNN and $\sim0.181$\AA{} ($\sim44km/s$) in the MCMC methodology. 
    This was noticed before in the literature and likely hints towards a shortcoming of the `shell model' \citep{Orlitova_2018,Li2021};
    
    \citet{Runnholm_2020} used a parametric method for redshift estimation in a slightly larger but mostly overlapping sample to ours. In comparison, our methodologies show a better accuracy in the redshift estimation. They find a sightly larger scatter ($\sim180\kms$, $\sim 0.7$\AA{}) around the true value. Also, a similar systemic shift of $\sim 34\kms$ ($\sim 0.14$\AA{}) was also obtained in \citet{Runnholm_2020}. As mentioned above, understanding whether this systemic shift has a true physical origin or is due to the simplicity of the `shell-model' is an interesting avenue for future work. Also, in \cite{Verhamme:2018aa}, authors developed an empirical methodology to estimate the systemic redshift using only the \Lya\ line profile. They reported an accuracy of $\sim 100\kms$ ($\sim 0.4$\AA{}) in a sample of 55 sources at different redshifts. 
    
    \subsubsection{ Comparison between  methodologies }\label{sss:comparison_metho}
    
    In Fig.~\ref{fig:comparison_NN_MC_props} we compare the outflow properties predicted by the MC DNN (horizontal axis) and the MCMC (vertical axis) methodologies. In the top row we show \vexp, \nh\ and \ta\ from left to right, while \ew, \w\ and \dlt\ are shown in the bottom panel from left to right. 
    In order to quantify the correlation between the MCMC and the MC DNN we used the Pearson correlation coefficient $\rho$, which ranges from -1 to 1. Large values of $|\rho|$ indicate tight correlations (if $\rho>0$) and anti-correlations (if $\rho<0$). Overall we find a good agreement between the properties predicted by the MC DNN methodology and those predicted by the MCMC approach ($\rho>0.8$). We find that most of the times the measurements are compatible in the 1-$\sigma$ confidence level with the one-to-one relations. 
    
    We also quantify the agreement between the predictions of both methods by computing the mean ($\mu$) and standard deviation ($\sigma$) of the distribution of the difference between the output of MC DNN and MCMC analysis. We show these quantities in their corresponding panel of Fig.~\ref{fig:comparison_NN_MC_props}. Overall, we find that, for all the properties, the means are compatible with zero if we consider their standard deviations. This indicates that, if the MC DNN and MCMC methodologies are biased, at least they are biased in the same way. We also find that \vexp\ is the outflow property that has a better agreement, with $\rm \mu(\Delta\;\log\;V_{exp})=0.08$ and $\rm \sigma(\Delta\;\log\;V_{exp})=0.23$. Meanwhile, the agreement in \dlt\ is remarkable, as  $\rm \mu(\Delta\;\Delta\lambda_0^{True})=-0.03$ and $\rm \sigma(\Delta\;\Delta\lambda_0^{True})=0.19$. 
    
    Considering that the quality of the observed data is usually comparable (see \S\ref{ss:data_props}) to our best mock sample (\dl=0.05, \w=0.1 and \sn=15), the agreement between the MC DNN and MCMC is consistent with the uncertainties derived for this level of spectral quality (see Fig.~\ref{fig:props_1_acc}). 
        
    When we compare the accuracy of the MCMC and the MC DNN in predicting the systemic redshift of these sources we find that both have a similar performance. In particular, the accuracy of the MC DNN methodology ($\sigma(\Delta\lambda_0^{\rm Zelda})=0.348$) is slightly better than that of the MCMC methodology ($\sigma(\Delta\lambda_0^{\rm Zelda})=0.366$), see Fig.\ref{fig:redshift_accuracy_obs}. It is noticeable that in the mock line profiles the MCMC methodology had a better accuracy than the MC DNN. This shows that the accuracy of both procedures is comparable, at least, in the redshift estimation.    
    
    \subsubsection{ Relations between model and observed properties  }\label{sss:obs_properties}
 
    In Fig.~\ref{fig:properties} we show the correlations of the outflow properties computed using the MC DNN methodology (blue), i.e., \vexp, \nh, \ta , \ew , \w, the redshift. We also show two extra parameters from the \lasd\ data base; the \lya\ luminosity \llya\ and the measured \lya\ equivalent width of the line in the rest frame of the galaxy \ewl (grey). In particular, in the diagonal we show the one dimensional PDF of each of these parameters. We find that the outflow parameters cover a wide dynamical range. In particular, if we compute the percentiles 5th and 90th of the \vexp\ distribution, 90\% of the line profiles are assigned \vexp\ values between 14.4${\rm km/s}$ and 246${\rm km/s}$ and the median of the PDF is 96.4${\rm km/s}$. In the same way, 90\% of the population in contained between \nh=$10^{18.0}{\rm cm}^{-2}$ and $10^{19.8}{\rm cm}^{-2}$ with median $10^{19.1}{\rm cm}^{-2}$. 
    For \ta, 90\% of the population is between $10^{-3.1}$ and $10^{-0.21}$ and the median is $10^{-1.2}$. For \ew, the 90\% is between 12.1\AA{} and 133\AA{} with median 54.1\AA{}. Finally, for \w, 90\% of the sample is between 0.15\AA{} and 1.1\AA{} with median 0.41\AA{}. The wide diversity of line profiles shapes shown in Fig.~\ref{fig:comparison_NN_MC_lines_0} and Fig.~\ref{fig:comparison_NN_MC_lines_1} is reflected in the large parameter volume that the observed sample spawns.  
    
    Also, in comparison with the volume covered by \zelda 's \lya\ line profile grid (\S\ref{ss:grid}) and the volume used to train the DNN (\S\ref{ss:training}), we find that most of the observed line profiles lie well inside our outflow parameter volume. Moreover, we find that none of the PDF of the outflow parameters end abruptly at its corresponding edge of the parameter range. This indicated that the volume covered by \zelda\ is large enough to explain, a priori, the majority of observed \lya\ line profiles. 
        
    Let us now focus on the panels of Fig.~\ref{fig:properties} comparing the behaviour of properties against other properties. 
    We confirm some possible correlations between different properties.  The Pearson coefficient computed using properties estimated with the  MC DNN approach are labeled as $\rho_{\rm NN}$, while those computed using the MCMC approach are labeled as $\rho_{\rm MC}$. We find that the variables that show a $|\rho|\ge0.4$ are: \nh\ with \llya\ ($\rho_{\rm NN}=-0.42$,$\rho_{\rm MC}=-0.22$) and \ewl\ ($\rho_{\rm NN}=-0.51$,$\rho_{\rm MC}=-0.36$). Then, \ew\ with \LLya\ ($\rho_{\rm NN}=0.40$, $\rho_{\rm MC}=0.37$) and  \ewl\ ($\rho_{\rm NN}=0.81$, $\rho_{\rm MC}=0.69$). 
    Finally, \LLya\ and $z$ ($\rho_{\rm NN}=0.62$) and \LLya\ and \ewl\ ($\rho=0.53$).
    
    Next, we estimate the significance of these trends by measuring the uncertainty in the Pearson's coefficients. 
    To do so, we draw $1000$ times a new set of outflow parameters from the posterior and re-compute the Pearson's correlation coefficient each time. Specifically, for the MCMC we pick randomly a walker within the region where the chains have already converge and or the MC DNN we pick a random realization. In this way, for both methodologies, the degeneracy between outflow parameters is kept within the individual line profiles. 
    
    In Fig.\ref{fig:pearson_perturbed}, we show the median Pearson coefficient and differences to the 16th and 84th percentile obtained from the procedure outlined above. We find that in general $\rho_{\rm MC}$ and $\rho_{\rm NN}$ are compatible with $1\sigma$, although there are a few exceptions. This indicates that the trends derived with the MCMC and the MC DNN methodologies are, in general, compatible. We list here the (anti)-correlations that we find significant:
    
    \begin{itemize}
    
        \item{ Correlation between the intrinsic equivalent width \ew\ and measured rest frame equivalent width \ewl\ ($\rho_{\rm NN}=0.77^{
        0.05}_{-0.05}$, $\rho_{\rm MC}=0.64^{+0.05}_{-0.06}$): This is the most clear correlation that we find in our \lya\ line profile sample. The ratio between these two properties can be written as as a function of the escape fraction of non-ionizing continuum photons $f_{\rm esc}^{\rm Cont}$ and the escape fraction of \lya\ photons $f_{\rm esc}^{\rm Ly\alpha}$ as
        
        \begin{equation}
            \frac{EW_{\rm in}}{EW_{\rm Ly\alpha}} = \frac{ f_{\rm esc}^{\rm Cont} }{f_{\rm esc}^{\rm Ly\alpha}}.
        \end{equation}
        
        We find that on average, \ew\ is larger than \ewl, which indicates that the \lya\ escape fraction is lower than the escape fraction of continuum photons.\\}
        
        \item{ Anti-correlation between \LLya\ and \nh\ and  ($\rho_{\rm NN}=-0.31^{+0.07}_{-0.07}$,$\rho_{\rm MC}=-0.22^{+0.05}_{-0.06}$): This anti-correlation can be explain by the dependence of the \lya\ escape fraction with \nh . In principle, for the same value of \ta, gas distributions with higher \nh\ would have lower escape fraction as the number of scattering events between \lya\ photons and neutral hydrogen would be larger, causing a larger dust attenuation. }
        
        \item{ Anti-correlation between \nh\ and \ewl\ ($\rho_{\rm NN}=-0.38^{+0.06}_{-0.07}$,$\rho_{\rm MC}=-0.30^{+0.08}_{-0.07}$): The equivalent width of a galaxy is a balance between \llya\ and continuum luminosity density. On one side of the balance, \LLya\ anti-correlates with \nh, as just discussed. On the other side, galaxies with higher \nh\ can be expected to be more massive and therefore to have a brighter continuum. In this case, we find that the flux reduction due to the \lya\ radiative transfer plays the determinant role in this trend and \ewl\ anti-correlates with \nh. Moreover, A similar trend, but with a much weaker correlation ($\rho\sim-0.21$), is observed for \nh\ and \ew .  }
        
        \item{ Correlation between \LLya\ and \ewl\ ($\rho\sim0.47$): This is given by the definition of these two properties, i.e., $EW_{\rm Ly\alpha}=L_{\rm Ly\alpha}/C_{\rm Ly\alpha}$, where $C_{\rm Ly\alpha}$ is the luminosity density of the continuum around \lya. }
        
        \item{ Correlation between \LLya\ and \w\ ($\rho_{\rm NN}=0.27^{+0.05}_{-0.06}$,$\rho_{\rm MC}=0.38^{+0.05}_{-0.05}$): In principle, the intrinsic width of the \lya\ line, \w, depends on the temperature of the HII region where the \lya\ photons are originated. The hotter the region, the larger \w. Also, in average, galaxy regions with higher star formation rate, therefore higher intrinsic \llya, would be hotter. this could explain this correlation.   }  
    
    \end{itemize}
    
    Interestingly, correlations of the spectral properties of an overlapping sample have been explored by \citet{Hayes2021}. Their findings (larger blue-to-red ratio as well as less shifted and narrower red peaks with luminosity and equivalent width) can be understood with the correlations we find here. Specifically, our finding versus the column density detailed above fit this scenario of less radiative transfer effects for fainter sources well.

\section{ Summary and conclusions}\label{s:conclusions}

In this work we have introduce \zelda\ (redshift Estimator for Line profiles of Distant Lyman-Alpha emitters), an open source \texttt{Python} module to improve the understanding of the \lya\ radiation based on \lyart\ \citep{orsi12} and \flareon\ \citep{GurungLopez_2019b}. \zelda\ has two principle functionalities: i) a fast computation of \lya\ line profiles and ii) several techniques to fit observed \lya\ line profiles to the outflow/inflow shell model usually used in the literature \citep[e.g.][]{zheng02,ahn04,verhamme06,orsi12,Gronke2017}. The first functionality is designed to produce large amounts of accurate \lya\ line profiles to populate cosmological volumes \citep{GurungLopez_2020a,gurung_2020b}, to make qualitative analysis of observed \lya\ line profiles \cite[e.g.][]{guaita_2017,guaita_2020} and to produce big data sets of mock \lya\ lines for neural network analysis. The second functionality is developed to find the most probable outflow model that reproduce a given observed \lya\ line profile. Several work in the literature have performed fitting analysis to \lya\ line profiles {\color{red}}. However, they relied in different outflow models and fitting techniques. We make \zelda\ publicly available so that the scientific community has a common methodology to interpret \lya\ line profiles and therefore, to make results between different works easier to compare.  

For the computation of \lya\ line profiles we use a similar approach to \flareon, in which the line profile are calculated from a pre-computed grid of models where the full radiative transfer is computed using \lyart. We have upgraded one of the outflow gas geometries already present at \flareon , the shell model. Now, this new model has two components: i) the macroscopic gas properties such as the bulk radial velocity \vexp, the $HI$ column density \nh\ and the optical depth of dust \ta\ and ii) the intrinsic \lya\ line profile, which is a Gaussian with width \w\ over a continuum with equivalent width \ew. For the first part we run the full radiative transfer using \lyart\ in a regular grid for configurations with $0\leq$\vexp$[km/s]\leq1000$, $10^{17}\leq$\nh$[cm^{-2}]\leq10^{21.5}$ and $10^{-4}\leq$\ta$\leq1$. Then, the different intrinsic spectrum are implemented in a post processing fashion in the range of $0.01\leq$\w[\AA{}]$\leq6$ and $0.1\leq$\ew[\AA{}]$\leq1000$. 

Then, in order to estimate the \lya\ spectrum for a given set of \{\vexp,\nh,\ta,\ew,\w\} inside the parameter range of the grid we use linear interpolation in five dimensions. In order to estimate the accuracy of this procedure we compute sample of \lya\ line profiles using \lyart\ in random locations inside the grid volume. We find that the interpolation approximation works really well, with typical Kolmogorov–Smirnov test values of 0.05. 

For the interpretation of observed line profiles we have developed a novel deep neural network approach. Specifically, we use mock \lya\ line profiles as the input of a deep neural network and the output is composed by the outflow/inflow properties and the location of the \lya\ wavelength. The mock \lya\ line profiles are embedded with the typical observational limitations; finite spectral resolution of full width half maximum \wg, wavelength binning \dl\ and noise with a ratio between signal and noise in the maximum of the line profile of \sn. 

Then we have presented the Monte Carlo Deep Neural Network methodology (MC DNN), which consists in performing MC perturbation of the \lya\ line profile for which we want to extract the outflow properties. Then, each of these perturbation passes as input to the DNN for produces a set of outflow variables. The {\it solution} parameters are then defined as the percentile 50th of their respective probability distribution function. We find that this methodology increases the accuracy of the DNN approach. Another advantage is that we can assigned accurate uncertainties to the outflow parameters by computing the corresponding percentiles of the PDFs. These uncertainties exhibit relative biases less than the 10\%. 

We have tested the performance of the MC DNN methodology on observed \lya\ line profiles gather in the {\it Lyman alpha Spectral Database}  \citep[\texttt{LASD}][]{Runnholm_2020}. In particular we have focused on line profiles from sources with a systematic measured a \lya\ independent approach. We find that \zelda's MC DNN approach manage to reproduce quite well the shape and systemic redshift of this sample. In particular, we find that typical error in determining the systemic \lya\ wavelength is about 0.3\AA{} in the rest frame of the source. This is a really good accuracy in comparison to other methodologies in the literature. 

Also, we illustrate the potential of \zelda\ to understand the \lya\ physics within galaxies. We have looked for possible correlations between the outflow properties derived by \zelda\ and the observed \lya\ luminosity \llya\ and measured rest frame equivalent width \ewl. We find several trends with a relatively good significant (Pearson correlation parameter > 0.4). For example, we find an anti-correlation between \nh\ and \llya\ that might be due to the fact that galaxies with higher \nh\ might be more prone to a larger number of \lya\ scattering events, which would reduce the \lya\ flux emerging from the galaxy as more photons would be absorbed by dust.

Future steps, building up on the pipeline presented here could be to introduce additional parameters (such as the effective temperature $T$ which can include effects of turbulent motion, or a parametrization of IGM transmission curves), or the inclusion of other geometries (such as a multiphase or anisotropic medium). Such additions can be implemented straightforwardly into \zelda.

We have verify the performance of \zelda\ on observed \lya\ line profiles by comparing the MC DNN methodology results to an Monte Carlo Markov Chain analysis which is commonly used in the literature \citep{Gronke_2015}. We find that the outflow and redshift estimated using both methodologies are compatible. An important perk of the MC DNN methodology is its low computational cost. Currently, computational time is not limiting as there are only a few hundreds of observed \lya\ line profiles with good enough quality to perform this kind of analysis. However, in the nearby future, with the launch the James Web Space Telescope \citep{James_Web} and other ground breaking experiments, the number of \lya\ emitters and \lya\ line profiles will increase several orders of magnitudes. In this scenario \zelda\ will be a extremely useful tool to analyse future data sets and to increase our knowledge about the \lya\ emitters population.

\section*{Acknowledgements}
 This research made use of matplotlib, a Python library for publication quality graphics \citep{Hunter:2007}, NumPy \citep{harris2020array} and SciPy \citep{Virtanen_2020}.

Authors acknowledge support from the Generalitat Valenciana project of excellence Prometeo/2020/085.

This work has made used of CEFCA's Scientific High Performance Computing system which has been funded by the Governments  of  Spain  and  Aragón  through the  Fondo  de  Inversiones  de  Teruel, and the Spanish Ministry of Economy and Competitivenes (MINECO-FEDER, grant AYA2012-30789) and also Project of excellence Prometeo/2020/085 from the Conselleria d'Innovaci\'o, Universitats, Ci\`encia i Societat Digital de la Generalitat Valenciana.
  
Authors acknowledge support from the Generalitat Valenciana project of excellence Prometeo/2020/085.

The authors acknowledge the support of the Spanish Ministerio de Economia y Competividad project No. AYA2015-66211-C2-P-2. 
 
SS was supported in part by World Premier International Research Center Initiative (WPI Initiative), MEXT, Japan. 
SS was also supported in part by the Munich Institute for Astro- and Particle Physics (MIAPP) which is funded by the Deutsche Forschungsgemeinschaft (DFG, German Research Foundation) under Germany's Excellence Strategy (EXC-2094-390783311). 

MG was supported by NASA through the NASA Hubble Fellowship grant HST-HF2-51409 and acknowledges support from HST grants HST-GO-15643.017-A, HST-AR15039.003-A, and XSEDE grant TG-AST180036.

\section*{Data Availability}

Most of the codes and data used in this work is publicly available. \zelda's software can be found at \url{https://github.com/sidgurun/Lya_zelda} and the documentation with an installation guide and several tutorials are at \url{http://zelda.rtfd.io}. Also, \lyart\ software \citep{orsi12} is stored at \url{https://github.com/aaorsi/LyaRT}. The mock \lya\ line profiles used to train and assess the quality of our different techniques is available upon request. Finally, the observed \lya\ line profiles were extracted from {\it Lyman Alpha Spectral Database}  \citep[\texttt{LASD}, \url{https://lasd.lyman-alpha.com} ][]{Runnholm_2020}




\bibliographystyle{mnras}
\bibliography{ref}

\begin{thebibliography}{}
\makeatletter
\relax
\def\mn@urlcharsother{\let\do\@makeother \do\$\do\&\do\#\do\^\do\_\do\%\do\~}
\def\mn@doi{\begingroup\mn@urlcharsother \@ifnextchar [ {\mn@doi@}
  {\mn@doi@[]}}
\def\mn@doi@[#1]#2{\def\@tempa{#1}\ifx\@tempa\@empty \href
  {http://dx.doi.org/#2} {doi:#2}\else \href {http://dx.doi.org/#2} {#1}\fi
  \endgroup}
\def\mn@eprint#1#2{\mn@eprint@#1:#2::\@nil}
\def\mn@eprint@arXiv#1{\href {http://arxiv.org/abs/#1} {{\tt arXiv:#1}}}
\def\mn@eprint@dblp#1{\href {http://dblp.uni-trier.de/rec/bibtex/#1.xml}
  {dblp:#1}}
\def\mn@eprint@#1:#2:#3:#4\@nil{\def\@tempa {#1}\def\@tempb {#2}\def\@tempc
  {#3}\ifx \@tempc \@empty \let \@tempc \@tempb \let \@tempb \@tempa \fi \ifx
  \@tempb \@empty \def\@tempb {arXiv}\fi \@ifundefined
  {mn@eprint@\@tempb}{\@tempb:\@tempc}{\expandafter \expandafter \csname
  mn@eprint@\@tempb\endcsname \expandafter{\@tempc}}}

\bibitem[\protect\citeauthoryear{{Ahn}}{{Ahn}}{2003}]{ahn03}
{Ahn} S.,  2003, Journal of Korean Astronomical Society, \href
  {http://adsabs.harvard.edu/abs/2003JKAS...36..145A} {36, 145}

\bibitem[\protect\citeauthoryear{{Ahn}}{{Ahn}}{2004}]{ahn04}
{Ahn} S.,  2004, \mn@doi [\apjl] {10.1086/381750}, \href
  {http://adsabs.harvard.edu/abs/2004ApJ...601L..25A} {601, L25}

\bibitem[\protect\citeauthoryear{{Bacon} et~al.,}{{Bacon}
  et~al.}{2010}]{bacon10}
{Bacon} R.,  et~al., 2010, in Society of Photo-Optical Instrumentation
  Engineers (SPIE) Conference Series. p.~8, \mn@doi{10.1117/12.856027}

\bibitem[\protect\citeauthoryear{{Bresolin}}{{Bresolin}}{2017}]{Bresolin_2017}
{Bresolin} F.,  2017, {Metallicities in the Outer Regions of Spiral Galaxies}.
p.~145, \mn@doi{10.1007/978-3-319-56570-5\_5}

\bibitem[\protect\citeauthoryear{{Byrohl} \& {Gronke}}{{Byrohl} \&
  {Gronke}}{2020}]{Byrohl2020}
{Byrohl} C.,  {Gronke} M.,  2020, \mn@doi [\aap] {10.1051/0004-6361/202038685},
  \href {https://ui.adsabs.harvard.edu/abs/2020A&A...642L..16B} {642, L16}

\bibitem[\protect\citeauthoryear{{Byrohl}, {Saito}  \& {Behrens}}{{Byrohl}
  et~al.}{2019}]{Byrohl_2019}
{Byrohl} C.,  {Saito} S.,   {Behrens} C.,  2019, \mn@doi [\mnras]
  {10.1093/mnras/stz2260}, \href
  {https://ui.adsabs.harvard.edu/abs/2019MNRAS.489.3472B} {489, 3472}

\bibitem[\protect\citeauthoryear{{Caruana} et~al.,}{{Caruana}
  et~al.}{2020}]{Caruana_2020}
{Caruana} J.,  et~al., 2020, VizieR Online Data Catalog, \href
  {https://ui.adsabs.harvard.edu/abs/2020yCat..74730030C} {p. J/MNRAS/473/30}

\bibitem[\protect\citeauthoryear{{Dijkstra}}{{Dijkstra}}{2017}]{dijkstra17}
{Dijkstra} M.,  2017, preprint, \href
  {http://adsabs.harvard.edu/abs/2017arXiv170403416D} {} (\mn@eprint {arXiv}
  {1704.03416})

\bibitem[\protect\citeauthoryear{{Dijkstra}, {Haiman}  \& {Spaans}}{{Dijkstra}
  et~al.}{2006}]{dijkstra06}
{Dijkstra} M.,  {Haiman} Z.,   {Spaans} M.,  2006, \mn@doi [\apj]
  {10.1086/506243}, \href {http://adsabs.harvard.edu/abs/2006ApJ...649...14D}
  {649, 14}

\bibitem[\protect\citeauthoryear{{Dijkstra}, {Gronke}  \&
  {Venkatesan}}{{Dijkstra} et~al.}{2016}]{Dijkstra_2016}
{Dijkstra} M.,  {Gronke} M.,   {Venkatesan} A.,  2016, \mn@doi [\apj]
  {10.3847/0004-637X/828/2/71}, \href
  {https://ui.adsabs.harvard.edu/abs/2016ApJ...828...71D} {828, 71}

\bibitem[\protect\citeauthoryear{{Erb}, {Steidel}  \& {Chen}}{{Erb}
  et~al.}{2018}]{Erb_2018}
{Erb} D.~K.,  {Steidel} C.~C.,   {Chen} Y.,  2018, \mn@doi [\apjl]
  {10.3847/2041-8213/aacff6}, \href
  {https://ui.adsabs.harvard.edu/abs/2018ApJ...862L..10E} {862, L10}

\bibitem[\protect\citeauthoryear{{Farrow} et~al.,}{{Farrow}
  et~al.}{2021}]{Farrow_2021}
{Farrow} D.~J.,  et~al., 2021, arXiv e-prints, \href
  {https://ui.adsabs.harvard.edu/abs/2021arXiv210404613F} {p. arXiv:2104.04613}

\bibitem[\protect\citeauthoryear{{Foreman-Mackey}, {Hogg}, {Lang}  \&
  {Goodman}}{{Foreman-Mackey} et~al.}{2013}]{Foreman_Mackey_2013}
{Foreman-Mackey} D.,  {Hogg} D.~W.,  {Lang} D.,   {Goodman} J.,  2013, \mn@doi
  [\pasp] {10.1086/670067}, \href
  {http://adsabs.harvard.edu/abs/2013PASP..125..306F} {125, 306}

\bibitem[\protect\citeauthoryear{{Gardner} et~al.,}{{Gardner}
  et~al.}{2006}]{James_Web}
{Gardner} J.~P.,  et~al., 2006, \mn@doi [\ssr] {10.1007/s11214-006-8315-7},
  \href {https://ui.adsabs.harvard.edu/abs/2006SSRv..123..485G} {123, 485}

\bibitem[\protect\citeauthoryear{{Garel}, {Blaizot}, {Guiderdoni}, {Schaerer},
  {Verhamme}  \& {Hayes}}{{Garel} et~al.}{2012}]{garel12}
{Garel} T.,  {Blaizot} J.,  {Guiderdoni} B.,  {Schaerer} D.,  {Verhamme} A.,
  {Hayes} M.,  2012, \mn@doi [\mnras] {10.1111/j.1365-2966.2012.20607.x}, \href
  {http://adsabs.harvard.edu/abs/2012MNRAS.422..310G} {422, 310}

\bibitem[\protect\citeauthoryear{{Granato}, {Lacey}, {Silva}, {Bressan},
  {Baugh}, {Cole}  \& {Frenk}}{{Granato} et~al.}{2000}]{granato00}
{Granato} G.~L.,  {Lacey} C.~G.,  {Silva} L.,  {Bressan} A.,  {Baugh} C.~M.,
  {Cole} S.,   {Frenk} C.~S.,  2000, \mn@doi [\apj] {10.1086/317032}, \href
  {http://adsabs.harvard.edu/abs/2000ApJ...542..710G} {542, 710}

\bibitem[\protect\citeauthoryear{{Green} et~al.,}{{Green}
  et~al.}{2012}]{Green_2012}
{Green} J.~C.,  et~al., 2012, \mn@doi [\apj] {10.1088/0004-637X/744/1/60},
  \href {https://ui.adsabs.harvard.edu/abs/2012ApJ...744...60G} {744, 60}

\bibitem[\protect\citeauthoryear{{Gronke}}{{Gronke}}{2017}]{Gronke2017}
{Gronke} M.,  2017, \mn@doi [\aap] {10.1051/0004-6361/201731791}, \href
  {http://adsabs.harvard.edu/abs/2017A%26A...608A.139G} {608, A139}

\bibitem[\protect\citeauthoryear{{Gronke}, {Bull}  \& {Dijkstra}}{{Gronke}
  et~al.}{2015}]{Gronke_2015}
{Gronke} M.,  {Bull} P.,   {Dijkstra} M.,  2015, \mn@doi [\apj]
  {10.1088/0004-637X/812/2/123}, \href
  {https://ui.adsabs.harvard.edu/abs/2015ApJ...812..123G} {812, 123}

\bibitem[\protect\citeauthoryear{{Gronke}, {Dijkstra}, {McCourt}  \&
  {Oh}}{{Gronke} et~al.}{2017}]{Gronke2017a}
{Gronke} M.,  {Dijkstra} M.,  {McCourt} M.,   {Oh} S.~P.,  2017, \mn@doi [\aap]
  {10.1051/0004-6361/201731013}, \href
  {https://ui.adsabs.harvard.edu/abs/2017A&A...607A..71G} {607, A71}

\bibitem[\protect\citeauthoryear{{Guaita} et~al.,}{{Guaita}
  et~al.}{2017}]{guaita_2017}
{Guaita} L.,  et~al., 2017, \mn@doi [\aap] {10.1051/0004-6361/201730603}, \href
  {https://ui.adsabs.harvard.edu/abs/2017A&A...606A..19G} {606, A19}

\bibitem[\protect\citeauthoryear{{Guaita} et~al.,}{{Guaita}
  et~al.}{2020}]{guaita_2020}
{Guaita} L.,  et~al., 2020, \mn@doi [\aap] {10.1051/0004-6361/201935855}, \href
  {https://ui.adsabs.harvard.edu/abs/2020A&A...640A.107G} {640, A107}

\bibitem[\protect\citeauthoryear{{Gurung-L{\'o}pez}, {Orsi}, {Bonoli}, {Baugh}
  \& {Lacey}}{{Gurung-L{\'o}pez} et~al.}{2019a}]{GurungLopez_2019a}
{Gurung-L{\'o}pez} S.,  {Orsi} {\'A}.~A.,  {Bonoli} S.,  {Baugh} C.~M.,
  {Lacey} C.~G.,  2019a, \mn@doi [\mnras] {10.1093/mnras/stz838}, \href
  {https://ui.adsabs.harvard.edu/abs/2019MNRAS.486.1882G} {486, 1882}

\bibitem[\protect\citeauthoryear{{Gurung-L{\'o}pez}, {Orsi}  \&
  {Bonoli}}{{Gurung-L{\'o}pez} et~al.}{2019b}]{GurungLopez_2019b}
{Gurung-L{\'o}pez} S.,  {Orsi} {\'A}.~A.,   {Bonoli} S.,  2019b, \mn@doi
  [\mnras] {10.1093/mnras/stz2591}, \href
  {https://ui.adsabs.harvard.edu/abs/2019MNRAS.490..733G} {490, 733}

\bibitem[\protect\citeauthoryear{{Gurung-L{\'o}pez}, {Orsi}, {Bonoli},
  {Padilla}, {Lacey}  \& {Baugh}}{{Gurung-L{\'o}pez}
  et~al.}{2020}]{GurungLopez_2020a}
{Gurung-L{\'o}pez} S.,  {Orsi} {\'A}.~A.,  {Bonoli} S.,  {Padilla} N.,  {Lacey}
  C.~G.,   {Baugh} C.~M.,  2020, \mn@doi [\mnras] {10.1093/mnras/stz3204},
  \href {https://ui.adsabs.harvard.edu/abs/2020MNRAS.491.3266G} {491, 3266}

\bibitem[\protect\citeauthoryear{{Gurung-L{\'o}pez}, {Saito}, {Baugh},
  {Bonoli}, {Lacey}  \& {Orsi}}{{Gurung-L{\'o}pez} et~al.}{2021}]{gurung_2020b}
{Gurung-L{\'o}pez} S.,  {Saito} S.,  {Baugh} C.~M.,  {Bonoli} S.,  {Lacey}
  C.~G.,   {Orsi} {\'A}.~A.,  2021, \mn@doi [\mnras] {10.1093/mnras/staa3269},
  \href {https://ui.adsabs.harvard.edu/abs/2021MNRAS.500..603G} {500, 603}

\bibitem[\protect\citeauthoryear{Harris et~al.,}{Harris
  et~al.}{2020}]{harris2020array}
Harris C.~R.,  et~al., 2020, \mn@doi [Nature] {10.1038/s41586-020-2649-2}, 585,
  357

\bibitem[\protect\citeauthoryear{{Hayes} et~al.,}{{Hayes}
  et~al.}{2014}]{Hayes_2014}
{Hayes} M.,  et~al., 2014, \mn@doi [\apj] {10.1088/0004-637X/782/1/6}, \href
  {https://ui.adsabs.harvard.edu/abs/2014ApJ...782....6H} {782, 6}

\bibitem[\protect\citeauthoryear{{Hayes}, {Runnholm}, {Gronke}  \&
  {Scarlata}}{{Hayes} et~al.}{2021}]{Hayes2021}
{Hayes} M.~J.,  {Runnholm} A.,  {Gronke} M.,   {Scarlata} C.,  2021, \mn@doi
  [\apj] {10.3847/1538-4357/abd246}, \href
  {https://ui.adsabs.harvard.edu/abs/2021ApJ...908...36H} {908, 36}

\bibitem[\protect\citeauthoryear{{Heckman} et~al.,}{{Heckman}
  et~al.}{2011}]{Heckman_2011}
{Heckman} T.~M.,  et~al., 2011, \mn@doi [\apj] {10.1088/0004-637X/730/1/5},
  \href {https://ui.adsabs.harvard.edu/abs/2011ApJ...730....5H} {730, 5}

\bibitem[\protect\citeauthoryear{{Heckman}, {Alexandroff}, {Borthakur},
  {Overzier}  \& {Leitherer}}{{Heckman} et~al.}{2015}]{Heckman_2015}
{Heckman} T.~M.,  {Alexandroff} R.~M.,  {Borthakur} S.,  {Overzier} R.,
  {Leitherer} C.,  2015, \mn@doi [\apj] {10.1088/0004-637X/809/2/147}, \href
  {https://ui.adsabs.harvard.edu/abs/2015ApJ...809..147H} {809, 147}

\bibitem[\protect\citeauthoryear{{Henry}, {Scarlata}, {Martin}  \&
  {Erb}}{{Henry} et~al.}{2015}]{Henry_2015}
{Henry} A.,  {Scarlata} C.,  {Martin} C.~L.,   {Erb} D.,  2015, \mn@doi [\apj]
  {10.1088/0004-637X/809/1/19}, \href
  {https://ui.adsabs.harvard.edu/abs/2015ApJ...809...19H} {809, 19}

\bibitem[\protect\citeauthoryear{{Herenz} et~al.,}{{Herenz}
  et~al.}{2017}]{Herenz2017}
{Herenz} E.~C.,  et~al., 2017, \mn@doi [\aap] {10.1051/0004-6361/201731055},
  \href {http://adsabs.harvard.edu/abs/2017A%26A...606A..12H} {606, A12}

\bibitem[\protect\citeauthoryear{{Hill} et~al.,}{{Hill} et~al.}{2008}]{hill08}
{Hill} G.~J.,  et~al., 2008, in {T.~Kodama, T.~Yamada, \& K.~Aoki} ed.,
  Astronomical Society of the Pacific Conference Series Vol. 399, Astronomical
  Society of the Pacific Conference Series. pp 115--+ (\mn@eprint {arXiv}
  {0806.0183})

\bibitem[\protect\citeauthoryear{{Hooker}, {Erhan}, {Kindermans}  \&
  {Kim}}{{Hooker} et~al.}{2018}]{Hooker_2018}
{Hooker} S.,  {Erhan} D.,  {Kindermans} P.-J.,   {Kim} B.,  2018, arXiv
  e-prints, \href {https://ui.adsabs.harvard.edu/abs/2018arXiv180610758H} {p.
  arXiv:1806.10758}

\bibitem[\protect\citeauthoryear{Hunter}{Hunter}{2007}]{Hunter:2007}
Hunter J.~D.,  2007, Computing In Science \& Engineering, 9, 90

\bibitem[\protect\citeauthoryear{{Izotov}, {Schaerer}, {Thuan}, {Worseck},
  {Guseva}, {Orlitov{\'a}}  \& {Verhamme}}{{Izotov} et~al.}{2016}]{Izotov_2016}
{Izotov} Y.~I.,  {Schaerer} D.,  {Thuan} T.~X.,  {Worseck} G.,  {Guseva} N.~G.,
   {Orlitov{\'a}} I.,   {Verhamme} A.,  2016, \mn@doi [\mnras]
  {10.1093/mnras/stw1205}, \href
  {https://ui.adsabs.harvard.edu/abs/2016MNRAS.461.3683I} {461, 3683}

\bibitem[\protect\citeauthoryear{{Izotov}, {Worseck}, {Schaerer}, {Guseva},
  {Thuan}, {Fricke}  \& {Orlitov{\'a}}}{{Izotov} et~al.}{2018}]{Izotov_2018}
{Izotov} Y.~I.,  {Worseck} G.,  {Schaerer} D.,  {Guseva} N.~G.,  {Thuan} T.~X.,
   {Fricke} Verhamme A.,   {Orlitov{\'a}} I.,  2018, \mn@doi [\mnras]
  {10.1093/mnras/sty1378}, \href
  {https://ui.adsabs.harvard.edu/abs/2018MNRAS.478.4851I} {478, 4851}

\bibitem[\protect\citeauthoryear{{Izotov}, {Schaerer}, {Worseck}, {Verhamme},
  {Guseva}, {Thuan}, {Orlitov{\'a}}  \& {Fricke}}{{Izotov}
  et~al.}{2020}]{Izotov_2020}
{Izotov} Y.~I.,  {Schaerer} D.,  {Worseck} G.,  {Verhamme} A.,  {Guseva} N.~G.,
   {Thuan} T.~X.,  {Orlitov{\'a}} I.,   {Fricke} K.~J.,  2020, \mn@doi [\mnras]
  {10.1093/mnras/stz3041}, \href
  {https://ui.adsabs.harvard.edu/abs/2020MNRAS.491..468I} {491, 468}

\bibitem[\protect\citeauthoryear{{Izotov}, {Worseck}, {Schaerer}, {Guseva},
  {Chisholm}, {Thuan}, {Fricke}  \& {Verhamme}}{{Izotov}
  et~al.}{2021}]{Izotov_2021}
{Izotov} Y.~I.,  {Worseck} G.,  {Schaerer} D.,  {Guseva} N.~G.,  {Chisholm} J.,
   {Thuan} T.~X.,  {Fricke} K.~J.,   {Verhamme} A.,  2021, \mn@doi [\mnras]
  {10.1093/mnras/stab612}, \href
  {https://ui.adsabs.harvard.edu/abs/2021MNRAS.503.1734I} {503, 1734}

\bibitem[\protect\citeauthoryear{{Jaskot} \& {Oey}}{{Jaskot} \&
  {Oey}}{2014}]{Jaskot_2014}
{Jaskot} A.~E.,  {Oey} M.~S.,  2014, \mn@doi [\apjl]
  {10.1088/2041-8205/791/2/L19}, \href
  {https://ui.adsabs.harvard.edu/abs/2014ApJ...791L..19J} {791, L19}

\bibitem[\protect\citeauthoryear{{Jaskot}, {Oey}, {Scarlata}  \&
  {Dowd}}{{Jaskot} et~al.}{2017}]{Jaskot_2017}
{Jaskot} A.~E.,  {Oey} M.~S.,  {Scarlata} C.,   {Dowd} T.,  2017, \mn@doi
  [\apjl] {10.3847/2041-8213/aa9d83}, \href
  {https://ui.adsabs.harvard.edu/abs/2017ApJ...851L...9J} {851, L9}

\bibitem[\protect\citeauthoryear{{Kakuma} et~al.,}{{Kakuma}
  et~al.}{2019}]{Kakuma_2019}
{Kakuma} R.,  et~al., 2019, arXiv e-prints, \href
  {https://ui.adsabs.harvard.edu/abs/2019arXiv190600173K} {p. arXiv:1906.00173}

\bibitem[\protect\citeauthoryear{{Kuleshov}, {Fenner}  \& {Ermon}}{{Kuleshov}
  et~al.}{2018}]{Kuleshov2018}
{Kuleshov} V.,  {Fenner} N.,   {Ermon} S.,  2018, arXiv e-prints, \href
  {https://ui.adsabs.harvard.edu/abs/2018arXiv180700263K} {p. arXiv:1807.00263}

\bibitem[\protect\citeauthoryear{{Laursen}, {Sommer-Larsen}  \&
  {Razoumov}}{{Laursen} et~al.}{2011}]{laursen11}
{Laursen} P.,  {Sommer-Larsen} J.,   {Razoumov} A.~O.,  2011, \mn@doi [\apj]
  {10.1088/0004-637X/728/1/52}, \href
  {http://adsabs.harvard.edu/abs/2011ApJ...728...52L} {728, 52}

\bibitem[\protect\citeauthoryear{{Leclercq} et~al.,}{{Leclercq}
  et~al.}{2017}]{Leclercq_2017}
{Leclercq} F.,  et~al., 2017, \mn@doi [\aap] {10.1051/0004-6361/201731480},
  \href {http://adsabs.harvard.edu/abs/2017A%26A...608A...8L} {608, A8}

\bibitem[\protect\citeauthoryear{{Li}, {Steidel}, {Gronke}, {Chen}  \&
  {Matsuda}}{{Li} et~al.}{2021a}]{Li2021}
{Li} Z.,  {Steidel} C.~C.,  {Gronke} M.,  {Chen} Y.,   {Matsuda} Y.,  2021a,
  arXiv e-prints, \href {https://ui.adsabs.harvard.edu/abs/2021arXiv210410682L}
  {p. arXiv:2104.10682}

\bibitem[\protect\citeauthoryear{{Li}, {Steidel}, {Gronke}  \& {Chen}}{{Li}
  et~al.}{2021b}]{Li2020}
{Li} Z.,  {Steidel} C.~C.,  {Gronke} M.,   {Chen} Y.,  2021b, \mn@doi [\mnras]
  {10.1093/mnras/staa3951}, \href
  {https://ui.adsabs.harvard.edu/abs/2021MNRAS.502.2389L} {502, 2389}

\bibitem[\protect\citeauthoryear{{Lundberg} \& {Lee}}{{Lundberg} \&
  {Lee}}{2017}]{Lundberg_2017}
{Lundberg} S.,  {Lee} S.-I.,  2017, arXiv e-prints, \href
  {https://ui.adsabs.harvard.edu/abs/2017arXiv170507874L} {p. arXiv:1705.07874}

\bibitem[\protect\citeauthoryear{{Martin}, {Moore}, {Morrissey}, {Matuszewski},
  {Rahman}, {Adkins}  \& {Epps}}{{Martin} et~al.}{2010}]{Martin_2010}
{Martin} C.,  {Moore} A.,  {Morrissey} P.,  {Matuszewski} M.,  {Rahman} S.,
  {Adkins} S.,   {Epps} H.,  2010, in {McLean} I.~S.,  {Ramsay} S.~K.,
  {Takami} H.,  eds,  Society of Photo-Optical Instrumentation Engineers (SPIE)
  Conference Series Vol. 7735, Ground-based and Airborne Instrumentation for
  Astronomy III. p. 77350M, \mn@doi{10.1117/12.858227}

\bibitem[\protect\citeauthoryear{Miranda}{Miranda}{2018}]{Miranda_2018}
Miranda L. J.~V.,  2018, \mn@doi [Journal of Open Source Software]
  {10.21105/joss.00433}, 3

\bibitem[\protect\citeauthoryear{{Muzahid} et~al.,}{{Muzahid}
  et~al.}{2019}]{Muzahid_2019}
{Muzahid} S.,  et~al., 2019, arXiv e-prints, \href
  {https://ui.adsabs.harvard.edu/abs/2019arXiv191003593M} {p. arXiv:1910.03593}

\bibitem[\protect\citeauthoryear{{Neufeld}}{{Neufeld}}{1990}]{neufeld90}
{Neufeld} D.~A.,  1990, \mn@doi [\apj] {10.1086/168375}, \href
  {http://adsabs.harvard.edu/abs/1990ApJ...350..216N} {350, 216}

\bibitem[\protect\citeauthoryear{{Orlitov{\'a}}, {Verhamme}, {Henry},
  {Scarlata}, {Jaskot}, {Oey}  \& {Schaerer}}{{Orlitov{\'a}}
  et~al.}{2018}]{Orlitova_2018}
{Orlitov{\'a}} I.,  {Verhamme} A.,  {Henry} A.,  {Scarlata} C.,  {Jaskot} A.,
  {Oey} M.~S.,   {Schaerer} D.,  2018, \mn@doi [\aap]
  {10.1051/0004-6361/201732478}, \href
  {http://adsabs.harvard.edu/abs/2018A%26A...616A..60O} {616, A60}

\bibitem[\protect\citeauthoryear{{Orsi}, {Lacey}  \& {Baugh}}{{Orsi}
  et~al.}{2012}]{orsi12}
{Orsi} A.,  {Lacey} C.~G.,   {Baugh} C.~M.,  2012, \mn@doi [\mnras]
  {10.1111/j.1365-2966.2012.21396.x}, \href
  {http://adsabs.harvard.edu/abs/2012MNRAS.425...87O} {425, 87}

\bibitem[\protect\citeauthoryear{{Ouchi} et~al.,}{{Ouchi}
  et~al.}{2018}]{Ouchi2018a}
{Ouchi} M.,  et~al., 2018, \mn@doi [\pasj] {10.1093/pasj/psx074}, \href
  {http://adsabs.harvard.edu/abs/2018PASJ...70S..13O} {70, S13}

\bibitem[\protect\citeauthoryear{Ouchi, Ono  \& Shibuya}{Ouchi
  et~al.}{2020}]{OuchiObservationsLymana2020}
Ouchi M.,  Ono Y.,   Shibuya T.,  2020, \mn@doi [Annual Review of Astronomy and
  Astrophysics] {10.1146/annurev-astro-032620-021859}, 58, 617

\bibitem[\protect\citeauthoryear{{Rauch}}{{Rauch}}{2015}]{Rauch_2015}
{Rauch} M.,  2015, in IAU General Assembly. p. 2258163

\bibitem[\protect\citeauthoryear{{Rivera-Thorsen} et~al.,}{{Rivera-Thorsen}
  et~al.}{2015}]{Rivera-Thorsen_2015}
{Rivera-Thorsen} T.~E.,  et~al., 2015, \mn@doi [\apj]
  {10.1088/0004-637X/805/1/14}, \href
  {https://ui.adsabs.harvard.edu/abs/2015ApJ...805...14R} {805, 14}

\bibitem[\protect\citeauthoryear{{Rudie}, {Steidel}  \& {Pettini}}{{Rudie}
  et~al.}{2012}]{Rudie_2012}
{Rudie} G.~C.,  {Steidel} C.~C.,   {Pettini} M.,  2012, \mn@doi [\apj]
  {10.1088/2041-8205/757/2/L30}, \href
  {https://ui.adsabs.harvard.edu/abs/2012ApJ...757L..30R} {757, L30}

\bibitem[\protect\citeauthoryear{{Runnholm}, {Gronke}  \& {Hayes}}{{Runnholm}
  et~al.}{2021}]{Runnholm_2020}
{Runnholm} A.,  {Gronke} M.,   {Hayes} M.,  2021, \mn@doi [\pasp]
  {10.1088/1538-3873/abe3ca}, \href
  {https://ui.adsabs.harvard.edu/abs/2021PASP..133c4507R} {133, 034507}

\bibitem[\protect\citeauthoryear{{Salzer} et~al.,}{{Salzer}
  et~al.}{2001}]{salzer_2001}
{Salzer} J.~J.,  et~al., 2001, \mn@doi [\aj] {10.1086/318040}, \href
  {https://ui.adsabs.harvard.edu/abs/2001AJ....121...66S} {121, 66}

\bibitem[\protect\citeauthoryear{{Schaerer} \& {Verhamme}}{{Schaerer} \&
  {Verhamme}}{2008}]{schaerer08}
{Schaerer} D.,  {Verhamme} A.,  2008, \mn@doi [\aap]
  {10.1051/0004-6361:20078913}, \href
  {http://adsabs.harvard.edu/abs/2008A%26A...480..369S} {480, 369}

\bibitem[\protect\citeauthoryear{{Schaerer}, {Hayes}, {Verhamme}  \&
  {Teyssier}}{{Schaerer} et~al.}{2011}]{Schaerer2011}
{Schaerer} D.,  {Hayes} M.,  {Verhamme} A.,   {Teyssier} R.,  2011, \mn@doi
  [\aap] {10.1051/0004-6361/201116709}, \href
  {https://ui.adsabs.harvard.edu/abs/2011A&A...531A..12S} {531, A12}

\bibitem[\protect\citeauthoryear{{Song}, {Seon}  \& {Hwang}}{{Song}
  et~al.}{2020}]{Song_2020}
{Song} H.,  {Seon} K.-I.,   {Hwang} H.~S.,  2020, \mn@doi [\apj]
  {10.3847/1538-4357/abac02}, \href
  {https://ui.adsabs.harvard.edu/abs/2020ApJ...901...41S} {901, 41}

\bibitem[\protect\citeauthoryear{{Songaila}, {Hu}, {Barger}, {Cowie},
  {Hasinger}, {Rosenwasser}  \& {Waters}}{{Songaila}
  et~al.}{2018}]{Songaila_2018}
{Songaila} A.,  {Hu} E.~M.,  {Barger} A.~J.,  {Cowie} L.~L.,  {Hasinger} G.,
  {Rosenwasser} B.,   {Waters} C.,  2018, \mn@doi [\apj]
  {10.3847/1538-4357/aac021}, \href
  {https://ui.adsabs.harvard.edu/abs/2018ApJ...859...91S} {859, 91}

\bibitem[\protect\citeauthoryear{{Spinoso} et~al.,}{{Spinoso}
  et~al.}{2020}]{spinoso_2020}
{Spinoso} D.,  et~al., 2020, \mn@doi [\aap] {10.1051/0004-6361/202038756},
  \href {https://ui.adsabs.harvard.edu/abs/2020A&A...643A.149S} {643, A149}

\bibitem[\protect\citeauthoryear{{Steidel}, {Erb}, {Shapley}, {Pettini},
  {Reddy}, {Bogosavljevi{\'c}}, {Rudie}  \& {Rakic}}{{Steidel}
  et~al.}{2010}]{steidel10}
{Steidel} C.~C.,  {Erb} D.~K.,  {Shapley} A.~E.,  {Pettini} M.,  {Reddy} N.,
  {Bogosavljevi{\'c}} M.,  {Rudie} G.~C.,   {Rakic} O.,  2010, \mn@doi [\apj]
  {10.1088/0004-637X/717/1/289}, \href
  {http://adsabs.harvard.edu/abs/2010ApJ...717..289S} {717, 289}

\bibitem[\protect\citeauthoryear{{Steidel}, {Bogosavljevi{\'c}}, {Shapley},
  {Kollmeier}, {Reddy}, {Erb}  \& {Pettini}}{{Steidel}
  et~al.}{2011}]{steidel11}
{Steidel} C.~C.,  {Bogosavljevi{\'c}} M.,  {Shapley} A.~E.,  {Kollmeier} J.~A.,
   {Reddy} N.~A.,  {Erb} D.~K.,   {Pettini} M.,  2011, \mn@doi [\apj]
  {10.1088/0004-637X/736/2/160}, \href
  {http://adsabs.harvard.edu/abs/2011ApJ...736..160S} {736, 160}

\bibitem[\protect\citeauthoryear{{Steidel}, {Bogosavljevi{\'c}}, {Shapley},
  {Reddy}, {Rudie}, {Pettini}, {Trainor}  \& {Strom}}{{Steidel}
  et~al.}{2018}]{Steidel_2018}
{Steidel} C.~C.,  {Bogosavljevi{\'c}} M.,  {Shapley} A.~E.,  {Reddy} N.~A.,
  {Rudie} G.~C.,  {Pettini} M.,  {Trainor} R.~F.,   {Strom} A.~L.,  2018,
  \mn@doi [\apj] {10.3847/1538-4357/aaed28}, \href
  {https://ui.adsabs.harvard.edu/abs/2018ApJ...869..123S} {869, 123}

\bibitem[\protect\citeauthoryear{Tumlinson, Peeples  \& Werk}{Tumlinson
  et~al.}{2017}]{Tumlinson2017}
Tumlinson J.,  Peeples M.~S.,   Werk J.~K.,  2017, \mn@doi [Annual Review of
  Astronomy and Astrophysics] {10.1146/annurev-astro-091916}, 55, 389

\bibitem[\protect\citeauthoryear{{Urrutia} et~al.,}{{Urrutia}
  et~al.}{2019}]{Urrutia2019A&A...624A.141U}
{Urrutia} T.,  et~al., 2019, \mn@doi [\aap] {10.1051/0004-6361/201834656},
  \href {https://ui.adsabs.harvard.edu/abs/2019A&A...624A.141U} {624, A141}

\bibitem[\protect\citeauthoryear{{Verhamme}, {Schaerer}  \&
  {Maselli}}{{Verhamme} et~al.}{2006}]{verhamme06}
{Verhamme} A.,  {Schaerer} D.,   {Maselli} A.,  2006, \mn@doi [\aap]
  {10.1051/0004-6361:20065554}, \href
  {http://adsabs.harvard.edu/abs/2006A%26A...460..397V} {460, 397}

\bibitem[\protect\citeauthoryear{{Verhamme}, {Schaerer}, {Atek}  \&
  {Tapken}}{{Verhamme} et~al.}{2007}]{Verhamme_2007}
{Verhamme} A.,  {Schaerer} D.,  {Atek} H.,   {Tapken} C.,  2007, in {Afonso}
  J.,  {Ferguson} H.~C.,  {Mobasher} B.,   {Norris} R.,  eds,  Astronomical
  Society of the Pacific Conference Series Vol. 380, Deepest Astronomical
  Surveys. p.~97

\bibitem[\protect\citeauthoryear{{Verhamme}, {Orlitov{\'a}}, {Schaerer}  \&
  {Hayes}}{{Verhamme} et~al.}{2015}]{verhamme_2014}
{Verhamme} A.,  {Orlitov{\'a}} I.,  {Schaerer} D.,   {Hayes} M.,  2015, \mn@doi
  [\aap] {10.1051/0004-6361/201423978}, \href
  {https://ui.adsabs.harvard.edu/abs/2015A&A...578A...7V} {578, A7}

\bibitem[\protect\citeauthoryear{{Verhamme} et~al.,}{{Verhamme}
  et~al.}{2018}]{Verhamme:2018aa}
{Verhamme} A.,  et~al., 2018, \mn@doi [\mnras] {10.1093/mnrasl/sly058}, \href
  {https://ui.adsabs.harvard.edu/abs/2018MNRAS.478L..60V} {478, L60}

\bibitem[\protect\citeauthoryear{{Vielfaure} et~al.,}{{Vielfaure}
  et~al.}{2020}]{Vielfaure_2020}
{Vielfaure} J.~B.,  et~al., 2020, \mn@doi [\aap] {10.1051/0004-6361/202038316},
  \href {https://ui.adsabs.harvard.edu/abs/2020A&A...641A..30V} {641, A30}

\bibitem[\protect\citeauthoryear{{Virtanen} et~al.,}{{Virtanen}
  et~al.}{2020}]{Virtanen_2020}
{Virtanen} P.,  et~al., 2020, \mn@doi [Nature Methods]
  {https://doi.org/10.1038/s41592-019-0686-2}, \href {https://rdcu.be/b08Wh}
  {17, 261}

\bibitem[\protect\citeauthoryear{{Weiss} et~al.,}{{Weiss}
  et~al.}{2021}]{Weiss_2021}
{Weiss} L.~H.,  et~al., 2021, \mn@doi [\apj] {10.3847/1538-4357/abedb9}, \href
  {https://ui.adsabs.harvard.edu/abs/2021ApJ...912..100W} {912, 100}

\bibitem[\protect\citeauthoryear{{Wisotzki} et~al.,}{{Wisotzki}
  et~al.}{2016}]{Wisotzki2016}
{Wisotzki} L.,  et~al., 2016, \mn@doi [\aap] {10.1051/0004-6361/201527384},
  \href {https://ui.adsabs.harvard.edu/abs/2016A&A...587A..98W} {587, A98}

\bibitem[\protect\citeauthoryear{{Wofford}, {Leitherer}  \& {Salzer}}{{Wofford}
  et~al.}{2013}]{Wofford_2013}
{Wofford} A.,  {Leitherer} C.,   {Salzer} J.,  2013, \mn@doi [\apj]
  {10.1088/0004-637X/765/2/118}, \href
  {https://ui.adsabs.harvard.edu/abs/2013ApJ...765..118W} {765, 118}

\bibitem[\protect\citeauthoryear{{Yang}, {Malhotra}, {Rhoads}  \&
  {Wang}}{{Yang} et~al.}{2017}]{Yang_2017}
{Yang} H.,  {Malhotra} S.,  {Rhoads} J.~E.,   {Wang} J.,  2017, \mn@doi [\apj]
  {10.3847/1538-4357/aa8809}, \href
  {https://ui.adsabs.harvard.edu/abs/2017ApJ...847...38Y} {847, 38}

\bibitem[\protect\citeauthoryear{{Zheng} \& {Miralda-Escud{\'e}}}{{Zheng} \&
  {Miralda-Escud{\'e}}}{2002}]{zheng02}
{Zheng} Z.,  {Miralda-Escud{\'e}} J.,  2002, \mn@doi [\apj] {10.1086/342400},
  \href {http://adsabs.harvard.edu/abs/2002ApJ...578...33Z} {578, 33}

\makeatother
\end{thebibliography}




\appendix

\section{Computation of \lya\ line profiles}\label{s:interpolation}
        
        In order to compute the \lya\ line profiles at an arbitrary point inside our grid range, we use linear interpolation between the closest nodes of the grid to that point. In particular we interpolate in the 5 dimensional space \{\vexp , $\log$\nh , $\log$\ta , $\log$\ew , \w \} . \\
        
        For simplicity, let's rename this space from \{\vexp , $\log$\nh , $\log$\ta , $\log$\ew , \w \} to \{$x_1$, $x_2$, $x_3$, $x_4$, $x_5$\}. Therefore, for a given combination of these parameters in which we want to compute the \lya\ line profile \{$V_{\rm exp}^{p}$ , $\log N_{\rm H}^{p}$ , $\log\tau_{\rm a}^{p}$, $\log EW_{\rm in}^{p}$ , $W_{\rm in}^{p}$ \} = \{$x_1^p$, $x_2^p$, $x_3^p$, $x_4^p$, $x_5^p$\}, first we find the closest neighbours in our grid that surround our target. Given that we are working in a 5 dimensional space there are a total of 32 neighbours, or grid nodes, surrounding the target. The coordinates of the neighbours are the 32 possible permutations of \{$x_1^{p,\phi}$, $x_2^{p,\phi}$, $x_3^{p,\phi}$, $x_4^{p,\phi}$, $x_5^{p,\phi}$\}, where $\phi$ can take two different values, $\downarrow$ and $\uparrow$ and they fulfil that $x_i^{p\downarrow}<x_i^p<x_i^{p\uparrow}$ for $i=1,2,3,4,5$. Each of these points is the corner of a hiper right parallelogrammic prism of 5 dimensions. Then, the \lya\ line profile in the target position $f(\lambda,x_1^p, x_2^p, x_3^p, x_4^p, x_5^p)$ is a weighed linear combination of the 32 \lya\ line profiles from the neighbours $f(\lambda,x_1^{p,\phi}, x_2^{p,\phi}, x_3^{p,\phi}, x_4^{p,\phi}, x_5^{p,\phi})$, i.e.,
        
        \begin{equation}
        f(x_1^p,\cdots, x_5^p) = \frac{ \displaystyle\sum\limits_{y_1}^{x_1^{p\downarrow} , x_1^{p\uparrow}} \cdots  \displaystyle\sum\limits_{y_5}^{x_5^{p\downarrow} , x_5^{p\uparrow}}   f( y_1, \cdots , y_5) w(y_1, \cdots , y_5) }
        {\displaystyle\sum\limits_{y_1}^{x_1^{p\downarrow} , x_1^{p\uparrow}} \cdots  \displaystyle\sum\limits_{y_5}^{x_5^{p\downarrow} , x_5^{p\uparrow}}  w(y_1, \cdots , y_5) },
        \end{equation}
        where $w(x_1^{p,\phi}, x_2^{p,\phi}, x_3^{p,\phi}, x_4^{p,\phi}, x_5^{p,\phi})$ is the weight of the neighbour located in \{$x_1^{p,\phi}$, $x_2^{p,\phi}$, $x_3^{p,\phi}$, $x_4^{p,\phi}$, $x_5^{p,\phi}$\} and it is equal to the hiper volume contained between the target position \{$x_1^p$, $x_2^p$, $x_3^p$, $x_4^p$, $x_5^p$\} and its opposite corner in the hiper right parallelogrammic prism, i.e., \{$x_1^{p,-\phi}$, $x_2^{p,-\phi}$, $x_3^{p,-\phi}$, $x_4^{p,-\phi}$, $x_5^{p,-\phi}$\}, where $\uparrow=-\downarrow$ and vice versa. Analytically, 
        
        \begin{equation}
        w(x_1^{p,\phi}, \cdots , x_5^{p,\phi}) = \displaystyle\prod\limits_{y}^{x_1^{p,-\phi}, \cdots , x_5^{p,-\phi}} L(y),
        \end{equation}
        where 
        
        \begin{equation}
        L(x_i^{p\uparrow}) = \frac{ x_i^p - x_i^{p\uparrow} }{ x_i^{p\uparrow} - x_i^{p\downarrow} } \; \; \; {\rm and} \; \; \; L(x_i^{p\downarrow})=1-L(x_i^{p\uparrow}).
        \end{equation}
        \\

\section{Inflow gas geometry}\label{s:inflow_making}
        
\begin{figure} 
        \includegraphics[width=3.4in]{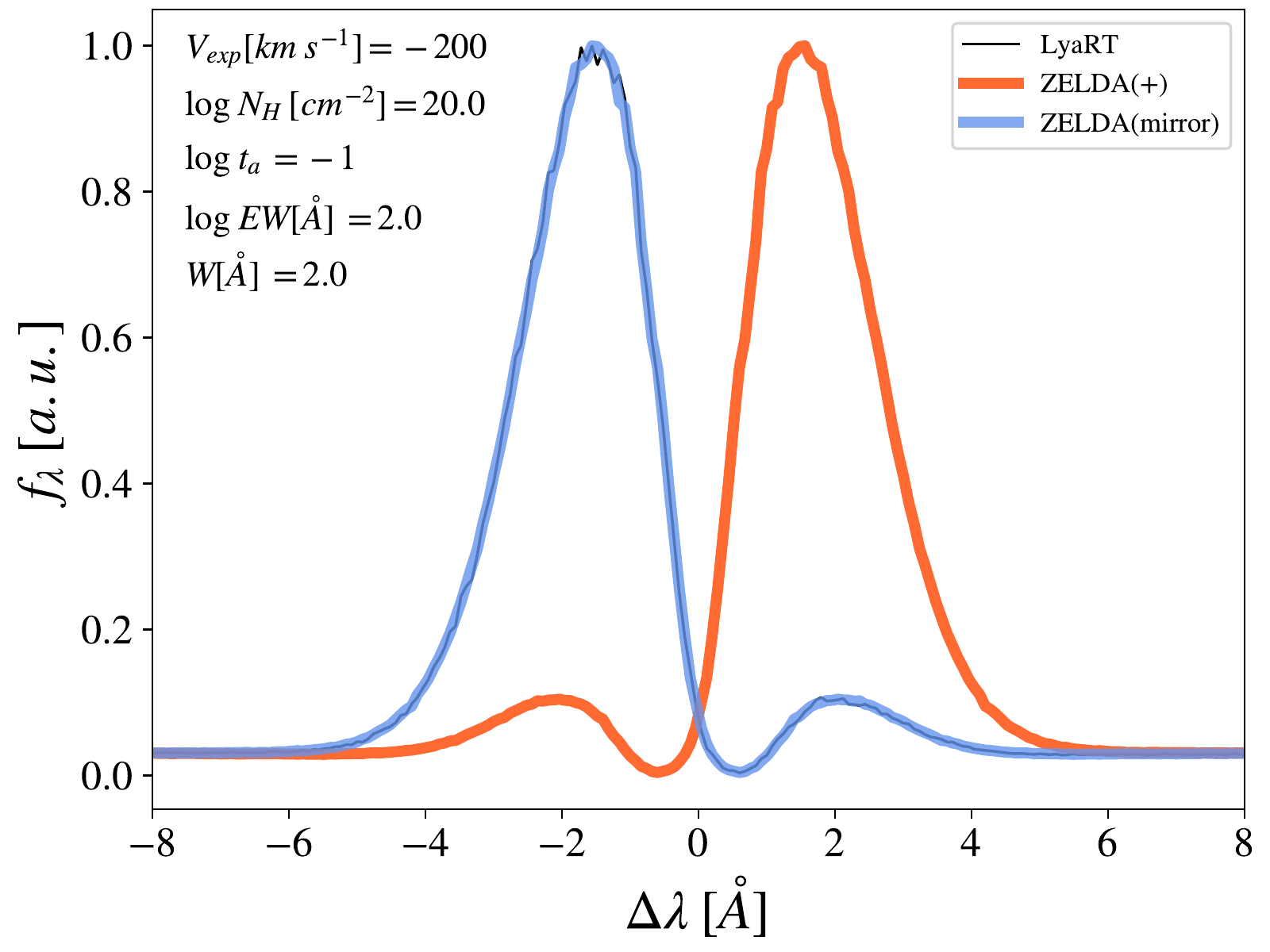}%
        \caption{ Illustration of how the line profiles emerging from inflows are computed in \zelda . In solid black, we show the full RT computation of a shell model with \vexp$=-200 km/s$. In red we show \zelda 's prediction for an outflow configuration with the same properties but \vexp$=+200 km/s$. In blue we display the inflow line profile predicted by \zelda\ after applying the procedure described in \S\ref{s:inflow_making}.     }
        \label{fig:inflow_example}
        \end{figure}
        
        Here we explain how the \lya\ line profiles from inflow are predicted in \zelda . 
        Because of the symmetry of the equations, the emergent spectrum of a shell with infall velocity $v_{\rm infall}=-v_{\rm exp}$ will be identical to the one for an outflow velocity $v_{\rm exp}$ but mirrored around the \lya\ wavelength \citep{neufeld90,dijkstra06,Schaerer2011}.
        Given the outflow line profile $f_{\lambda ,\;  \rm Out} ^{\rm Ly\alpha}$ for a given configuration of \{\vexp>0 , \nh , \ta , \ew \w \}, then, the line profile of the inflow with  \{ - \vexp , \nh , \ta , \ew \w \} is computed by doing the mapping 
        
        \begin{equation}
            f_{\lambda ,\;  \rm Out} ^{\rm Ly\alpha} ( \lambda - \lambda _{\rm Ly\alpha}) \rightarrow  f_{\lambda ,\; \rm Out} ^{\rm Ly\alpha} ( \lambda _{\rm Ly\alpha} - \lambda ) = f_{\lambda ,\;  \rm In} ^{\rm Ly\alpha} ( \lambda - \lambda _{\rm Ly\alpha}).
        \end{equation}
        This can be seen as inversion of the wavelength around the \lya\ wavelength. 
        
        In Fig. \ref{fig:inflow_example} we show an example of a line profile emerging from an inflow. In black we show the full RT computation made by \lyart\ and using $V_{\rm exp}=-200 {\rm km/s}$ running $2\times10^6$ photons. In red we show the \zelda 's prediction of the line profile of an outflow with the same properties by  $V_{\rm exp}=+200 {\rm km/s}$. In blue we show our computation of the inflow line profile, computed as described above from the outflow line profile. The line profiles computed using \lyart\ and our scheme agree well. We have tried that several inflow configurations with \lyart\ and found that this method always works well. 
    
\section{Comparison of measured accuracy in different mock line profile samples.}\label{s:accuracy_comparisons}

\begin{figure*}
    \includegraphics[width=6.9in]{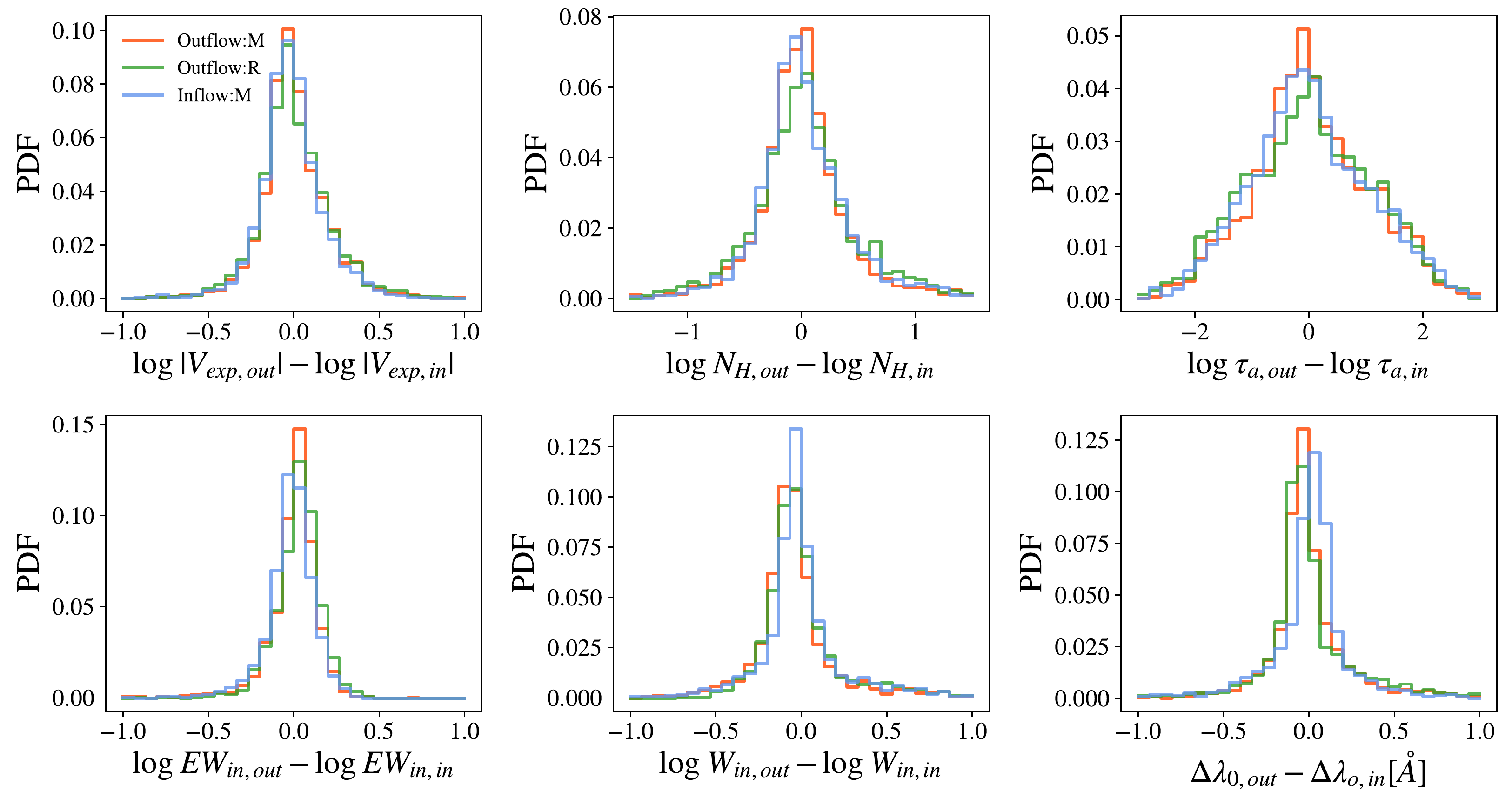}%
    \caption{ Accuracy of DNN predictions. 2000 mock line profiles computed making interpolation in the grid with random outflow parameters and from redshift 0.0001 to 4.0 with quality \wg=0.5\AA{} , \dl=0.1\AA{} and \sn = 10 .}
    \label{fig:NN_comparison}
    \end{figure*} 
    
In this section we demonstrate that the accuracy for the DNN approaches in inflows (\vexp$<0$) is the same than in outflows (\vexp$>0$). In \S \ref{s:DNN_build} we have measured the accuracy of you DNN approaches with the samples described in \S\ref{ss:sample}, which is composed by the line profiles of outflows directly produced by \lyart. We will refer to this sample as \outr. 

In Fig.~\ref{fig:NN_comparison} we compare the accuracy of the MC DNN approach in different samples. All with quality \w=0.5\AA{}, \dl=0.1\AA{} and \sn=10. In the first place, \outr is shown in green. Then, \outm (red) is a sample of 2000 line profiles from outflows populating the same parameter space volume than \outr. \outm line profiles are produced by the interpolation scheme described in \ref{s:modeling}, in contrast with \outr that uses the full RT computation of \lyart. We find that the PDF of the difference between the true parameter value (e.g. $V_{\rm exp,in}$) and the predicted by the MC DNN approach ($V_{\rm exp,out}$) are similar. This shows that the accuracy of this methodology in line profiles produced by \lyart\ and those predicted by \zelda\ is almost the same. We have checked that this is the case also for the other quality configurations, but we decided to focus on one for clarity.

Then, in order to asses the accuracy in inflows we produced a sample of line profiles for inflows that is the same as \outm, but using inverting the sign of \vexp. We call this sample \inm (shown in blue). We find that the accuracy is the same in \outm as in \inm. In conjunction with the fact that the MC DNN approach behaves similarly in line profiles from \lyart\ and from \zelda, this shows that the accuracy in of the MC DNN is the same in outflows than in inflows.

\section{Excluded line profiles.}\label{s:excluded}

\begin{figure*} 
\includegraphics[width=6.8in]{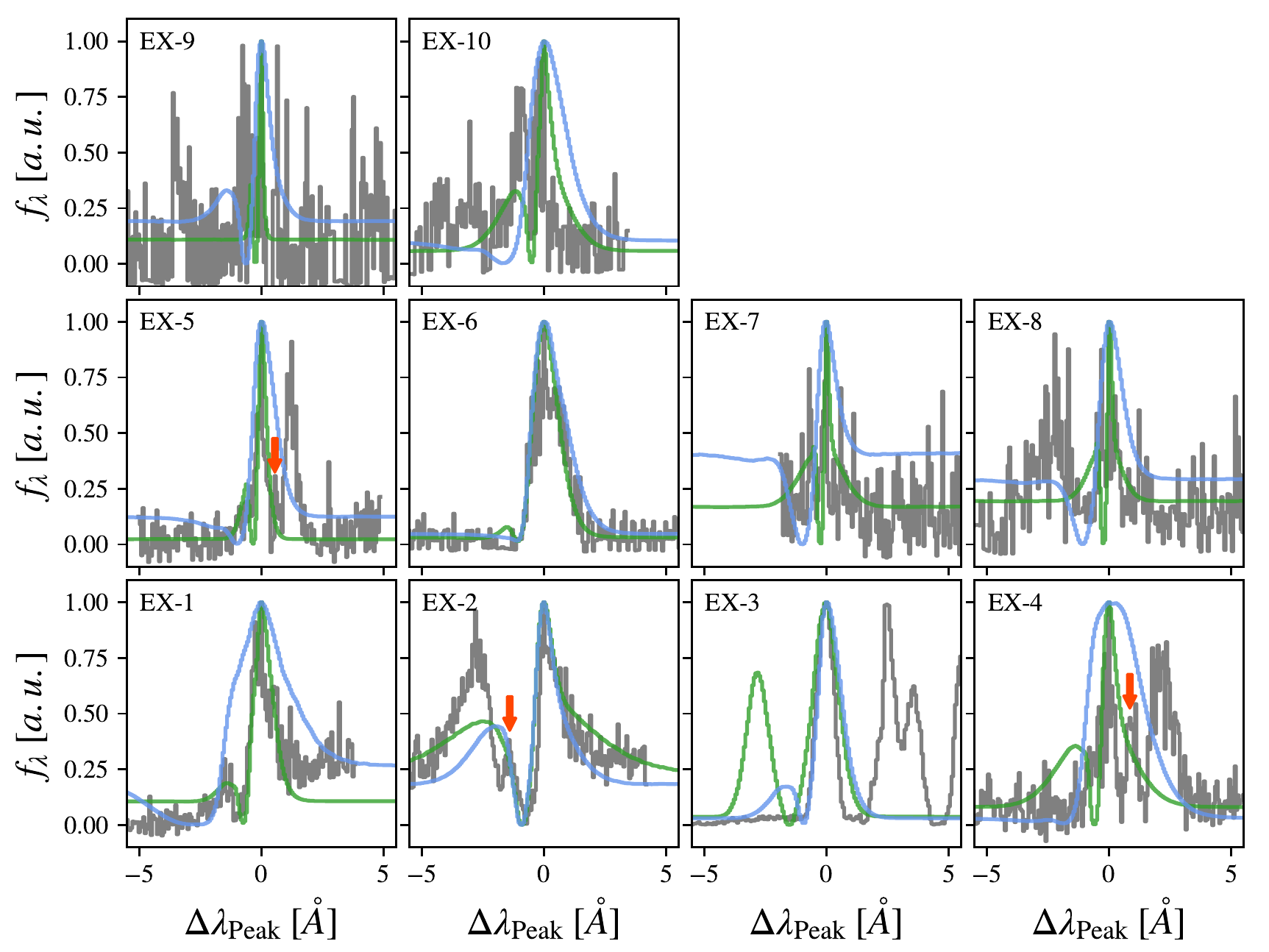}
\caption{ Line profiles excluded from the analysis performed in \S\ref{s:observation}. In grey we display the observed spectrum while in blue and in green we show the best fits of the MC DNN and MCMC methodologies respectively. In the top left we label each spectrum. In the case of EX-2, EX-4 and EX-5 we drew a red arrow to mark an extra component in the line profile. }
\label{fig:excluded_line_profile}
\end{figure*}

Here we describe the observed spectrum that were excluded from the analysis performed in \S\ref{s:observation}. The \lasd\ data base contains a total of 107 local line profiles with the systemic redshift calculated using a \lya\ independent analysis. From this set, 97 were used in \S\ref{s:observation}. The remaining 10 line profiles are shown in Fig.\ref{fig:excluded_line_profile} (grey) and the source name and systemic redshift are listed on Table \ref{tab:obs_excluded}. The best fit of the MCMC (green) and MC DNN (blue) is also displayed. Overall, we find that for these line profiles \zelda\ fits lack accuracy. 

Depending on the line profile, the cause behind the low quality fit change. First, for Ex-7, EX-9 and EX-10, the line profiles exhibit low \sn\ values (5.40, 4.89 and 5.16 respectively). Second, line profiles with more complex components, in particular, EX-2 , EX-4 and EX-5. These sources exhibit the typical double peak line profiles, but they also show an extra peak with lower amplitude (marked with a red arrow) between the main red peak and blue peak. This fainter peak is not reproduced by the shell model. The presence of this feature might be due to the fact that the spectrum might contain information of more than one source, each with their unique \lya\ line profile. Then, EX-3 has high \sn\ but there are clearly there emission lines in the regions where \lya\ should be given the systemic redshift. 

The source EX-1 exhibit a steep continuum with increasing flux towards redder wavelengths. As described in \S\ref{ss:gas_geometry}, the intrinsic Line profile that we inject in the \ThinShell\ is a Gaussian centered in \lya\ on top of a flat continuum. As neither the MCMC or the MC DNN methodologies find a suitable \lya\ in our \ThinShell\ model, the steep continuum of EX-1 might be intrinsic to the galaxy continuum and not associated with the \lya\ RT. Therefore, we exclude this galaxy from our studied sample.  

Next, EX-8 exhibits a double peak line profile in which the blue peak is wider than the red peak. In the \ThinShell\ model, normally, the width of the peaks is very similar, which could caused the low quality of the fit. We also tried to fit EX-8 line profile using an inflow instead of an outflow but the fit still lacked accuracy, probably for the same reason. 

\renewcommand{\arraystretch}{1.0} 
\begin{table}
\caption{ Name, systemic redshift, \lya\ luminosity, observed equivalent width and \sn\ for the excluded observed galaxies. The column {\it Label} indicates the number that appears with the spectra in Fig.\ref{fig:examples_observed}, \ref{fig:comparison_NN_MC_lines_0} and \ref{fig:comparison_NN_MC_lines_1}.  }
\label{tab:obs_excluded}
\begin{center}
\begin{tabular}{cccc}
Label & Name   & $z^{\rm Sys}$ & \sn   \\ \hline
EX-1 & SDSSJ1113+2930 & 0.1751 & 8.2 \\
EX-2 & SDSSJ0921+4509 & 0.235 & 10.59 \\
EX-3 & SDSSJ1525+0757 & 0.0758 & 50.51 \\
EX-4 & J1032+4919 & 0.0442 & 10.27 \\
EX-5 & J0007+0226 & 0.0636 & 11.1 \\
EX-6 & GALEX1717+5944 & 0.1979 & 12.6 \\
EX-7 & GP0749+3337 & 0.2732 & 5.4 \\
EX-8 & GP1032+2717 & 0.1925 & 6.92 \\
EX-9 & GP1205+2620 & 0.3426 & 4.89 \\
EX-10 & GP1543+3446 & 0.1873 & 5.16 \\
\end{tabular}
\end{center}
\end{table}
    
\begin{figure*} 
    \includegraphics[width=3.4in]{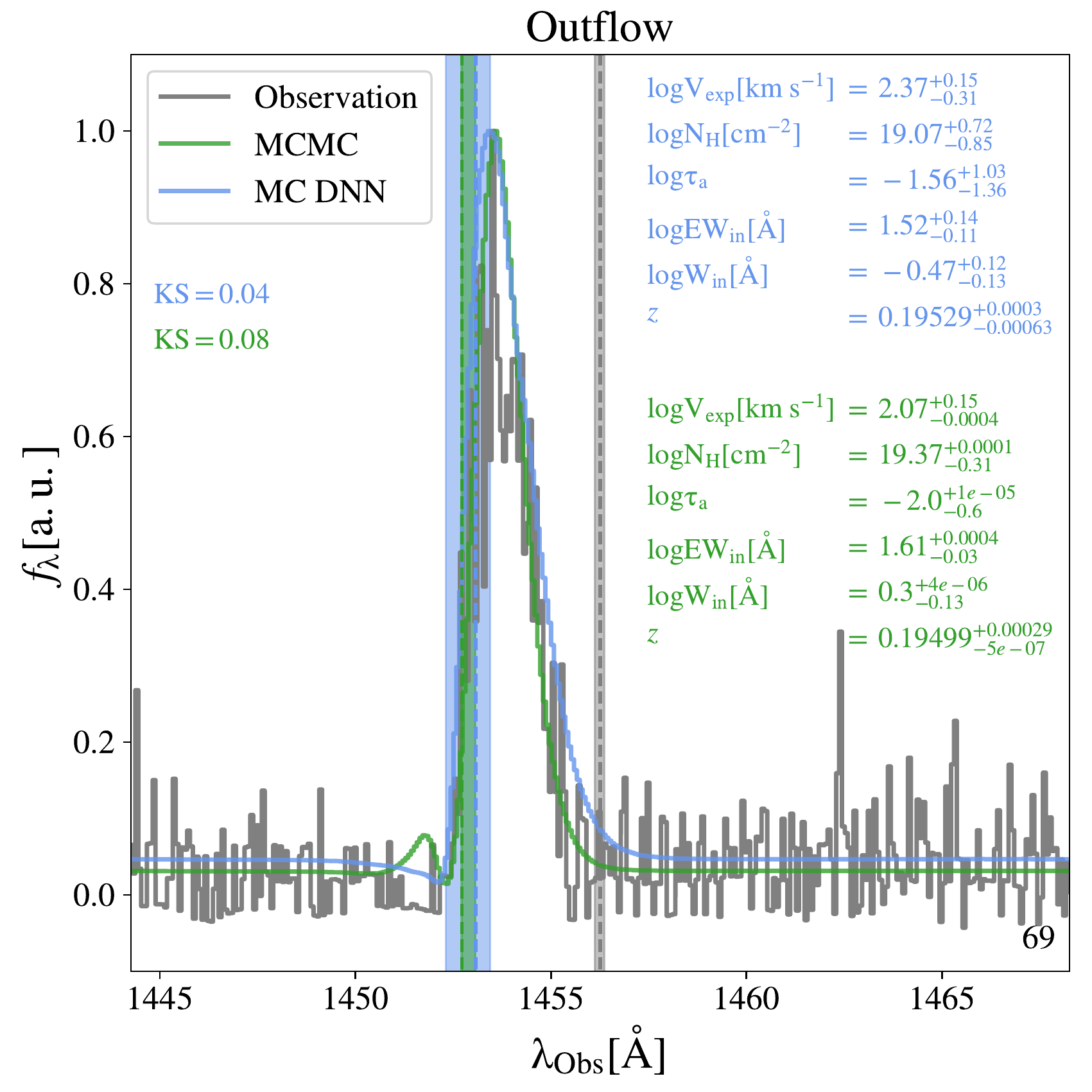}%
    \includegraphics[width=3.4in]{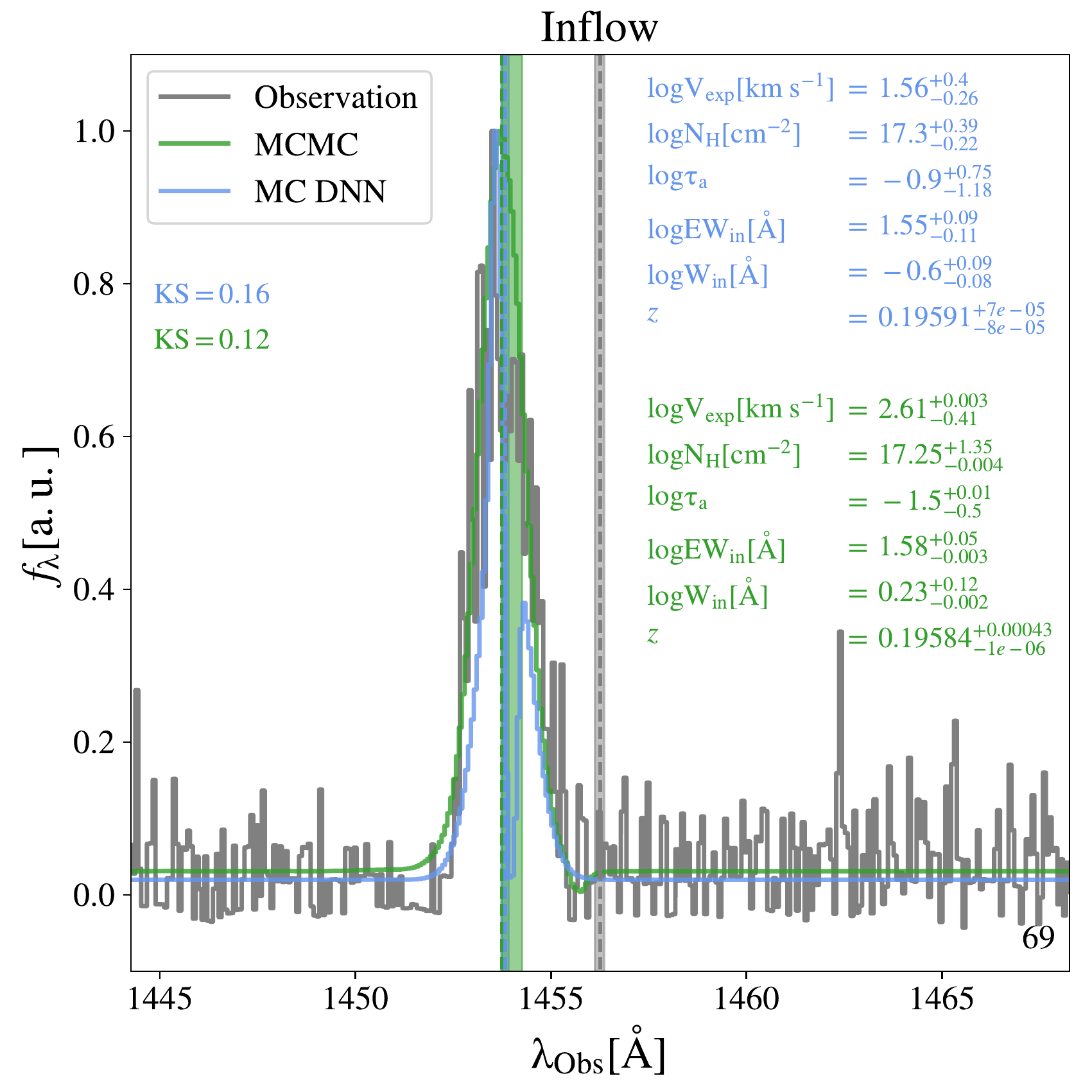} 
    \caption{ Out layer line profile EX-6, fitted by the outflow (left) and inflow (right) models. The model lines given by the MC DNN (MCMC) methodology are shown in blue (green) and the value of the outflow/inflow parameters and the 1-$\sigma$ uncertainties are indicated in the same color in the top right (middle right). The vertical dashed lines and shaded regions mark the value and the 1-$\sigma$ uncertainty of the true \lya\ wavelength of the observation (grey) and the \lya\ wavelength predicted by the MC DNN (blue) and MCMC (green). The KS estimator is given on the left of each panel. On top, for the MC DNN and on the bottom for the MCMC approach. }
    \label{fig:outlayer_line_profile}
    \end{figure*}
    
In Fig.\ref{fig:outlayer_line_profile} we show the observed line profile (grey) of EX-6 and the best fitting models for the outflow (left) and inflow (right) geometries. The results for the MCMC are shown in green while those for the MC DNN approach are shown in blue. Also the KS estimator values for the fits are shown in the left of each panel (top for MC DNN and bottom for MCMC). For this particular case, we decided to repeat the analysis assuming that the line profile was caused by an inflow due to the fact that, we observed that the \lya\ frequency given by the non-\lya\ redshift estimators was redder than the majority of the flux of the emission line. 

Overall, we find that both, the inflow and outflow models, with both methodologies, reproduce relatively well the line profile shape (KS<0.2). Also, neither the best fitting model for the outflow nor inflow provide a \lya\ wavelength close to that given by observations. In fact, within the models model (inflow or outflow) the predicted redshift is compatible between the MCMC and the MC DNN methodologies. In particular, the redshift predicted for the outflow model is $\sim 0.195$, for the inflow $\sim 0.196$ and in the observation $\sim 0.198$ with typical uncertainty of $10^{-4}$. The inflow model predicts a  \lya\ wavelength closer to the provided by observations. However, it is still $\sim 2.3$\AA{} ($\sim 1.9$\AA{}) displaced from the true \lya\ wavelength in the observed (rest) frame (470$km/s$). For the outflow model this difference is even larger, and the displacement is $\sim 3.5$\AA{} in the observed frame and in rest frame $\sim 2.9$\AA{} (715$km/s$). The wavelength shift in the outflow model is about 2.5 times the maximum displacement found for the other line profiles (see Fig.\ref{fig:redshift_accuracy_obs}). In contrast with the fact that the inflow model gives a better redshift estimation, the outflow model reproduces better the shape of the line profile. In fact, both, the MCMC (KS=0.08) and MC DNN (KS=0.04) methodologies fit the line profile better than any of the methodologies in the inflow model (KS=0.12 for the MCMC and KS=0.16 for the MC DNN). 

The disparity between the observed systemic redshift and the predictions given by \zelda\ might come from several facts. One scenario could be that the systemic redshift given by the \lya\ independent redshift estimator is not properly constrained. However, usually, these estimators give a good redshift accuracy. Other possibility is that the thin shell model might be incomplete and not able to produce these kind of line with large shifts between the \lya\ wavelength and the bulk of the line profile. Although, we find that the outflow model reproduces quite well the shape of the line (KS$\sim 0.06$). Another possible explanation is that the \lya\ radiation might come from a different galaxy component than the radiation used for measuring the systemic redshift. 

Overall, the reason behind the tension between the systemic redshift given by the \lya\ independent redshift estimator and \zelda\ remains unknown. Therefore, we decided to exclude this particular line profile from the analysis performed in \S\ref{s:observation}.

\section{Best fitting parameters of the observed line profiles.}

In Tab.\ref{tab:real_mcmc_LUM_0} and \ref{tab:real_mcmc_LUM_1} we  list the systemic redshift obtained by studying  different features than the \lya\ line $z^{\rm Sys}$, the name of the sources, \LLya and \ewl. These last two were also obtained from the {\it Lyman alpha Spectral Database}  \citep[\texttt{LASD}][]{Runnholm_2020}. In addition, we display the name and the 'Label' of each source. This 'Label' matches with the numbers shown in the bottom right of Fig.~\ref{fig:comparison_NN_MC_props} and in the top left of Fig.~\ref{fig:comparison_NN_MC_lines_0} and \ref{fig:comparison_NN_MC_lines_1}. 

In Tab.\ref{tab:real_mcmc_0}, \ref{tab:real_mcmc_1} and \ref{tab:real_mcmc_2} we list all the best fitting parameters for the 97 observed line profiles analyzed in \S\ref{s:observation} for both, the MCMC and the MC DNN approaches. The model parameters are the systemic redshift $z^{\rm Zelda}$, the outflow expansion velocity \vexp, the neural hydrogen column density \nh, the dust optical depth \ta, the rest-frame injected equivalent width of the \lya\ line \ew\ and its rest frame intrinsic width \w.  

\renewcommand{\arraystretch}{1.0}

\begin{table}
\caption{ Name, systemic redshift, \lya\ luminosity and observed equivalent width for the observed galaxies. The column {\it Label} indicates the number that appears with the spectra in Fig.\ref{fig:examples_observed}, \ref{fig:comparison_NN_MC_lines_0} and \ref{fig:comparison_NN_MC_lines_1}.  }
\label{tab:real_mcmc_LUM_0}
\resizebox{3.4in}{!}{%
\begin{tabular}{ccccc}
Label & Name   & $z^{\rm Sys}$ &  $\rm \log$\Lya & $\rm \log$\ewl   \\
 & & & [$\rm erg/s$] & [\AA{}] \\ \hline
1 & SDSSJ0213+1259 & 0.219 & 41.7 & 0.84 \\
2 & SDSSJ1025+3622 & 0.1265 & 42.31 & 1.39 \\
3 & SDSSJ0150+1308 & 0.1467 & 41.37 & 0.59 \\
4 & SDSSJ0055-0021 & 0.1674 & 41.45 & 0.67 \\
5 & SDSSJ1112+5503 & 0.1316 & 41.88 & 1.02 \\
6 & SDSSJ1144+4012 & 0.127 & 40.71 & 0.32 \\
7 & SDSSJ1414+0540 & 0.0819 & 40.5 & 0.5 \\
8 & SDSSJ0808+3948 & 0.0912 & 42.01 & 1.17 \\
9 & SDSSJ1429+0643 & 0.1735 & 42.73 & 1.56 \\
10 & SDSSJ1416+1223 & 0.1232 & 41.2 & 0.24 \\
11 & SDSSJ1521+0759 & 0.0943 & 41.63 & 0.73 \\
12 & SDSSJ1428+1653 & 0.1817 & 42.46 & 1.29 \\
13 & SDSSJ1612+0817 & 0.1491 & 42.3 & 1.26 \\
14 & SDSSJ0926+4427 & 0.1807 & 42.78 & 1.57 \\
15 & GP0303–0759 & 0.1649 & 41.94 & 0.95 \\
16 & J0159+0751 & 0.0611 & 41.59 & 2.17 \\
17 & SDSSJ0938+5428 & 0.1021 & 41.35 & 0.55 \\
18 & SDSSJ0021+0052 & 0.0984 & 42.59 & 1.41 \\
19 & SDSSJ2103-0728 & 0.1369 & 42.21 & 1.25 \\
20 & HARO11 & 0.0206 & 41.26 & 0.93 \\
21 & GP0911+1831 & 0.2622 & 42.8 & 1.68 \\
22 & GP1219+1526 & 0.1956 & 43.19 & 2.12 \\
23 & GP1133+6514 & 0.2414 & 42.58 & 1.51 \\
24 & GP1054+5238 & 0.2526 & 42.54 & 1.15 \\
25 & GP1137+3524 & 0.1944 & 42.62 & 1.53 \\
26 & GP1244+0216 & 0.2394 & 42.56 & 1.69 \\
27 & GP1249+1234 & 0.2634 & 43.07 & 1.96 \\
28 & GP1424+4217 & 0.1848 & 42.93 & 1.88 \\
29 & J0925+1403 & 0.3012 & 42.83 & 1.82 \\
30 & J0820+5431 & 0.0386 & 40.47 & 1.81 \\
31 & J1205+4551 & 0.0654 & 41.65 & 2.41 \\
32 & J1355+4651 & 0.0278 & 40.95 & 2.15 \\
33 & J1242+4851 & 0.0623 & 41.44 & 2.28 \\
34 & J1152+3400 & 0.3419 & 43.0 & 1.82 \\
35 & J1503+3644 & 0.3557 & 42.86 & 1.75 \\
36 & J1333+6246 & 0.3181 & 42.75 & 1.88 \\
37 & J1442-0209 & 0.2937 & 43.16 & 1.94 \\
38 & J0901+2119 & 0.2993 & 42.48 & 2.16 \\
39 & J1154+2443 & 0.3689 & 42.87 & 2.01 \\
40 & J1011+1947 & 0.3322 & 42.64 & 2.17 \\
41 & J1243+4646 & 0.4317 & 43.08 & 1.89 \\
42 & J1256+4509 & 0.353 & 42.54 & 2.01 \\
43 & J1248+4259 & 0.3629 & 42.8 & 2.41 \\
44 & SDSSJ1457+2232 & 0.1486 & 41.24 & 1.04 \\
45 & SDSSJ0815+2156 & 0.141 & 42.34 & 1.79 \\
46 & GALEX1417+5228 & 0.2083 & 42.06 & 2.03 \\
47 & J0213+0056 & 0.0399 & 41.33 & 1.77 \\
48 & GALEX1001+0233 & 0.3824 & 42.36 & 2.09 \\
49 & GALEX1417+5305 & 0.2671 & 41.39 & 1.79 \\
\end{tabular}}
\end{table}

\begin{table}
\caption{ Name, systemic redshift, \lya\ luminosity and observed equivalent width for the observed galaxies. The column {\it Label} indicates the number that appears with the spectra in Fig.\ref{fig:examples_observed}, \ref{fig:comparison_NN_MC_lines_0} and \ref{fig:comparison_NN_MC_lines_1}.  }
\label{tab:real_mcmc_LUM_1}
\resizebox{3.4in}{!}{%
\begin{tabular}{ccccc}
Label & Name   & $z^{\rm Sys}$ &  $\rm \log$\Lya & $\rm \log$\ewl   \\
 & & & [$\rm erg/s$] & [\AA{}] \\ \hline
50 & J0240-0828 & 0.0822 & 42.32 & 2.27 \\
51 & GALEX1423+5246 & 0.3431 & 41.68 & 1.76 \\
52 & J0808+1728 & 0.0442 & 41.4 & 1.44 \\
53 & GALEX1418+5217 & 0.2398 & 40.81 & 1.25 \\
54 & GALEX1419+5315 & 0.2637 & 41.25 & 1.54 \\
55 & GALEX1418+5307 & 0.2034 & 41.63 & 1.58 \\
56 & GALEX1418+5218 & 0.2388 & 41.21 & 1.82 \\
57 & GALEX1420+5243 & 0.247 & 41.07 & 1.07 \\
58 & GALEX1434+3532 & 0.1946 & 41.27 & 1.51 \\
59 & J0851+5840 & 0.0919 & 41.64 & 1.7 \\
60 & GALEX1436+3456 & 0.2684 & 41.72 & 1.65 \\
61 & GALEX1437+3445 & 0.3237 & 41.88 & 1.44 \\
62 & J1200+2719 & 0.0819 & 42.43 & 1.98 \\
63 & KISSR1084 & 0.0321 & 39.46 & 0.52 \\
64 & J1226+0415 & 0.0942 & 42.02 & 1.88 \\
65 & KISSR1578 & 0.028 & 41.16 & 1.0 \\
66 & KISSR1567 & 0.0426 & 39.05 & 0.93 \\
67 & J1311-0038 & 0.0811 & 42.03 & 1.86 \\
68 & J1509+3731 & 0.0325 & 41.0 & 1.57 \\
69 & J1608+3528 & 0.0327 & 41.17 & 2.39 \\
70 & J1735+5703 & 0.0472 & 42.17 & 1.88 \\
71 & J2302+0049 & 0.0331 & 41.3 & 1.85 \\
72 & GP0822+2241 & 0.2162 & 42.36 & 1.66 \\
73 & GP0751+1638 & 0.2647 & 41.59 & 1.2 \\
74 & GP0917+3152 & 0.3004 & 42.59 & 1.26 \\
75 & GP1009+2916 & 0.2219 & 42.34 & 1.79 \\
76 & GP0927+1740 & 0.2883 & 41.62 & 1.07 \\
77 & GP1018+4106 & 0.237 & 41.89 & 1.48 \\
78 & GP1122+6154 & 0.2046 & 42.26 & 1.73 \\
79 & GP1339+1516 & 0.192 & 41.83 & 1.42 \\
80 & GP1440+4619 & 0.3008 & 42.83 & 1.48 \\
81 & GP1514+3852 & 0.3326 & 42.74 & 1.49 \\
82 & GP1454+4528 & 0.2685 & 42.25 & 1.45 \\
83 & GP1559+0841 & 0.297 & 42.57 & 1.77 \\
84 & GP2237+1336 & 0.2935 & 42.26 & 1.13 \\
85 & KISSR242 & 0.0378 & 41.37 & 1.34 \\
86 & LARS04 & 0.0325 & 39.96 & 0.66 \\
87 & LARS02 & 0.0298 & 41.03 & 1.61 \\
88 & LARS03 & 0.0307 & 39.9 & 0.77 \\
89 & LARS08 & 0.0382 & 40.14 & 0.53 \\
90 & LARS11 & 0.0844 & 40.7 & 0.76 \\
91 & GALEX0330-2816 & 0.2813 & 41.65 & 1.2 \\
92 & LARS05 & 0.0338 & 41.28 & 1.19 \\
93 & GALEX0333-2821 & 0.2471 & 41.51 & 1.34 \\
94 & GALEX0332-2801 & 0.2155 & 41.5 & 1.52 \\
95 & GALEX0331-2814 & 0.2803 & 42.02 & 2.03 \\
96 & GALEX0332-2811 & 0.2043 & 42.07 & 1.56 \\
97 & GALEX1000+0157 & 0.2647 & 42.11 & 1.38 \\
\end{tabular}}
\end{table}

\renewcommand{\arraystretch}{1.5}
\begin{landscape}
\begin{table}
\caption{ Parameters associated with the line profiles. The column {\it Label} indicates the number that appears with the spectra in Fig.\ref{fig:examples_observed}, \ref{fig:comparison_NN_MC_lines_0} and \ref{fig:comparison_NN_MC_lines_1}.  }
\label{tab:real_mcmc_0}
\resizebox{9.5in}{!}{%
\begin{tabular}{ccccccccccccc}
& \multicolumn{6}{c}{MCMC} & \multicolumn{6}{c}{DNN}  \\ \cmidrule(r){2-7} \cmidrule(l){8-13}
Label &  $z^{\rm Zelda}$ & $\log$\vexp & $\log$\nh & $\log$\ta & $\log$\ew & $\rm \log$\w & $z^{\rm Zelda}$ & $\log$\vexp & $\log$\nh & $\log$\ta & $\log$\ew & $\rm \log$\w   \\
 & & [$\rm km/s$] & [$\rm cm^{-2}$] & & [\AA{}] & [\AA{}] & & [$\rm km/s$] & [$\rm cm^{-2}$] & & [\AA{}] & [\AA{}] \\ \hline
1 & $0.21867^{+1.6e-04}_{-4.9e-07}$ & $1.34^{+4.4e-01}_{-1.4e-03}$ & $19.21^{+1.5e-03}_{-7.1e-01}$ & $-3.86^{+1.2e+00}_{-1.4e-01}$ & $1.13^{+1.3e-03}_{-9.3e-03}$ & $0.04^{+6.8e-04}_{-3.8e-04}$ & $0.21872^{+5.0e-05}_{-3.1e-05}$ & $1.52^{+1.0e-01}_{-9.8e-02}$ & $18.68^{+2.0e-01}_{-2.1e-01}$ & $-3.56^{+3.1e-01}_{-1.6e-01}$ & $1.16^{+5.0e-02}_{-6.5e-02}$ & $-0.05^{+5.0e-02}_{-4.1e-02}$ \\
2 & $0.12659^{+2.6e-06}_{-6.5e-08}$ & $2.1^{+3.3e-04}_{-8.2e-05}$ & $19.35^{+1.5e-04}_{-6.7e-03}$ & $-0.54^{+1.7e-03}_{-6.9e-03}$ & $1.58^{+4.5e-05}_{-5.8e-03}$ & $-0.19^{+2.9e-04}_{-1.0e-03}$ & $0.12671^{+7.0e-05}_{-7.6e-05}$ & $2.2^{+2.7e-02}_{-4.8e-02}$ & $18.85^{+1.5e-01}_{-1.5e-01}$ & $-3.02^{+2.6e-01}_{-2.8e-01}$ & $1.36^{+1.9e-02}_{-1.9e-02}$ & $-0.27^{+2.8e-02}_{-2.9e-02}$ \\
3 & $0.147^{+6.4e-07}_{-3.5e-08}$ & $2.18^{+7.4e-08}_{-5.4e-05}$ & $19.66^{+5.4e-05}_{-1.0e-03}$ & $-0.06^{+4.6e-05}_{-2.4e-03}$ & $1.15^{+3.9e-05}_{-1.3e-03}$ & $-0.7^{+1.9e-07}_{-4.8e-05}$ & $0.14714^{+1.6e-04}_{-4.0e-04}$ & $2.2^{+9.9e-02}_{-1.7e-01}$ & $19.65^{+3.6e-01}_{-1.9e-01}$ & $-0.5^{+3.7e-01}_{-4.3e-01}$ & $1.01^{+2.9e-01}_{-2.6e-01}$ & $-0.72^{+2.1e-01}_{-1.6e-01}$ \\
4 & $0.16825^{+3.2e-04}_{-5.3e-04}$ & $2.25^{+1.3e-01}_{-1.8e-01}$ & $19.54^{+4.7e-01}_{-2.6e-01}$ & $-2.5^{+1.5e+00}_{-5.8e-03}$ & $0.8^{+2.5e-01}_{-3.3e-03}$ & $-0.4^{+7.1e-01}_{-9.3e-02}$ & $0.16819^{+4.1e-04}_{-4.8e-04}$ & $2.4^{+1.4e-01}_{-2.3e-01}$ & $19.64^{+2.8e-01}_{-3.5e-01}$ & $-2.62^{+1.1e+00}_{-8.7e-01}$ & $1.08^{+1.6e-01}_{-4.0e-01}$ & $-0.08^{+2.6e-01}_{-2.4e-01}$ \\
5 & $0.13193^{+4.8e-06}_{-1.1e-06}$ & $2.34^{+2.0e-03}_{-9.0e-04}$ & $19.56^{+3.4e-04}_{-7.7e-03}$ & $-0.2^{+1.5e-03}_{-5.9e-02}$ & $1.36^{+3.4e-04}_{-7.0e-03}$ & $-0.69^{+3.4e-02}_{-2.2e-04}$ & $0.1317^{+1.6e-04}_{-1.7e-04}$ & $2.23^{+1.0e-01}_{-9.1e-02}$ & $19.67^{+1.6e-01}_{-1.3e-01}$ & $-0.89^{+3.0e-01}_{-3.5e-01}$ & $1.09^{+1.1e-01}_{-7.5e-02}$ & $-0.53^{+7.7e-02}_{-1.3e-01}$ \\
6 & $0.12701^{+5.7e-05}_{-1.7e-04}$ & $1.75^{+9.7e-02}_{-1.0e-01}$ & $20.3^{+2.4e-01}_{-5.6e-02}$ & $-0.09^{+4.5e-02}_{-3.2e-01}$ & $1.74^{+2.8e-01}_{-2.0e-01}$ & $-0.65^{+1.4e-01}_{-9.7e-01}$ & $0.12642^{+4.8e-04}_{-4.1e-04}$ & $1.77^{+2.0e-01}_{-2.3e-01}$ & $20.58^{+2.3e-01}_{-5.3e-01}$ & $-0.36^{+2.9e-01}_{-3.0e-01}$ & $1.45^{+3.3e-01}_{-3.9e-01}$ & $-0.57^{+3.8e-01}_{-1.6e-01}$ \\
7 & $0.08187^{+6.0e-04}_{-7.2e-03}$ & $1.68^{+5.1e-01}_{-4.0e-01}$ & $19.16^{+7.8e-01}_{-4.1e-01}$ & $-1.41^{+1.1e+00}_{-1.1e+00}$ & $0.83^{+1.8e-01}_{-2.9e-01}$ & $0.15^{+1.2e-01}_{-1.2e+00}$ & $0.08204^{+3.3e-04}_{-9.0e-04}$ & $1.79^{+3.7e-01}_{-4.3e-01}$ & $19.6^{+7.7e-01}_{-6.5e-01}$ & $-1.3^{+9.3e-01}_{-1.6e+00}$ & $0.94^{+4.9e-01}_{-3.6e-01}$ & $0.24^{+2.4e-01}_{-4.2e-01}$ \\
8 & $0.09138^{+4.8e-06}_{-2.0e-07}$ & $2.56^{+5.3e-02}_{-2.0e-04}$ & $18.45^{+1.6e-03}_{-4.7e-01}$ & $-2.0^{+1.6e+00}_{-3.7e-04}$ & $1.28^{+5.2e-05}_{-7.2e-03}$ & $-0.79^{+3.3e-04}_{-2.2e-02}$ & $0.09137^{+9.5e-06}_{-1.8e-05}$ & $2.59^{+6.9e-02}_{-4.4e-02}$ & $18.1^{+9.6e-02}_{-8.7e-02}$ & $-1.9^{+1.7e-01}_{-1.8e-01}$ & $1.25^{+1.6e-02}_{-1.8e-02}$ & $-0.79^{+1.7e-02}_{-1.4e-02}$ \\
9 & $0.17389^{+5.0e-08}_{-3.6e-10}$ & $1.95^{+1.2e-04}_{-1.9e-06}$ & $19.25^{+2.2e-06}_{-1.1e-09}$ & $-0.69^{+6.8e-04}_{-5.9e-06}$ & $1.73^{+1.8e-04}_{-3.7e-06}$ & $0.3^{+2.1e-06}_{-4.7e-05}$ & $0.17398^{+3.6e-05}_{-5.6e-05}$ & $2.06^{+4.8e-02}_{-1.6e-02}$ & $19.08^{+6.9e-02}_{-8.9e-02}$ & $-0.81^{+4.3e-01}_{-2.9e-01}$ & $1.78^{+5.0e-02}_{-3.2e-02}$ & $0.3^{+4.2e-02}_{-1.6e-02}$ \\
10 & $0.12327^{+1.2e-04}_{-1.3e-05}$ & $0.99^{+5.8e-01}_{-1.1e-01}$ & $19.9^{+6.5e-02}_{-4.4e-01}$ & $-0.67^{+3.5e-01}_{-1.2e-01}$ & $1.01^{+3.8e-02}_{-1.2e-01}$ & $0.07^{+7.3e-02}_{-2.1e-02}$ & $0.12325^{+7.6e-05}_{-1.6e-04}$ & $1.21^{+7.9e-02}_{-6.9e-02}$ & $19.51^{+2.0e-01}_{-2.4e-01}$ & $-0.38^{+1.4e-01}_{-3.1e-01}$ & $1.02^{+2.3e-01}_{-2.3e-01}$ & $0.06^{+1.3e-01}_{-1.6e-01}$ \\
11 & $0.09423^{+1.4e-04}_{-8.4e-08}$ & $1.99^{+5.3e-02}_{-9.9e-04}$ & $19.56^{+1.4e-03}_{-1.6e-01}$ & $-0.2^{+2.5e-04}_{-2.6e-02}$ & $1.21^{+5.7e-05}_{-8.4e-02}$ & $-0.48^{+3.4e-04}_{-2.6e-01}$ & $0.09425^{+9.9e-05}_{-8.8e-05}$ & $2.17^{+5.0e-02}_{-9.6e-02}$ & $19.46^{+1.5e-01}_{-1.4e-01}$ & $-0.14^{+6.2e-02}_{-2.9e-01}$ & $1.27^{+1.1e-01}_{-1.6e-01}$ & $-0.74^{+9.0e-02}_{-8.3e-02}$ \\
12 & $0.18209^{+5.5e-05}_{-2.9e-04}$ & $2.25^{+7.2e-02}_{-1.5e-01}$ & $18.25^{+6.4e-01}_{-1.6e-03}$ & $-2.5^{+1.7e+00}_{-2.7e-03}$ & $1.26^{+1.1e-01}_{-2.9e-04}$ & $-0.81^{+6.1e-02}_{-1.9e-03}$ & $0.18149^{+4.3e-04}_{-1.3e-04}$ & $2.12^{+1.7e-01}_{-2.4e-01}$ & $19.49^{+1.7e-01}_{-6.4e-01}$ & $-1.51^{+3.4e-01}_{-1.5e+00}$ & $1.41^{+1.4e-01}_{-1.4e-01}$ & $-0.63^{+6.1e-02}_{-1.5e-01}$ \\
13 & $0.14936^{+2.4e-08}_{-1.2e-09}$ & $2.24^{+7.3e-06}_{-2.5e-06}$ & $19.43^{+6.6e-07}_{-7.7e-05}$ & $-0.0^{+2.4e-09}_{-1.4e-06}$ & $1.68^{+1.6e-06}_{-6.2e-05}$ & $-1.05^{+6.5e-05}_{-2.6e-06}$ & $0.14922^{+1.1e-04}_{-4.4e-05}$ & $2.27^{+4.9e-02}_{-3.4e-02}$ & $19.76^{+7.6e-02}_{-1.9e-01}$ & $-0.03^{+1.8e-02}_{-1.9e-02}$ & $1.69^{+8.3e-02}_{-1.2e-01}$ & $-0.83^{+2.6e-02}_{-6.7e-02}$ \\
14 & $0.18093^{+7.8e-09}_{-4.1e-07}$ & $2.16^{+2.2e-05}_{-4.9e-04}$ & $18.94^{+4.1e-04}_{-6.9e-05}$ & $-1.5^{+1.5e-07}_{-4.9e-04}$ & $1.62^{+6.5e-06}_{-7.3e-04}$ & $0.06^{+2.0e-05}_{-1.9e-04}$ & $0.18117^{+3.4e-05}_{-1.4e-04}$ & $2.36^{+3.9e-02}_{-8.3e-02}$ & $18.53^{+2.9e-01}_{-9.6e-02}$ & $-3.14^{+2.0e+00}_{-2.0e-01}$ & $1.6^{+4.7e-02}_{-1.2e-02}$ & $-0.26^{+4.0e-02}_{-2.7e-02}$ \\
15 & $0.1648^{+1.4e-07}_{-1.3e-09}$ & $1.81^{+1.7e-04}_{-2.6e-06}$ & $19.59^{+1.6e-06}_{-1.8e-04}$ & $-2.5^{+6.0e-06}_{-9.7e-10}$ & $0.95^{+6.2e-05}_{-6.8e-07}$ & $-0.32^{+5.2e-06}_{-1.0e-04}$ & $0.16462^{+1.5e-04}_{-1.6e-04}$ & $1.97^{+1.3e-01}_{-1.3e-01}$ & $19.71^{+2.1e-01}_{-2.6e-01}$ & $-2.86^{+5.7e-01}_{-3.9e-01}$ & $0.94^{+3.7e-02}_{-3.2e-02}$ & $-0.41^{+8.1e-02}_{-1.0e-01}$ \\
16 & $0.06097^{+1.6e-06}_{-2.3e-07}$ & $1.78^{+9.9e-05}_{-1.3e-03}$ & $19.25^{+4.5e-04}_{-4.5e-03}$ & $-0.0^{+1.0e-04}_{-2.0e-03}$ & $2.45^{+7.6e-03}_{-1.5e-03}$ & $-0.83^{+5.7e-04}_{-3.1e-03}$ & $0.06102^{+1.7e-05}_{-3.2e-06}$ & $1.62^{+1.0e-01}_{-8.3e-03}$ & $19.09^{+2.0e-02}_{-8.4e-02}$ & $-0.84^{+5.1e-01}_{-7.2e-02}$ & $2.09^{+7.7e-02}_{-5.7e-03}$ & $-0.91^{+2.2e-02}_{-8.0e-03}$ \\
17 & $0.10287^{+2.6e-05}_{-5.2e-07}$ & $2.07^{+2.3e-02}_{-2.4e-04}$ & $18.98^{+3.1e-03}_{-5.3e-02}$ & $-1.0^{+2.9e-04}_{-2.5e+00}$ & $0.96^{+4.1e-03}_{-2.8e-02}$ & $0.52^{+8.2e-02}_{-3.7e-03}$ & $0.10265^{+3.0e-04}_{-2.3e-04}$ & $1.7^{+3.5e-01}_{-4.3e-01}$ & $19.47^{+5.4e-01}_{-6.2e-01}$ & $-1.84^{+1.0e+00}_{-1.2e+00}$ & $0.63^{+1.7e-01}_{-2.1e-01}$ & $0.27^{+1.5e-01}_{-1.4e-01}$ \\
18 & $0.09889^{+2.6e-07}_{-2.4e-06}$ & $2.32^{+6.0e-05}_{-5.1e-04}$ & $18.89^{+2.6e-03}_{-1.4e-04}$ & $-2.5^{+3.8e-04}_{-1.5e+00}$ & $1.42^{+5.2e-05}_{-2.2e-03}$ & $-0.8^{+1.3e-03}_{-5.2e-04}$ & $0.09865^{+6.9e-05}_{-7.9e-05}$ & $2.34^{+5.5e-02}_{-7.2e-02}$ & $19.04^{+1.0e-01}_{-1.3e-01}$ & $-1.84^{+5.4e-01}_{-2.5e-01}$ & $1.48^{+4.8e-02}_{-1.8e-02}$ & $-0.51^{+2.6e-02}_{-2.4e-02}$ \\
19 & $0.1372^{+7.9e-09}_{-1.2e-07}$ & $2.4^{+1.2e-05}_{-1.3e-04}$ & $19.0^{+2.8e-08}_{-1.1e-05}$ & $-1.15^{+7.9e-04}_{-2.9e-05}$ & $1.46^{+1.3e-04}_{-5.2e-06}$ & $-0.66^{+1.5e-04}_{-1.0e-05}$ & $0.13701^{+1.0e-04}_{-1.9e-04}$ & $2.36^{+4.3e-02}_{-7.9e-02}$ & $19.08^{+1.9e-01}_{-8.9e-02}$ & $-0.25^{+1.1e-01}_{-1.1e+00}$ & $1.57^{+2.9e-02}_{-5.0e-02}$ & $-0.21^{+1.6e-01}_{-1.8e-01}$ \\
20 & $0.02089^{+1.2e-08}_{-2.0e-07}$ & $2.34^{+1.1e-04}_{-1.4e-05}$ & $18.98^{+7.2e-04}_{-1.9e-05}$ & $-0.5^{+3.1e-05}_{-5.1e-08}$ & $1.13^{+1.2e-04}_{-6.3e-06}$ & $-0.85^{+5.6e-06}_{-3.1e-04}$ & $0.02088^{+2.2e-05}_{-2.7e-05}$ & $2.44^{+4.2e-02}_{-3.1e-02}$ & $19.29^{+8.5e-02}_{-9.5e-02}$ & $-1.17^{+5.7e-01}_{-4.6e-01}$ & $1.08^{+6.0e-02}_{-2.9e-02}$ & $-0.83^{+5.4e-02}_{-2.1e-02}$ \\
21 & $0.26218^{+4.1e-07}_{-2.2e-06}$ & $1.9^{+2.1e-04}_{-2.7e-02}$ & $18.79^{+1.7e-03}_{-4.3e-02}$ & $-0.5^{+2.2e-03}_{-2.5e+00}$ & $1.87^{+1.1e-03}_{-1.1e-01}$ & $-0.2^{+1.1e-01}_{-3.2e-04}$ & $0.26216^{+2.1e-05}_{-2.1e-05}$ & $1.96^{+1.8e-02}_{-1.5e-02}$ & $18.77^{+3.8e-02}_{-3.9e-02}$ & $-1.62^{+9.0e-02}_{-9.0e-02}$ & $1.73^{+1.3e-02}_{-1.2e-02}$ & $-0.52^{+1.6e-02}_{-1.4e-02}$ \\
22 & $0.19599^{+3.7e-09}_{-1.1e-08}$ & $1.89^{+9.5e-06}_{-8.6e-06}$ & $18.02^{+1.2e-04}_{-2.2e-05}$ & $-1.5^{+1.0e-07}_{-6.4e-05}$ & $2.17^{+3.9e-05}_{-8.5e-04}$ & $0.05^{+1.5e-05}_{-8.5e-06}$ & $0.19592^{+4.8e-05}_{-4.9e-05}$ & $1.81^{+9.2e-03}_{-2.2e-02}$ & $17.94^{+1.1e-01}_{-6.2e-02}$ & $-1.01^{+1.1e-01}_{-9.2e-02}$ & $2.14^{+6.2e-03}_{-6.2e-03}$ & $-0.0^{+2.5e-02}_{-5.8e-02}$ \\
23 & $0.24198^{+1.4e-08}_{-4.9e-08}$ & $1.73^{+3.8e-04}_{-3.2e-06}$ & $18.44^{+3.0e-04}_{-5.4e-06}$ & $-0.07^{+6.9e-04}_{-9.7e-06}$ & $1.65^{+2.6e-05}_{-4.5e-05}$ & $-0.01^{+2.5e-06}_{-2.2e-04}$ & $0.242^{+2.6e-05}_{-2.9e-05}$ & $1.82^{+9.4e-02}_{-7.6e-02}$ & $18.06^{+7.3e-02}_{-7.2e-02}$ & $-0.08^{+8.3e-02}_{-1.1e-01}$ & $1.65^{+2.1e-02}_{-2.3e-02}$ & $-0.08^{+1.7e-02}_{-1.8e-02}$ \\
24 & $0.25296^{+2.7e-07}_{-5.1e-08}$ & $2.0^{+2.7e-06}_{-1.3e-04}$ & $19.25^{+3.4e-04}_{-1.0e-05}$ & $-0.0^{+2.4e-05}_{-7.9e-04}$ & $1.53^{+4.8e-05}_{-5.2e-04}$ & $-0.27^{+2.5e-04}_{-1.4e-03}$ & $0.25293^{+6.6e-05}_{-8.9e-05}$ & $2.07^{+1.0e-01}_{-8.4e-02}$ & $18.98^{+1.6e-01}_{-1.2e-01}$ & $-1.06^{+4.3e-01}_{-4.5e-01}$ & $1.2^{+6.0e-02}_{-5.8e-02}$ & $-0.4^{+4.8e-02}_{-5.8e-02}$ \\
25 & $0.19454^{+6.5e-07}_{-1.4e-08}$ & $2.16^{+2.9e-04}_{-9.6e-05}$ & $19.25^{+1.0e-06}_{-8.6e-04}$ & $-0.17^{+1.2e-03}_{-7.8e-03}$ & $1.8^{+6.0e-05}_{-5.5e-03}$ & $-0.07^{+1.2e-05}_{-5.5e-04}$ & $0.19432^{+3.9e-04}_{-1.2e-04}$ & $2.23^{+6.9e-02}_{-1.0e-01}$ & $19.45^{+1.8e-01}_{-5.9e-01}$ & $-1.67^{+3.0e-01}_{-7.4e-01}$ & $1.62^{+4.6e-02}_{-9.3e-02}$ & $-0.35^{+5.6e-02}_{-7.5e-02}$ \\
26 & $0.23979^{+1.8e-08}_{-8.8e-10}$ & $2.02^{+5.6e-05}_{-6.0e-06}$ & $19.25^{+2.3e-08}_{-9.7e-06}$ & $-0.12^{+7.3e-05}_{-9.4e-06}$ & $1.91^{+1.6e-05}_{-1.6e-06}$ & $0.19^{+1.5e-06}_{-8.2e-06}$ & $0.2399^{+6.0e-05}_{-9.1e-05}$ & $2.1^{+3.0e-02}_{-4.6e-02}$ & $19.01^{+1.0e-01}_{-8.7e-02}$ & $0.09^{+1.6e-01}_{-3.8e-01}$ & $1.87^{+4.3e-02}_{-3.1e-02}$ & $0.28^{+3.0e-02}_{-2.7e-02}$ \\
27 & $0.26279^{+5.7e-04}_{-1.7e-06}$ & $2.13^{+2.2e-01}_{-2.7e-03}$ & $19.67^{+2.2e-03}_{-5.6e-01}$ & $-3.48^{+1.1e+00}_{-2.2e-02}$ & $1.97^{+1.4e-02}_{-1.4e-02}$ & $-0.34^{+6.5e-02}_{-1.1e-01}$ & $0.26349^{+1.0e-04}_{-6.3e-04}$ & $2.41^{+3.5e-02}_{-1.5e-01}$ & $18.54^{+9.7e-01}_{-1.4e-01}$ & $-3.11^{+2.3e-01}_{-2.8e-01}$ & $2.03^{+2.4e-02}_{-3.2e-02}$ & $-0.34^{+1.5e-01}_{-4.6e-02}$ \\
28 & $0.18518^{+6.0e-06}_{-1.1e-03}$ & $1.4^{+2.0e-02}_{-1.1e-01}$ & $19.2^{+1.3e-03}_{-1.4e+00}$ & $-1.14^{+5.3e-03}_{-1.4e+00}$ & $2.0^{+4.4e-04}_{-3.0e-01}$ & $0.2^{+2.1e-03}_{-1.7e-01}$ & $0.18513^{+1.2e-04}_{-1.8e-04}$ & $1.41^{+4.5e-01}_{-1.8e-01}$ & $19.25^{+3.7e-01}_{-6.4e-01}$ & $-2.09^{+1.2e+00}_{-1.1e+00}$ & $2.09^{+1.7e-01}_{-5.8e-01}$ & $0.06^{+1.8e-01}_{-3.2e-01}$ \\
29 & $0.30134^{+8.7e-10}_{-1.9e-08}$ & $1.48^{+1.7e-08}_{-2.0e-06}$ & $19.25^{+8.0e-09}_{-6.5e-07}$ & $-0.79^{+1.7e-04}_{-1.5e-05}$ & $2.05^{+2.6e-04}_{-8.8e-06}$ & $-0.18^{+3.4e-06}_{-8.7e-05}$ & $0.30137^{+2.3e-05}_{-3.9e-05}$ & $1.49^{+7.4e-02}_{-4.2e-02}$ & $19.19^{+1.0e-01}_{-1.2e-01}$ & $-1.01^{+6.9e-01}_{-6.3e-01}$ & $2.09^{+4.6e-02}_{-6.7e-02}$ & $-0.15^{+7.6e-02}_{-1.2e-01}$ \\
30 & $0.03847^{+7.6e-08}_{-1.2e-07}$ & $0.91^{+6.1e-05}_{-5.5e-05}$ & $19.36^{+1.4e-04}_{-5.0e-05}$ & $-3.0^{+6.6e-07}_{-2.5e-05}$ & $1.75^{+6.9e-04}_{-2.0e-05}$ & $-0.4^{+4.2e-05}_{-1.0e-04}$ & $0.03855^{+2.5e-05}_{-2.9e-05}$ & $1.14^{+3.6e-02}_{-2.7e-02}$ & $18.72^{+1.5e-01}_{-1.6e-01}$ & $-3.47^{+3.5e-01}_{-2.5e-01}$ & $1.96^{+2.5e-02}_{-2.7e-02}$ & $-0.47^{+2.9e-02}_{-2.9e-02}$ \\
31 & $0.06533^{+3.7e-09}_{-1.5e-07}$ & $1.11^{+6.5e-05}_{-1.5e-04}$ & $19.0^{+9.2e-05}_{-4.3e-07}$ & $-3.5^{+3.3e-03}_{-5.4e-06}$ & $2.1^{+2.1e-04}_{-1.8e-04}$ & $-0.19^{+3.0e-05}_{-4.8e-04}$ & $0.06531^{+9.0e-06}_{-4.3e-06}$ & $1.06^{+6.7e-03}_{-6.8e-03}$ & $19.14^{+2.0e-02}_{-1.3e-02}$ & $-3.33^{+9.3e-02}_{-7.6e-01}$ & $1.75^{+9.2e-03}_{-6.7e-02}$ & $-0.42^{+1.2e-02}_{-1.7e-02}$ \\
32 & $0.02813^{+4.2e-09}_{-3.6e-07}$ & $1.69^{+7.7e-05}_{-1.9e-03}$ & $18.75^{+1.7e-07}_{-1.8e-05}$ & $-0.0^{+8.5e-07}_{-1.6e-04}$ & $2.52^{+3.5e-04}_{-2.1e-04}$ & $-0.57^{+3.0e-05}_{-8.2e-04}$ & $0.02803^{+3.8e-06}_{-4.0e-06}$ & $1.45^{+8.5e-03}_{-9.0e-03}$ & $19.22^{+1.1e-02}_{-1.1e-02}$ & $-0.5^{+2.9e-02}_{-6.0e-02}$ & $2.18^{+1.8e-03}_{-2.3e-03}$ & $-0.54^{+4.0e-03}_{-4.2e-03}$ \\
33 & $0.0622^{+1.3e-06}_{-2.8e-05}$ & $1.25^{+8.7e-04}_{-1.5e-01}$ & $19.0^{+2.1e-01}_{-1.4e-04}$ & $-0.39^{+2.1e-03}_{-1.1e-01}$ & $2.47^{+6.4e-02}_{-2.3e-03}$ & $-0.65^{+1.2e-03}_{-8.0e-02}$ & $0.06216^{+3.8e-06}_{-4.9e-05}$ & $1.05^{+1.4e-02}_{-3.0e-02}$ & $19.16^{+7.4e-02}_{-1.3e-02}$ & $-1.87^{+4.0e-02}_{-4.4e-02}$ & $2.11^{+1.1e-02}_{-5.9e-03}$ & $-0.54^{+2.8e-02}_{-9.1e-02}$ \\
\end{tabular}}
\end{table}
\end{landscape}

\renewcommand{\arraystretch}{1.5}
\begin{landscape}
\begin{table}
\caption{ Parameters associated with the line profiles. The column {\it Label} indicates the number that appears with the spectra in Fig.\ref{fig:examples_observed}, \ref{fig:comparison_NN_MC_lines_0} and \ref{fig:comparison_NN_MC_lines_1}.  }
\label{tab:real_mcmc_1}
\resizebox{9.5in}{!}{%
\begin{tabular}{ccccccccccccc}
& \multicolumn{6}{c}{MCMC} & \multicolumn{6}{c}{DNN}  \\ \cmidrule(r){2-7} \cmidrule(l){8-13}
Label &  $z^{\rm Zelda}$ & $\log$\vexp & $\log$\nh & $\log$\ta & $\log$\ew & $\rm \log$\w & $z^{\rm Zelda}$ & $\log$\vexp & $\log$\nh & $\log$\ta & $\log$\ew & $\rm \log$\w   \\
 & & [$\rm km/s$] & [$\rm cm^{-2}$] & & [\AA{}] & [\AA{}] & & [$\rm km/s$] & [$\rm cm^{-2}$] & & [\AA{}] & [\AA{}] \\ \hline
34 & $0.34225^{+3.4e-10}_{-3.8e-11}$ & $1.71^{+5.8e-07}_{-1.6e-05}$ & $18.46^{+4.2e-07}_{-4.7e-05}$ & $-0.25^{+2.2e-05}_{-2.2e-07}$ & $1.93^{+2.6e-06}_{-2.4e-06}$ & $0.15^{+1.4e-07}_{-1.2e-05}$ & $0.34229^{+1.6e-05}_{-1.2e-05}$ & $1.77^{+5.0e-02}_{-4.2e-02}$ & $18.01^{+6.1e-02}_{-7.3e-02}$ & $-0.03^{+5.5e-02}_{-6.1e-02}$ & $2.04^{+3.0e-02}_{-3.5e-02}$ & $0.07^{+3.0e-02}_{-2.7e-02}$ \\
35 & $0.35558^{+2.4e-08}_{-3.9e-07}$ & $2.0^{+5.1e-06}_{-2.9e-04}$ & $19.25^{+1.9e-05}_{-1.8e-07}$ & $-0.0^{+3.0e-06}_{-1.4e-04}$ & $2.26^{+2.2e-04}_{-2.5e-03}$ & $-0.33^{+7.5e-05}_{-2.3e-04}$ & $0.35546^{+4.3e-05}_{-7.3e-05}$ & $1.89^{+3.4e-02}_{-3.6e-02}$ & $19.24^{+1.0e-01}_{-6.4e-02}$ & $-0.47^{+1.5e-01}_{-2.3e-01}$ & $2.05^{+5.6e-02}_{-6.0e-02}$ & $-0.25^{+2.2e-02}_{-2.5e-02}$ \\
36 & $0.31828^{+1.1e-06}_{-3.3e-06}$ & $2.32^{+1.1e-02}_{-4.2e-03}$ & $18.8^{+5.6e-02}_{-1.4e-02}$ & $-3.97^{+2.5e+00}_{-2.9e-02}$ & $1.84^{+2.3e-02}_{-6.1e-03}$ & $-0.87^{+1.5e-03}_{-1.2e-03}$ & $0.31817^{+8.5e-05}_{-1.1e-04}$ & $2.36^{+4.1e-02}_{-5.2e-02}$ & $18.81^{+1.7e-01}_{-1.6e-01}$ & $-1.95^{+3.9e-01}_{-3.8e-01}$ & $1.82^{+2.5e-02}_{-2.6e-02}$ & $-0.66^{+2.8e-02}_{-2.2e-02}$ \\
37 & $0.29356^{+4.7e-08}_{-2.4e-10}$ & $2.12^{+1.8e-07}_{-2.0e-05}$ & $19.11^{+8.3e-07}_{-2.0e-04}$ & $-0.72^{+1.0e-06}_{-4.0e-04}$ & $2.03^{+8.6e-07}_{-7.7e-05}$ & $-0.18^{+3.8e-07}_{-2.8e-06}$ & $0.29353^{+5.2e-05}_{-4.4e-05}$ & $2.18^{+3.6e-02}_{-3.1e-02}$ & $18.9^{+8.6e-02}_{-9.1e-02}$ & $-2.04^{+3.1e-01}_{-2.1e-01}$ & $2.01^{+1.3e-02}_{-1.4e-02}$ & $-0.28^{+1.7e-02}_{-1.9e-02}$ \\
38 & $0.29948^{+4.9e-08}_{-1.7e-08}$ & $1.87^{+2.1e-05}_{-2.0e-04}$ & $18.75^{+8.7e-08}_{-1.6e-05}$ & $-0.5^{+1.5e-05}_{-2.8e-03}$ & $2.33^{+2.3e-04}_{-8.1e-05}$ & $-0.18^{+1.7e-05}_{-8.1e-05}$ & $0.29953^{+9.2e-06}_{-1.1e-05}$ & $2.0^{+2.1e-02}_{-2.2e-02}$ & $18.54^{+6.6e-02}_{-5.6e-02}$ & $-1.22^{+1.4e-01}_{-1.4e-01}$ & $2.12^{+2.0e-02}_{-2.0e-02}$ & $-0.03^{+2.0e-02}_{-2.1e-02}$ \\
39 & $0.36933^{+4.7e-11}_{-5.2e-09}$ & $1.74^{+1.1e-05}_{-3.0e-07}$ & $17.32^{+7.2e-05}_{-2.1e-06}$ & $-1.0^{+6.2e-06}_{-3.7e-09}$ & $2.1^{+2.6e-06}_{-2.4e-05}$ & $-0.24^{+2.8e-07}_{-2.0e-05}$ & $0.36922^{+6.5e-05}_{-6.9e-05}$ & $1.63^{+3.3e-02}_{-4.9e-02}$ & $17.53^{+1.4e-01}_{-9.5e-02}$ & $-1.01^{+2.7e-01}_{-2.9e-01}$ & $2.08^{+2.1e-02}_{-3.5e-02}$ & $-0.28^{+4.7e-02}_{-1.6e-02}$ \\
40 & $0.33214^{+3.7e-08}_{-1.1e-09}$ & $1.27^{+1.4e-04}_{-4.2e-06}$ & $19.25^{+5.2e-10}_{-7.5e-07}$ & $-0.85^{+3.9e-06}_{-6.9e-04}$ & $2.29^{+7.6e-06}_{-4.2e-04}$ & $-0.44^{+1.6e-04}_{-1.4e-06}$ & $0.33214^{+1.6e-05}_{-1.5e-05}$ & $1.22^{+4.0e-02}_{-4.1e-02}$ & $19.26^{+4.8e-02}_{-5.8e-02}$ & $-1.34^{+2.6e-01}_{-2.5e-01}$ & $2.13^{+3.1e-02}_{-3.0e-02}$ & $-0.32^{+2.0e-02}_{-1.9e-02}$ \\
41 & $0.43204^{+1.4e-07}_{-6.2e-05}$ & $1.52^{+2.9e-04}_{-3.7e-01}$ & $17.0^{+2.3e-02}_{-2.3e-04}$ & $-0.05^{+4.4e-02}_{-1.4e-02}$ & $1.96^{+1.1e-02}_{-1.5e-03}$ & $-0.18^{+1.2e-02}_{-1.1e-03}$ & $0.432^{+4.2e-05}_{-7.9e-05}$ & $1.35^{+1.5e-01}_{-1.4e-01}$ & $17.25^{+2.1e-01}_{-1.6e-01}$ & $-1.35^{+3.9e-01}_{-4.6e-01}$ & $2.01^{+3.6e-02}_{-3.4e-02}$ & $-0.2^{+6.2e-02}_{-4.3e-02}$ \\
42 & $0.3533^{+1.8e-08}_{-3.4e-09}$ & $1.68^{+3.6e-06}_{-2.2e-04}$ & $18.75^{+1.3e-07}_{-2.3e-05}$ & $-0.5^{+9.0e-07}_{-5.3e-07}$ & $2.1^{+4.3e-05}_{-1.1e-03}$ & $-0.47^{+1.9e-05}_{-3.2e-04}$ & $0.35324^{+4.1e-05}_{-5.4e-05}$ & $1.57^{+4.5e-02}_{-4.5e-02}$ & $18.85^{+1.7e-01}_{-1.7e-01}$ & $-0.88^{+5.0e-01}_{-3.5e-01}$ & $2.08^{+3.2e-02}_{-4.9e-02}$ & $-0.52^{+3.9e-02}_{-3.1e-02}$ \\
43 & $0.36314^{+1.1e-07}_{-4.7e-09}$ & $1.72^{+2.7e-04}_{-1.8e-05}$ & $18.57^{+2.6e-05}_{-1.1e-04}$ & $-0.0^{+3.2e-07}_{-2.4e-04}$ & $2.59^{+9.3e-05}_{-1.4e-03}$ & $-0.1^{+1.2e-04}_{-2.6e-06}$ & $0.36317^{+6.4e-05}_{-2.6e-05}$ & $1.88^{+5.0e-02}_{-6.3e-02}$ & $18.14^{+1.3e-01}_{-1.1e-01}$ & $0.04^{+7.9e-02}_{-8.7e-02}$ & $2.17^{+1.4e-02}_{-1.4e-02}$ & $-0.07^{+3.1e-02}_{-4.8e-02}$ \\
44 & $0.1488^{+3.6e-05}_{-9.0e-05}$ & $1.08^{+1.1e-01}_{-3.2e-02}$ & $20.34^{+2.3e-01}_{-1.4e-01}$ & $-0.83^{+1.0e-01}_{-2.4e+00}$ & $1.6^{+6.5e-02}_{-6.5e-01}$ & $-0.03^{+2.6e-02}_{-8.9e-02}$ & $0.14875^{+1.3e-04}_{-8.8e-05}$ & $1.11^{+1.4e-01}_{-5.8e-02}$ & $20.27^{+1.3e-01}_{-2.3e-01}$ & $-1.79^{+6.2e-01}_{-9.8e-01}$ & $1.28^{+1.8e-01}_{-2.0e-01}$ & $-0.09^{+1.7e-01}_{-2.4e-01}$ \\
45 & $0.14112^{+4.2e-06}_{-6.6e-05}$ & $1.75^{+6.5e-03}_{-1.0e-01}$ & $18.54^{+3.3e-01}_{-4.2e-04}$ & $-0.05^{+6.3e-04}_{-2.3e+00}$ & $2.06^{+2.0e-04}_{-1.8e-01}$ & $-0.25^{+1.5e-01}_{-1.1e-04}$ & $0.14103^{+7.2e-05}_{-6.7e-05}$ & $1.68^{+1.2e-01}_{-1.6e-01}$ & $18.52^{+4.7e-01}_{-2.3e-01}$ & $-1.35^{+9.7e-01}_{-1.2e+00}$ & $2.17^{+8.8e-02}_{-3.6e-01}$ & $-0.15^{+8.2e-02}_{-1.1e-01}$ \\
46 & $0.20811^{+4.2e-08}_{-6.0e-10}$ & $1.7^{+1.3e-07}_{-8.9e-11}$ & $18.8^{+6.7e-06}_{-2.4e-04}$ & $-0.06^{+3.1e-06}_{-1.3e-04}$ & $2.4^{+3.0e-05}_{-2.8e-05}$ & $-0.85^{+1.7e-05}_{-1.4e-06}$ & $0.20804^{+7.7e-06}_{-8.0e-06}$ & $1.53^{+1.3e-02}_{-1.2e-02}$ & $19.06^{+5.4e-02}_{-5.4e-02}$ & $-1.13^{+2.1e-01}_{-1.8e-01}$ & $2.12^{+1.2e-02}_{-1.2e-02}$ & $-0.88^{+1.1e-02}_{-1.1e-02}$ \\
47 & $0.03997^{+3.5e-07}_{-3.9e-08}$ & $1.86^{+3.1e-04}_{-4.1e-05}$ & $19.3^{+3.7e-04}_{-2.3e-04}$ & $-0.0^{+1.5e-06}_{-4.8e-04}$ & $2.27^{+5.5e-04}_{-1.6e-04}$ & $-0.46^{+1.3e-04}_{-1.0e-04}$ & $0.04001^{+9.5e-06}_{-1.7e-05}$ & $1.84^{+3.6e-02}_{-2.0e-02}$ & $19.21^{+5.6e-02}_{-3.0e-02}$ & $-0.26^{+1.2e-01}_{-7.0e-02}$ & $2.14^{+6.0e-02}_{-3.8e-02}$ & $-0.55^{+9.0e-03}_{-9.0e-03}$ \\
48 & $0.38269^{+5.8e-10}_{-2.0e-08}$ & $2.01^{+4.0e-06}_{-5.2e-05}$ & $19.09^{+5.7e-05}_{-2.5e-06}$ & $-3.0^{+1.3e-03}_{-2.5e-06}$ & $1.69^{+1.5e-04}_{-1.7e-05}$ & $-0.26^{+7.2e-05}_{-7.7e-06}$ & $0.38277^{+3.2e-04}_{-3.3e-04}$ & $2.17^{+1.9e-01}_{-1.9e-01}$ & $19.06^{+4.7e-01}_{-6.4e-01}$ & $-1.85^{+9.5e-01}_{-1.4e+00}$ & $1.82^{+1.5e-01}_{-1.3e-01}$ & $-0.34^{+1.8e-01}_{-2.0e-01}$ \\
49 & $0.26721^{+1.3e-04}_{-2.0e-04}$ & $1.84^{+8.8e-01}_{-1.4e-01}$ & $18.67^{+1.6e-01}_{-3.1e-01}$ & $-1.49^{+1.1e+00}_{-4.0e-01}$ & $2.02^{+9.3e-02}_{-4.8e-02}$ & $-1.3^{+3.1e-01}_{-1.8e-01}$ & $0.26731^{+1.3e-04}_{-2.6e-04}$ & $2.29^{+4.0e-01}_{-3.8e-01}$ & $18.43^{+8.3e-01}_{-9.1e-01}$ & $-1.33^{+1.1e+00}_{-9.5e-01}$ & $1.99^{+1.4e-01}_{-1.8e-01}$ & $-0.85^{+1.2e-01}_{-1.1e-01}$ \\
50 & $0.08227^{+8.2e-08}_{-2.3e-09}$ & $1.51^{+2.7e-06}_{-4.9e-05}$ & $19.0^{+1.0e-07}_{-1.0e-09}$ & $-1.47^{+2.4e-03}_{-8.6e-06}$ & $2.34^{+2.7e-04}_{-7.6e-06}$ & $-0.11^{+2.4e-06}_{-3.8e-04}$ & $0.08224^{+3.7e-06}_{-3.0e-06}$ & $1.4^{+1.4e-02}_{-1.1e-02}$ & $19.22^{+1.2e-02}_{-1.6e-02}$ & $-1.1^{+1.2e-01}_{-9.3e-02}$ & $2.21^{+5.0e-03}_{-4.6e-03}$ & $-0.27^{+9.0e-03}_{-7.4e-03}$ \\
51 & $0.3438^{+8.6e-08}_{-2.7e-07}$ & $1.3^{+1.1e-03}_{-1.2e-02}$ & $17.5^{+9.3e-05}_{-2.8e-03}$ & $-3.5^{+1.6e-02}_{-3.2e-04}$ & $1.62^{+6.5e-02}_{-2.0e-03}$ & $-0.07^{+3.2e-02}_{-8.3e-04}$ & $0.3435^{+5.8e-05}_{-6.6e-05}$ & $1.17^{+8.1e-02}_{-4.8e-02}$ & $19.16^{+2.5e-01}_{-1.7e-01}$ & $-3.77^{+1.4e+00}_{-4.6e-01}$ & $1.61^{+1.5e-01}_{-1.3e-01}$ & $0.22^{+1.3e-01}_{-1.4e-01}$ \\
52 & $0.04436^{+2.4e-10}_{-1.3e-07}$ & $2.08^{+1.1e-06}_{-1.1e-04}$ & $18.56^{+1.9e-04}_{-2.0e-06}$ & $-1.0^{+4.7e-08}_{-1.0e-04}$ & $1.49^{+2.3e-04}_{-3.7e-06}$ & $-1.21^{+2.7e-04}_{-6.7e-06}$ & $0.04429^{+7.9e-06}_{-6.0e-06}$ & $2.03^{+1.0e-02}_{-9.5e-03}$ & $18.24^{+3.8e-02}_{-4.7e-02}$ & $-0.23^{+4.5e-02}_{-3.4e-02}$ & $1.64^{+1.2e-02}_{-1.2e-02}$ & $-0.82^{+5.6e-03}_{-5.9e-03}$ \\
53 & $0.24031^{+4.7e-06}_{-1.3e-04}$ & $1.78^{+6.8e-02}_{-1.3e-02}$ & $18.44^{+7.3e-01}_{-1.1e-01}$ & $-2.0^{+1.5e+00}_{-1.5e-02}$ & $1.4^{+1.6e-01}_{-8.5e-02}$ & $-0.59^{+9.3e-02}_{-5.4e-01}$ & $0.24045^{+1.5e-02}_{-6.0e-04}$ & $1.71^{+4.0e-01}_{-4.6e-01}$ & $18.65^{+8.8e-01}_{-1.4e+00}$ & $-1.45^{+1.0e+00}_{-1.4e+00}$ & $1.05^{+3.8e-01}_{-6.1e-01}$ & $-0.42^{+5.7e-01}_{-3.5e-01}$ \\
54 & $0.26394^{+2.3e-04}_{-2.4e-05}$ & $1.55^{+1.2e+00}_{-2.8e-02}$ & $17.0^{+1.5e+00}_{-1.2e-03}$ & $-0.73^{+4.5e-02}_{-9.0e-01}$ & $1.62^{+1.8e-02}_{-1.3e-02}$ & $-0.46^{+4.4e-02}_{-4.6e-01}$ & $0.26396^{+1.3e-04}_{-2.1e-04}$ & $2.51^{+2.5e-01}_{-5.2e-01}$ & $18.2^{+8.9e-01}_{-8.8e-01}$ & $-3.0^{+1.6e+00}_{-7.6e-01}$ & $1.74^{+1.5e-01}_{-1.5e-01}$ & $-0.59^{+1.3e-01}_{-1.3e-01}$ \\
55 & $0.2035^{+4.2e-08}_{-1.9e-08}$ & $1.89^{+4.7e-06}_{-7.5e-06}$ & $18.73^{+5.6e-05}_{-1.9e-04}$ & $-2.5^{+4.8e-09}_{-1.5e-07}$ & $1.63^{+4.5e-06}_{-6.9e-06}$ & $-0.17^{+3.3e-06}_{-1.5e-05}$ & $0.20358^{+6.8e-05}_{-1.1e-04}$ & $2.0^{+4.7e-02}_{-6.2e-02}$ & $18.37^{+3.2e-01}_{-2.9e-01}$ & $-2.72^{+7.3e-01}_{-5.6e-01}$ & $1.72^{+5.5e-02}_{-5.2e-02}$ & $-0.19^{+4.6e-02}_{-4.8e-02}$ \\
56 & $0.2394^{+9.8e-07}_{-8.9e-05}$ & $1.18^{+5.2e-01}_{-4.0e-03}$ & $17.0^{+1.5e+00}_{-2.2e-03}$ & $-2.97^{+1.2e+00}_{-2.6e-02}$ & $1.55^{+5.1e-02}_{-5.4e-03}$ & $-0.5^{+2.7e-03}_{-1.0e-01}$ & $0.23929^{+2.0e-04}_{-4.7e-04}$ & $2.05^{+2.9e-01}_{-2.4e-01}$ & $18.49^{+7.9e-01}_{-7.6e-01}$ & $-1.89^{+1.2e+00}_{-1.3e+00}$ & $1.62^{+1.4e-01}_{-1.2e-01}$ & $-0.55^{+1.6e-01}_{-1.2e-01}$ \\
57 & $0.24814^{+1.8e-02}_{-4.4e-06}$ & $0.92^{+5.2e-01}_{-1.2e-02}$ & $17.1^{+7.5e-01}_{-1.0e-01}$ & $-2.04^{+2.3e-01}_{-6.4e-01}$ & $0.9^{+1.6e-02}_{-1.4e-01}$ & $-0.4^{+8.0e-03}_{-4.7e-01}$ & $0.24729^{+8.4e-04}_{-9.4e-04}$ & $2.07^{+3.5e-01}_{-3.6e-01}$ & $19.33^{+8.8e-01}_{-1.3e+00}$ & $-1.5^{+8.9e-01}_{-1.2e+00}$ & $1.09^{+2.9e-01}_{-2.5e-01}$ & $-0.39^{+3.1e-01}_{-2.4e-01}$ \\
58 & $0.19494^{+9.5e-04}_{-5.1e-05}$ & $2.18^{+2.5e-04}_{-1.5e+00}$ & $19.25^{+4.5e-02}_{-1.1e+00}$ & $-1.1^{+2.6e-01}_{-3.3e-01}$ & $1.48^{+1.8e-02}_{-1.3e-01}$ & $-0.01^{+9.7e-03}_{-2.7e-01}$ & $0.19534^{+3.3e-04}_{-3.7e-04}$ & $2.27^{+2.4e-01}_{-5.3e-01}$ & $18.5^{+4.9e-01}_{-5.4e-01}$ & $-0.89^{+6.7e-01}_{-1.4e+00}$ & $1.38^{+1.5e-01}_{-1.2e-01}$ & $-0.18^{+1.7e-01}_{-1.5e-01}$ \\
59 & $0.09199^{+1.3e-08}_{-9.6e-08}$ & $1.76^{+3.9e-05}_{-2.9e-04}$ & $19.25^{+2.0e-07}_{-7.1e-06}$ & $-0.0^{+5.2e-07}_{-3.1e-05}$ & $2.13^{+9.4e-05}_{-3.7e-05}$ & $-0.37^{+2.2e-05}_{-2.6e-04}$ & $0.09199^{+8.3e-06}_{-9.2e-06}$ & $1.7^{+7.6e-03}_{-8.0e-03}$ & $19.23^{+2.9e-02}_{-2.6e-02}$ & $-0.29^{+1.0e-01}_{-1.0e-01}$ & $2.12^{+4.5e-02}_{-4.7e-02}$ & $-0.29^{+7.4e-03}_{-7.5e-03}$ \\
60 & $0.26901^{+1.4e-06}_{-1.6e-04}$ & $2.06^{+1.6e-03}_{-5.8e-02}$ & $18.99^{+1.0e-01}_{-4.4e-02}$ & $-2.5^{+1.6e+00}_{-6.7e-02}$ & $1.45^{+1.7e-01}_{-1.3e-02}$ & $-1.66^{+9.8e-01}_{-1.7e-01}$ & $0.26931^{+1.9e-04}_{-4.4e-04}$ & $2.43^{+1.5e-01}_{-2.0e-01}$ & $18.85^{+5.2e-01}_{-5.2e-01}$ & $-1.99^{+1.4e+00}_{-1.4e+00}$ & $1.65^{+1.1e-01}_{-1.2e-01}$ & $-0.58^{+1.4e-01}_{-8.8e-02}$ \\
61 & $0.32333^{+3.0e-04}_{-1.7e-06}$ & $2.18^{+7.2e-04}_{-1.6e-01}$ & $19.75^{+1.2e-03}_{-6.1e-01}$ & $-0.01^{+8.8e-03}_{-1.9e+00}$ & $2.02^{+4.0e-03}_{-5.3e-01}$ & $-0.52^{+1.9e-02}_{-1.3e-01}$ & $0.32345^{+4.3e-04}_{-6.0e-04}$ & $2.2^{+2.7e-01}_{-3.1e-01}$ & $19.36^{+7.2e-01}_{-5.2e-01}$ & $-2.3^{+1.4e+00}_{-1.2e+00}$ & $1.44^{+1.6e-01}_{-1.3e-01}$ & $-0.4^{+2.5e-01}_{-1.9e-01}$ \\
62 & $0.08194^{+2.5e-10}_{-1.1e-07}$ & $2.11^{+8.1e-07}_{-1.4e-05}$ & $18.85^{+2.9e-04}_{-2.4e-06}$ & $-0.37^{+3.0e-04}_{-7.9e-06}$ & $2.2^{+2.3e-04}_{-5.7e-07}$ & $-1.04^{+1.1e-05}_{-1.6e-06}$ & $0.08181^{+4.4e-05}_{-1.6e-05}$ & $2.09^{+3.1e-02}_{-7.3e-03}$ & $19.23^{+5.9e-02}_{-1.7e-01}$ & $-0.28^{+1.6e-01}_{-7.4e-02}$ & $2.14^{+3.4e-02}_{-2.9e-02}$ & $-0.8^{+1.0e-02}_{-1.5e-02}$ \\
63 & $0.03215^{+1.8e-05}_{-1.0e-05}$ & $1.92^{+7.7e-02}_{-3.9e-02}$ & $19.57^{+3.1e-02}_{-6.8e-02}$ & $-0.05^{+3.6e-02}_{-4.7e-01}$ & $1.29^{+7.2e-02}_{-1.3e-01}$ & $-0.87^{+7.0e-02}_{-1.3e-01}$ & $0.03185^{+4.2e-04}_{-2.8e-04}$ & $1.83^{+3.6e-01}_{-2.4e-01}$ & $20.0^{+3.0e-01}_{-5.6e-01}$ & $-1.12^{+6.3e-01}_{-9.0e-01}$ & $0.93^{+4.1e-01}_{-3.5e-01}$ & $-0.62^{+2.1e-01}_{-2.3e-01}$ \\
64 & $0.09439^{+2.8e-07}_{-2.3e-05}$ & $2.02^{+1.9e-04}_{-3.0e-02}$ & $18.57^{+1.0e-01}_{-7.7e-04}$ & $-3.0^{+2.2e+00}_{-8.9e-04}$ & $1.94^{+4.7e-02}_{-3.5e-03}$ & $-0.15^{+3.6e-04}_{-2.6e-02}$ & $0.09424^{+3.7e-05}_{-3.0e-05}$ & $1.97^{+2.0e-02}_{-2.1e-02}$ & $18.97^{+6.5e-02}_{-5.9e-02}$ & $-2.99^{+1.1e-01}_{-8.7e-02}$ & $2.06^{+1.1e-02}_{-9.6e-03}$ & $-0.18^{+1.3e-02}_{-1.0e-02}$ \\
65 & $0.02813^{+1.7e-06}_{-3.5e-04}$ & $2.2^{+5.7e-04}_{-2.0e-01}$ & $19.16^{+4.0e-01}_{-8.0e-03}$ & $-0.41^{+2.4e-01}_{-9.2e-03}$ & $1.26^{+3.1e-01}_{-4.1e-03}$ & $-0.93^{+5.2e-01}_{-6.1e-03}$ & $0.02817^{+8.1e-05}_{-1.4e-04}$ & $2.29^{+3.7e-02}_{-4.6e-02}$ & $18.98^{+1.4e-01}_{-6.1e-02}$ & $-1.18^{+3.3e-01}_{-2.5e-01}$ & $1.08^{+4.3e-02}_{-3.4e-02}$ & $-0.71^{+4.9e-02}_{-4.0e-02}$ \\
66 & $0.04313^{+2.2e-07}_{-3.6e-07}$ & $1.57^{+6.9e-03}_{-5.7e-04}$ & $18.39^{+5.8e-04}_{-5.5e-03}$ & $-3.5^{+7.7e-05}_{-4.5e-02}$ & $0.69^{+1.6e-02}_{-2.2e-04}$ & $-1.3^{+1.1e-05}_{-6.0e-04}$ & $0.04301^{+3.0e-04}_{-2.2e-04}$ & $1.85^{+2.7e-01}_{-3.4e-01}$ & $18.85^{+6.0e-01}_{-6.7e-01}$ & $-0.95^{+7.0e-01}_{-9.3e-01}$ & $1.14^{+2.4e-01}_{-2.3e-01}$ & $-0.5^{+2.7e-01}_{-2.0e-01}$ \\
\end{tabular}}
\end{table}
\end{landscape}

\renewcommand{\arraystretch}{1.5}
\begin{landscape}
\begin{table}
\caption{ Parameters associated with the line profiles. The column {\it Label} indicates the number that appears with the spectra in Fig.\ref{fig:examples_observed}, \ref{fig:comparison_NN_MC_lines_0} and \ref{fig:comparison_NN_MC_lines_1}.  }
\label{tab:real_mcmc_2}
\resizebox{9.5in}{!}{%
\begin{tabular}{ccccccccccccc}
& \multicolumn{6}{c}{MCMC} & \multicolumn{6}{c}{DNN}  \\ \cmidrule(r){2-7} \cmidrule(l){8-13}
Label &  $z^{\rm Zelda}$ & $\log$\vexp & $\log$\nh & $\log$\ta & $\log$\ew & $\rm \log$\w & $z^{\rm Zelda}$ & $\log$\vexp & $\log$\nh & $\log$\ta & $\log$\ew & $\rm \log$\w   \\
 & & [$\rm km/s$] & [$\rm cm^{-2}$] & & [\AA{}] & [\AA{}] & & [$\rm km/s$] & [$\rm cm^{-2}$] & & [\AA{}] & [\AA{}] \\ \hline
67 & $0.08126^{+5.1e-08}_{-2.8e-10}$ & $1.91^{+1.9e-05}_{-1.4e-06}$ & $18.51^{+4.0e-06}_{-1.5e-04}$ & $-2.5^{+2.3e-06}_{-4.7e-04}$ & $1.91^{+2.6e-05}_{-1.5e-05}$ & $-0.33^{+4.0e-06}_{-9.8e-05}$ & $0.0812^{+9.4e-05}_{-1.7e-05}$ & $1.9^{+7.5e-02}_{-2.1e-02}$ & $18.59^{+8.6e-02}_{-3.2e-01}$ & $-2.2^{+2.7e-01}_{-1.8e-01}$ & $2.0^{+1.4e-02}_{-1.3e-02}$ & $-0.43^{+9.8e-03}_{-9.4e-03}$ \\
68 & $0.0324^{+1.8e-07}_{-1.3e-06}$ & $1.25^{+1.9e-02}_{-1.4e-03}$ & $20.12^{+2.7e-03}_{-1.0e-02}$ & $-0.73^{+1.1e-03}_{-1.2e-02}$ & $2.41^{+5.7e-03}_{-5.9e-02}$ & $-1.96^{+9.0e-01}_{-4.3e-02}$ & $0.0325^{+1.5e-05}_{-3.2e-05}$ & $1.6^{+5.6e-02}_{-1.1e-01}$ & $19.5^{+1.1e-01}_{-4.9e-02}$ & $-1.5^{+4.2e-01}_{-1.8e-01}$ & $1.66^{+7.2e-02}_{-5.5e-02}$ & $-0.48^{+1.1e-02}_{-1.1e-02}$ \\
69 & $0.03277^{+1.4e-08}_{-2.6e-07}$ & $1.54^{+5.3e-05}_{-4.4e-04}$ & $18.5^{+1.9e-06}_{-1.9e-04}$ & $-0.0^{+1.3e-05}_{-1.4e-03}$ & $2.7^{+4.6e-04}_{-5.9e-03}$ & $-0.86^{+3.8e-05}_{-2.5e-04}$ & $0.03267^{+1.1e-06}_{-1.1e-06}$ & $1.39^{+2.6e-03}_{-2.4e-03}$ & $18.82^{+6.2e-03}_{-6.2e-03}$ & $-1.65^{+1.5e-02}_{-4.0e-02}$ & $2.13^{+7.5e-04}_{-6.1e-04}$ & $-0.86^{+2.0e-03}_{-2.2e-03}$ \\
70 & $0.04731^{+2.1e-07}_{-3.1e-09}$ & $1.92^{+2.3e-04}_{-3.8e-06}$ & $19.25^{+2.0e-05}_{-1.6e-07}$ & $-0.43^{+9.4e-04}_{-6.4e-05}$ & $2.14^{+2.5e-04}_{-2.3e-05}$ & $-0.24^{+7.4e-06}_{-3.3e-04}$ & $0.04739^{+2.7e-05}_{-6.2e-06}$ & $1.95^{+1.7e-02}_{-8.3e-03}$ & $18.92^{+2.2e-02}_{-1.2e-01}$ & $-0.55^{+7.1e-02}_{-2.2e-01}$ & $1.99^{+8.5e-03}_{-1.8e-02}$ & $-0.36^{+6.4e-03}_{-7.0e-03}$ \\
71 & $0.03321^{+4.6e-05}_{-3.3e-06}$ & $1.6^{+2.0e-01}_{-5.0e-02}$ & $18.85^{+9.3e-03}_{-3.2e-01}$ & $-0.88^{+1.6e-01}_{-1.8e+00}$ & $1.92^{+2.4e-02}_{-9.8e-02}$ & $-0.14^{+5.3e-02}_{-3.8e-02}$ & $0.03318^{+8.3e-06}_{-8.3e-05}$ & $1.66^{+2.5e-02}_{-2.2e-02}$ & $18.75^{+1.9e-01}_{-5.7e-02}$ & $-2.65^{+2.0e-01}_{-1.6e-01}$ & $2.06^{+1.1e-02}_{-1.3e-02}$ & $-0.11^{+8.7e-03}_{-4.4e-02}$ \\
72 & $0.21646^{+4.9e-09}_{-6.7e-08}$ & $1.6^{+7.9e-05}_{-6.8e-07}$ & $19.5^{+3.3e-06}_{-1.3e-08}$ & $-1.0^{+7.3e-08}_{-1.7e-05}$ & $1.85^{+3.0e-05}_{-3.7e-04}$ & $0.27^{+5.1e-05}_{-4.2e-04}$ & $0.21647^{+1.7e-05}_{-2.0e-05}$ & $1.74^{+6.2e-02}_{-5.5e-02}$ & $19.25^{+7.6e-02}_{-5.2e-02}$ & $-0.47^{+6.7e-02}_{-1.2e-01}$ & $1.93^{+4.1e-02}_{-3.8e-02}$ & $0.2^{+3.1e-02}_{-4.0e-02}$ \\
73 & $0.26491^{+6.8e-05}_{-5.0e-05}$ & $1.86^{+9.7e-02}_{-1.8e-01}$ & $18.75^{+3.5e-02}_{-9.7e-03}$ & $-2.0^{+1.8e-02}_{-2.4e-02}$ & $1.37^{+4.8e-02}_{-9.6e-03}$ & $0.03^{+4.5e-02}_{-5.3e-02}$ & $0.26514^{+4.3e-04}_{-7.0e-04}$ & $1.88^{+4.5e-01}_{-4.3e-01}$ & $19.21^{+8.4e-01}_{-9.4e-01}$ & $-1.92^{+9.4e-01}_{-1.0e+00}$ & $2.04^{+3.1e-01}_{-8.0e-01}$ & $-0.29^{+3.5e-01}_{-2.5e-01}$ \\
74 & $0.30045^{+3.3e-07}_{-2.8e-08}$ & $1.78^{+7.3e-06}_{-7.7e-03}$ & $18.75^{+1.6e-05}_{-1.3e-06}$ & $-2.5^{+5.0e-01}_{-6.7e-04}$ & $1.46^{+3.9e-03}_{-1.2e-04}$ & $0.03^{+8.3e-05}_{-4.3e-03}$ & $0.30046^{+1.4e-05}_{-2.1e-05}$ & $1.94^{+2.9e-02}_{-4.6e-02}$ & $18.5^{+8.5e-02}_{-7.0e-02}$ & $-1.4^{+3.8e-01}_{-6.7e-01}$ & $1.55^{+6.4e-02}_{-4.8e-02}$ & $0.0^{+2.9e-02}_{-3.9e-02}$ \\
75 & $0.22224^{+3.5e-09}_{-1.6e-08}$ & $1.95^{+1.3e-06}_{-2.5e-09}$ & $18.74^{+1.4e-05}_{-1.3e-04}$ & $-0.03^{+4.9e-05}_{-1.8e-03}$ & $2.01^{+1.8e-05}_{-4.9e-04}$ & $-0.36^{+8.6e-06}_{-2.4e-05}$ & $0.22217^{+3.3e-05}_{-1.8e-05}$ & $2.01^{+2.8e-02}_{-3.2e-02}$ & $18.56^{+7.1e-02}_{-7.0e-02}$ & $-0.5^{+1.6e-01}_{-3.1e-01}$ & $1.92^{+3.1e-02}_{-3.5e-02}$ & $-0.23^{+2.7e-02}_{-4.4e-02}$ \\
76 & $0.2899^{+4.3e-04}_{-8.8e-06}$ & $0.84^{+1.4e-01}_{-9.4e-01}$ & $17.25^{+1.3e-01}_{-2.3e-01}$ & $-2.63^{+6.2e-01}_{-3.7e-01}$ & $1.31^{+1.0e-01}_{-2.4e-01}$ & $-0.13^{+5.9e-02}_{-3.2e-01}$ & $0.28997^{+3.2e-04}_{-1.1e-03}$ & $1.43^{+4.3e-01}_{-2.2e-01}$ & $18.41^{+8.9e-01}_{-8.0e-01}$ & $-1.09^{+4.8e-01}_{-9.5e-01}$ & $0.78^{+2.6e-01}_{-2.2e-01}$ & $-0.4^{+2.6e-01}_{-2.2e-01}$ \\
77 & $0.237^{+2.7e-05}_{-2.7e-06}$ & $1.6^{+6.5e-03}_{-2.8e-02}$ & $19.75^{+6.3e-05}_{-1.3e-03}$ & $-0.46^{+4.5e-02}_{-2.4e+00}$ & $1.97^{+3.5e-02}_{-4.5e-01}$ & $-0.24^{+1.6e-01}_{-1.2e-03}$ & $0.2368^{+7.3e-05}_{-7.4e-05}$ & $1.67^{+8.9e-02}_{-9.6e-02}$ & $19.74^{+1.4e-01}_{-1.6e-01}$ & $-2.98^{+9.2e-01}_{-6.5e-01}$ & $1.5^{+6.1e-02}_{-5.5e-02}$ & $0.04^{+4.8e-02}_{-6.2e-02}$ \\
78 & $0.20493^{+4.1e-09}_{-2.4e-07}$ & $1.68^{+8.1e-05}_{-2.1e-05}$ & $18.36^{+9.8e-04}_{-4.1e-05}$ & $-0.0^{+1.2e-07}_{-4.7e-05}$ & $1.86^{+7.0e-04}_{-4.3e-05}$ & $-0.1^{+7.0e-05}_{-1.9e-05}$ & $0.20492^{+1.1e-05}_{-2.1e-05}$ & $1.48^{+3.4e-02}_{-4.0e-02}$ & $18.32^{+1.3e-01}_{-8.7e-02}$ & $0.05^{+4.5e-02}_{-8.7e-02}$ & $2.1^{+1.1e-02}_{-1.9e-02}$ & $-0.28^{+3.1e-02}_{-2.2e-02}$ \\
79 & $0.19218^{+1.4e-08}_{-5.1e-10}$ & $2.15^{+4.5e-07}_{-4.7e-05}$ & $19.13^{+1.7e-06}_{-5.1e-05}$ & $-0.49^{+1.9e-04}_{-7.5e-06}$ & $1.7^{+4.3e-05}_{-2.0e-06}$ & $0.34^{+5.6e-09}_{-4.2e-06}$ & $0.19209^{+1.8e-04}_{-1.0e-04}$ & $2.14^{+1.1e-01}_{-1.4e-01}$ & $19.23^{+1.9e-01}_{-3.1e-01}$ & $-0.65^{+2.9e-01}_{-7.0e-01}$ & $1.6^{+5.9e-02}_{-6.2e-02}$ & $0.36^{+5.2e-02}_{-3.9e-02}$ \\
80 & $0.30088^{+2.9e-05}_{-2.2e-07}$ & $2.35^{+1.6e-02}_{-4.9e-03}$ & $18.93^{+1.4e-03}_{-3.1e-02}$ & $-2.0^{+4.2e-04}_{-2.0e+00}$ & $1.47^{+5.1e-04}_{-1.3e-02}$ & $-0.85^{+9.9e-03}_{-1.3e-03}$ & $0.3008^{+7.0e-05}_{-6.6e-05}$ & $2.43^{+3.4e-02}_{-2.8e-02}$ & $19.28^{+4.9e-02}_{-1.5e-01}$ & $-1.15^{+4.5e-01}_{-9.7e-01}$ & $1.42^{+3.9e-02}_{-3.3e-02}$ & $-0.61^{+4.1e-02}_{-7.2e-02}$ \\
81 & $0.33275^{+2.3e-06}_{-8.5e-08}$ & $1.85^{+8.8e-05}_{-3.7e-06}$ & $18.78^{+4.8e-04}_{-1.4e-02}$ & $-2.0^{+3.5e-04}_{-1.2e+00}$ & $1.72^{+2.1e-04}_{-3.5e-03}$ & $0.06^{+7.4e-03}_{-9.4e-04}$ & $0.33271^{+7.1e-05}_{-9.6e-05}$ & $2.06^{+5.2e-02}_{-7.3e-02}$ & $18.7^{+1.9e-01}_{-1.4e-01}$ & $-1.57^{+5.9e-01}_{-1.2e+00}$ & $1.71^{+4.9e-02}_{-4.4e-02}$ & $0.01^{+4.8e-02}_{-5.9e-02}$ \\
82 & $0.26927^{+7.6e-08}_{-1.6e-08}$ & $1.08^{+1.0e-03}_{-3.8e-05}$ & $20.0^{+3.5e-05}_{-8.1e-08}$ & $-0.5^{+2.9e-07}_{-8.5e-05}$ & $2.17^{+3.1e-04}_{-4.6e-05}$ & $-0.08^{+9.3e-04}_{-2.4e-05}$ & $0.26919^{+6.8e-05}_{-7.5e-05}$ & $1.13^{+7.0e-02}_{-5.1e-02}$ & $19.9^{+1.3e-01}_{-1.5e-01}$ & $-1.34^{+5.7e-01}_{-3.9e-01}$ & $1.65^{+1.6e-01}_{-1.2e-01}$ & $0.06^{+6.3e-02}_{-6.7e-02}$ \\
83 & $0.29686^{+6.1e-06}_{-1.2e-07}$ & $1.78^{+3.8e-05}_{-1.9e-03}$ & $19.74^{+4.1e-04}_{-1.4e-03}$ & $-0.0^{+1.4e-04}_{-1.4e-01}$ & $2.58^{+2.8e-04}_{-2.8e-01}$ & $-0.38^{+1.9e-04}_{-1.8e-03}$ & $0.2968^{+1.3e-04}_{-4.8e-05}$ & $1.88^{+5.9e-02}_{-5.1e-02}$ & $19.61^{+7.5e-02}_{-1.7e-01}$ & $-0.92^{+7.9e-02}_{-3.1e-01}$ & $2.01^{+7.1e-02}_{-8.3e-02}$ & $-0.22^{+8.0e-02}_{-3.9e-02}$ \\
84 & $0.29329^{+7.5e-04}_{-4.9e-06}$ & $2.16^{+3.4e-01}_{-1.3e-02}$ & $19.75^{+2.5e-03}_{-4.5e-01}$ & $-0.0^{+3.1e-03}_{-9.6e-01}$ & $1.65^{+8.4e-03}_{-4.9e-01}$ & $-0.52^{+1.9e-04}_{-3.6e-01}$ & $0.29344^{+2.9e-04}_{-3.5e-04}$ & $2.39^{+9.6e-02}_{-1.8e-01}$ & $19.78^{+2.4e-01}_{-2.0e-01}$ & $-0.61^{+3.3e-01}_{-5.1e-01}$ & $1.2^{+1.1e-01}_{-1.2e-01}$ & $-0.42^{+1.3e-01}_{-1.8e-01}$ \\
85 & $0.03804^{+9.3e-08}_{-1.2e-08}$ & $2.24^{+1.0e-05}_{-8.3e-05}$ & $19.47^{+3.2e-05}_{-3.1e-04}$ & $-0.0^{+2.0e-07}_{-1.7e-04}$ & $1.68^{+1.9e-05}_{-2.0e-04}$ & $-0.78^{+6.4e-05}_{-7.5e-06}$ & $0.03819^{+1.8e-05}_{-2.3e-05}$ & $2.37^{+2.7e-02}_{-1.0e-01}$ & $19.21^{+3.5e-02}_{-6.4e-02}$ & $-0.52^{+1.4e-01}_{-2.9e-01}$ & $1.43^{+2.4e-02}_{-2.8e-02}$ & $-0.81^{+1.6e-02}_{-3.3e-02}$ \\
86 & $0.03203^{+5.1e-05}_{-8.3e-05}$ & $1.49^{+1.1e-01}_{-1.7e-01}$ & $21.12^{+1.9e-01}_{-2.5e-02}$ & $-0.27^{+1.6e-01}_{-3.0e-01}$ & $2.54^{+1.9e-01}_{-1.5e-01}$ & $-0.02^{+4.7e-02}_{-1.6e-01}$ & $0.03206^{+1.6e-04}_{-1.2e-04}$ & $1.48^{+1.6e-01}_{-1.4e-01}$ & $20.98^{+1.5e-01}_{-2.1e-01}$ & $-0.71^{+3.2e-01}_{-2.6e-01}$ & $1.76^{+3.8e-01}_{-2.7e-01}$ & $-0.3^{+2.8e-01}_{-3.7e-01}$ \\
87 & $0.03018^{+5.6e-08}_{-7.0e-07}$ & $2.36^{+2.4e-05}_{-2.1e-04}$ & $18.06^{+2.9e-03}_{-2.0e-04}$ & $-2.1^{+1.6e-02}_{-1.0e-03}$ & $1.58^{+3.3e-04}_{-1.1e-05}$ & $-0.7^{+5.3e-05}_{-4.4e-08}$ & $0.03008^{+4.8e-05}_{-1.8e-04}$ & $2.31^{+3.1e-02}_{-1.0e-01}$ & $18.53^{+3.4e-01}_{-2.1e-01}$ & $-1.24^{+2.7e-01}_{-3.1e-01}$ & $1.66^{+4.5e-02}_{-3.4e-02}$ & $-0.71^{+2.2e-02}_{-2.4e-02}$ \\
88 & $0.03045^{+3.5e-07}_{-1.4e-04}$ & $1.78^{+3.2e-04}_{-2.7e-01}$ & $20.57^{+2.4e-01}_{-3.4e-04}$ & $-0.17^{+2.6e-04}_{-3.9e-01}$ & $2.1^{+9.6e-04}_{-1.1e-01}$ & $-0.52^{+1.9e-02}_{-1.0e-01}$ & $0.03041^{+1.0e-04}_{-1.6e-04}$ & $1.64^{+1.6e-01}_{-2.1e-01}$ & $20.55^{+1.1e-01}_{-1.2e-01}$ & $-0.46^{+1.9e-01}_{-2.1e-01}$ & $1.77^{+2.5e-01}_{-1.7e-01}$ & $-0.63^{+1.5e-01}_{-1.4e-01}$ \\
89 & $0.03751^{+4.6e-06}_{-8.9e-08}$ & $1.85^{+3.5e-05}_{-7.8e-04}$ & $20.25^{+9.4e-05}_{-6.5e-05}$ & $-0.04^{+2.0e-04}_{-2.6e-03}$ & $1.96^{+2.1e-03}_{-7.0e-04}$ & $-1.3^{+8.5e-04}_{-2.8e-02}$ & $0.03759^{+1.6e-04}_{-1.7e-04}$ & $1.96^{+1.3e-01}_{-1.5e-01}$ & $20.25^{+2.7e-01}_{-1.9e-01}$ & $-0.22^{+1.2e-01}_{-2.6e-01}$ & $1.68^{+1.9e-01}_{-2.1e-01}$ & $-0.63^{+1.1e-01}_{-1.2e-01}$ \\
90 & $0.08488^{+8.7e-06}_{-1.2e-03}$ & $2.74^{+4.4e-03}_{-7.8e-01}$ & $19.68^{+7.9e-02}_{-5.7e-01}$ & $-0.09^{+8.4e-02}_{-2.6e+00}$ & $0.91^{+6.6e-01}_{-2.7e-02}$ & $-1.07^{+5.5e-01}_{-1.9e-02}$ & $0.08471^{+1.9e-04}_{-1.1e-03}$ & $2.61^{+1.3e-01}_{-2.2e-01}$ & $19.47^{+5.7e-01}_{-3.0e-01}$ & $-2.58^{+1.9e+00}_{-1.0e+00}$ & $1.35^{+5.5e-01}_{-3.1e-01}$ & $-0.59^{+1.8e-01}_{-1.4e-01}$ \\
91 & $0.28123^{+1.2e-07}_{-4.3e-08}$ & $1.85^{+2.4e-06}_{-9.8e-05}$ & $19.25^{+7.7e-06}_{-4.1e-04}$ & $-0.0^{+5.0e-06}_{-2.6e-04}$ & $1.75^{+2.0e-04}_{-1.6e-04}$ & $-0.56^{+8.1e-05}_{-3.5e-04}$ & $0.28131^{+1.8e-04}_{-2.0e-04}$ & $2.01^{+1.4e-01}_{-1.7e-01}$ & $19.22^{+2.1e-01}_{-2.3e-01}$ & $-0.22^{+1.6e-01}_{-4.0e-01}$ & $1.54^{+2.0e-01}_{-1.4e-01}$ & $-0.56^{+1.9e-01}_{-1.6e-01}$ \\
92 & $0.03307^{+7.6e-06}_{-3.6e-05}$ & $1.73^{+2.6e-02}_{-1.3e-02}$ & $20.64^{+3.2e-02}_{-3.0e-02}$ & $-0.12^{+2.0e-02}_{-4.2e-02}$ & $2.67^{+1.5e-01}_{-2.2e-02}$ & $-0.6^{+1.4e-01}_{-1.4e+00}$ & $0.03344^{+1.0e-04}_{-8.8e-05}$ & $2.11^{+2.1e-02}_{-2.6e-01}$ & $19.94^{+1.4e-01}_{-1.1e-01}$ & $-0.04^{+1.3e-02}_{-1.5e-01}$ & $1.8^{+9.1e-02}_{-6.1e-02}$ & $-0.59^{+1.5e-01}_{-8.4e-02}$ \\
93 & $0.24833^{+3.3e-07}_{-7.9e-05}$ & $1.44^{+1.2e-01}_{-1.3e-03}$ & $17.01^{+1.0e+00}_{-6.0e-03}$ & $-3.0^{+1.3e-01}_{-2.0e-03}$ & $1.17^{+1.3e-02}_{-1.2e-03}$ & $-0.55^{+9.4e-04}_{-3.4e-02}$ & $0.24813^{+1.3e-04}_{-2.7e-04}$ & $1.8^{+2.2e-01}_{-1.8e-01}$ & $18.01^{+7.5e-01}_{-5.9e-01}$ & $-2.9^{+1.1e+00}_{-5.5e-01}$ & $1.1^{+8.5e-02}_{-7.3e-02}$ & $-0.44^{+1.8e-01}_{-1.2e-01}$ \\
94 & $0.21511^{+7.0e-08}_{-2.1e-06}$ & $1.32^{+1.3e-02}_{-4.9e-05}$ & $20.55^{+1.1e-04}_{-8.8e-03}$ & $-1.03^{+3.9e-04}_{-8.5e-03}$ & $2.17^{+8.5e-05}_{-2.3e-02}$ & $-0.7^{+7.3e-02}_{-4.6e-04}$ & $0.21506^{+2.0e-04}_{-9.4e-05}$ & $1.55^{+3.0e-01}_{-1.5e-01}$ & $20.31^{+1.1e-01}_{-4.1e-01}$ & $-2.5^{+1.0e+00}_{-8.4e-01}$ & $1.53^{+1.5e-01}_{-1.1e-01}$ & $-0.42^{+2.1e-01}_{-2.6e-01}$ \\
95 & $0.28026^{+1.6e-05}_{-8.1e-05}$ & $2.78^{+2.0e-01}_{-7.9e-02}$ & $17.75^{+2.5e-01}_{-2.6e-01}$ & $-1.48^{+4.8e-01}_{-1.5e+00}$ & $2.08^{+2.7e-02}_{-3.4e-02}$ & $-0.35^{+1.4e-02}_{-6.5e-03}$ & $0.28017^{+9.4e-05}_{-9.9e-05}$ & $2.75^{+6.7e-02}_{-1.0e-01}$ & $17.47^{+6.2e-01}_{-3.7e-01}$ & $-2.84^{+1.9e+00}_{-9.6e-01}$ & $2.07^{+7.1e-02}_{-8.4e-02}$ & $-0.17^{+6.7e-02}_{-7.6e-02}$ \\
96 & $0.20497^{+2.8e-08}_{-1.6e-10}$ & $2.22^{+1.4e-05}_{-6.7e-08}$ & $19.33^{+7.5e-07}_{-7.1e-05}$ & $-0.55^{+3.4e-06}_{-2.6e-04}$ & $1.63^{+2.4e-07}_{-4.2e-05}$ & $-0.8^{+3.5e-05}_{-1.9e-07}$ & $0.20513^{+9.3e-05}_{-1.6e-04}$ & $2.42^{+4.6e-02}_{-1.3e-01}$ & $19.22^{+1.4e-01}_{-1.5e-01}$ & $-0.49^{+3.5e-01}_{-5.1e-01}$ & $1.61^{+6.6e-02}_{-8.9e-02}$ & $-0.63^{+6.9e-02}_{-3.6e-02}$ \\
97 & $0.26491^{+3.1e-08}_{-4.3e-07}$ & $2.18^{+1.6e-05}_{-4.6e-07}$ & $19.26^{+8.1e-05}_{-1.2e-03}$ & $-0.53^{+2.3e-03}_{-5.7e-02}$ & $1.62^{+3.5e-04}_{-1.4e-02}$ & $0.08^{+2.0e-03}_{-1.2e-04}$ & $0.26484^{+2.4e-04}_{-2.1e-04}$ & $2.12^{+1.4e-01}_{-2.0e-01}$ & $19.33^{+3.2e-01}_{-4.7e-01}$ & $-1.99^{+8.0e-01}_{-1.2e+00}$ & $1.36^{+8.8e-02}_{-7.9e-02}$ & $0.2^{+9.1e-02}_{-1.6e-01}$ \\
\end{tabular}}
\end{table}
\end{landscape}

\bsp	
\label{lastpage}
\end{document}